\documentclass[english,aip,letter,superscriptaddress,twocolumn,floatfix,jcp]{revtex4}
\usepackage[utf8]{inputenc}
\usepackage[T1]{fontenc} 
\usepackage{graphicx}
\usepackage{color}
\usepackage{pstricks,pst-node}
\usepackage{psfrag}
\usepackage{verbatim}
\usepackage{bm}
\usepackage{lscape}
\usepackage{stmaryrd}
\usepackage{amsfonts}
\usepackage{amsmath}
\usepackage{amssymb}
\usepackage{endnotes}
\usepackage{ifthen}
\usepackage{bm}
\usepackage{setspace}
\usepackage{endnotes}
\usepackage{fancyhdr}
\usepackage{fancybox}
\usepackage{makeidx}
\usepackage{listings}
\definecolor{shadecolor}{rgb}{0,.7,0}
\definecolor{shadecolor}{gray}{0.7}
\definecolor{shadecolor}{gray}{0.95}

\newcommand{\norm}{|\!|}
\newcommand{\bra}{\llbracket}
\newcommand{\ket}{\rrbracket}
\newcommand{\llangle}{\left\langle}
\newcommand{\rrangle}{\right\rangle}
\newcommand{\esc}{\!\cdot\!}

\global\long\def\V#1{\boldsymbol{#1}}
\global\long\def\M#1{\boldsymbol{#1}}

\global\long\def\D#1{\Delta#1}

\newcommand{\Pepmodified}[1]{#1}
\newcommand{\incite}[1]{[\onlinecite{#1}]}
\newcommand\W{\boldsymbol{\mathcal{W}}} 
\newcommand\Vav {\hat{\bf v}^{\rm hydro}} 
\newcommand\vb{{\bf v}} 

\begin{document}

\title{Coupling a nano-particle with isothermal  fluctuating hydrodynamics:\\Coarse-graining from microscopic to mesoscopic dynamics}

\author{Pep  Espa\~{n}ol$^{1}$, Aleksandar Donev$^{2}$}

\affiliation{$^1$Dept.   F\'{\i}sica Fundamental, Universidad  Nacional de
  Educaci\'on  a  Distancia,  Aptdo.   60141 E-28080,  Madrid,  Spain \\}
\affiliation{$^2$Courant Institute of Mathematical Sciences, New York University \\
 251 Mercer Street, New York, NY 10012}

\date{\today}

\begin{abstract} 
  We  derive  a  coarse-grained  description  of  the  dynamics  of  a
  nanoparticle immersed in an isothermal  simple fluid by performing a
  systematic coarse  graining of the underlying  microscopic dynamics.
  As coarse-grained  or relevant variables  we select the  position of
  the  nanoparticle and  the  \emph{total} mass  and momentum  density
  field  of the  fluid,  which are  locally  conserved slow  variables
  because  they  are  defined  to  include  the  contribution  of  the
  nanoparticle.  The theory  of coarse graining based  on the Zwanzing
  projection   operator   leads  us   to   a   system  of   stochastic
  \emph{ordinary} differential  equations (SODEs)  that are  closed in
  the  relevant   variables.  
  We  demonstrate  that  our  discrete  coarse-grained  equations  are
  consistent with a Petrov-Galerkin finite-element discretization of a
  system  of formal  stochastic \emph{partial}  differential equations
  (SPDEs) which resemble previously-used phenomenological models based
  on fluctuating  hydrodynamics.  
  Key  to  this   connection  between  our
  ``bottom-up'' and previous ``top-down'' approaches is the use of the
  same dual  orthogonal set  of linear  basis functions  familiar from
  finite  element methods  (FEM), both  as a  way to  coarse-grain the
  microscopic  degrees of  freedom, and  as  a way  to discretize  the
  equations of  fluctuating hydrodynamics.  Another key  ingredient is
  the use of a ``linear  for spiky'' weak approximation which replaces
  microscopic  ``fields''   with  a   linear  FE   interpolant  inside
  expectation values.   For the irreversible or  dissipative dynamics,
  we  approximate  the  constrained  Green-Kubo  expressions  for  the
  dissipation  coefficients with  their  equilibrium averages.   Under
  suitable  approximations we  obtain \emph{closed}  approximations of
  the coarse-grained  dynamics in  a manner which  gives them  a clear
  physical  interpretation, and  provides \emph{explicit}  microscopic
  expressions for  all of the  coefficients appearing in  the closure.
  Our work leads  to a model
    for  dilute  nanocolloidal  suspensions  that  can  be  simulated
  effectively using  feasibly short molecular dynamics  simulations as
  input to a FEM fluctuating hydrodynamic solver.
\end{abstract}
\maketitle
\section{Introduction}

The study of the Brownian motion of rigid  particles suspended in a viscous solvent
is one of the oldest  subjects in nonequilibrium statistical mechanics
since        the       pioneering        work       of        Einstein
\cite{Einstein1905}. Nevertheless, it was not until the seventies that
it    was   realized    how   subtle    diffusion   in    liquids   is
\cite{Faxen_FluctuatingHydro,VACF_FluctHydro,BrownianCompressibility_Zwanzig,DiffusionRenormalization_I,DiffusionRenormalization_II,DiffusionRenormalization_III,VACF_Langevin,LangevinDynamics_Theory},
and  to this  day there  remain open  fundamental questions  about the
collective  diffusion  in  colloidal  suspensions.  For  example,  the
validity  of  Fick's macroscopic  law  is  questioned for  suspensions
confined  to  a  two dimensions  \cite{ConfinedDiffusion_2D},  and  it
remains as a substantial mathematical  challenge to prove that a local
Fickian equation is the law of large numbers in three dimensions, even
for dilute suspensions \cite{DiffusionJSTAT}.  These questions are not
of purely academic  interest since diffusion is  of crucial importance
in  a number  of applications  in chemical  engineering and  materials
science,  such as  the  study of  the dynamics  of  passive or  active
\cite{ActiveSuspensions,Nanomotors_Kapral}  particles  in  suspension,
the       dynamics       of       biomolecules       in       solution
\cite{HYDROPRO_Globular,RotationalBD_Torre},   the  design   of  novel
nanocolloidal                                              suspensions
\cite{Nanofluids_Review,Nanofluidics_Review,NanoparticlesAtInterface},
and  others.   The  importance  of coarse-graining  to  the  study  of
diffusion  in nanocolloidal  suspensions  is easy  to appreciate;  the
number of  degrees of  freedom necessary  to simulate  Brownian motion
directly using  Molecular Dynamics (MD)  is large enough to  make this
approach  prohibitively  expensive.  In  this  paper,  we derive  from
``first  principles''  a  coarse-grained   dynamic  equation  for  the
position of  a nanoparticle immersed  in a simple fluid,  fully taking
into account hydrodynamic effects.

The  key source  of difficulty  in the  theoretical and  computational
modeling of colloidal diffusion is the presence of viscous dissipation
in the surrounding fluid. This hydrodynamic dissipation in the solvent
induces long-ranged hydrodynamic fields that  couple the motion of the
solute particles to  boundaries and to other  particles. These effects
are termed hydrodynamic interactions in  the literature, but it should
be kept  in mind that  these ``interactions'' are different  in nature
from  direct  interactions such  as  steric  repulsion or  long-ranged
attractions  among  the  colloids.   The  well-known  Smoluchowski  or
Brownian                         Dynamics                         (BD)
\cite{BrownianDynamics_OrderNlogN,BrownianDynamics_OrderN}    approach
captures the effect  of the solvent through a mobility  matrix that is
approximated using  hydrodynamic models based on  assumptions that are
of questionable  validity for  nanoscopic particles. In  particular, a
gold  nanocolloid and  a biomolecule  such as  a protein  can only  be
distinguished in BD based on an effective hydrodynamic no-slip surface
but not  based on the  nature of  their interaction with  the solvent.
This makes BD unsuitable for capturing multiscale effects such as slip
on  the surface  of the  particle, layering  of the  solvent molecules
around  the  colloid,  transient  hydrogen bond  networks  around  the
protein, etc.

The fluctuation-dissipation balance principle  informs us that viscous
dissipation  is  intimately  related  to  fluctuations  of  the  fluid
velocity.   It is  well-known that  diffusion in  liquids is  strongly
affected    by   advection    by    thermal   velocity    fluctuations
\cite{DiffusionRenormalization_I,ExtraDiffusion_Vailati,DiffusionRenormalization,Nanopore_Fluctuations},
and that  nonequilibrium diffusive mixing is  accompanied by ``giant''
long-range          correlated           thermal          fluctuations
\cite{FluctHydroNonEq_Book,LongRangeCorrelations_MD,GiantFluctuations_Universal,FractalDiffusion_Microgravity}.
As            explained             in            detail            in
Refs.    \cite{SIBM_Brownian,BrownianBlobs,DiffusionJSTAT,DDFT_Hydro},
there is a direct relation between these unusual properties of thermal
fluctuations in liquid solutions  and Brownian Dynamics. Specifically,
a    simplified    model    of   colloidal    diffusion    based    on
\emph{incompressible   fluctuating   hydrodynamics}  can   be   mapped
one-to-one  to  the  equations  of  BD  and  related  Dynamic  Density
Functional       Theories        (DDFT)       with       hydrodynamics
\cite{DDFT_Hydro_Lowen,DDFT_Pep};    this   derivation    shows   that
hydrodynamic   interactions   are   nothing    more  nor   less   than
\emph{hydrodynamic correlations}  induced by the  thermal fluctuations
in   the    solvent.    Such   a   fluctuating    hydrodynamic   model
\cite{SIBM_Brownian,BrownianBlobs,DiffusionJSTAT,DDFT_Hydro}  explains
the   appearance  of   giant   nonequilibrium   fluctuations  in   the
concentration  of colloidal  particles, justifies  the Stokes-Einstein
relation in the limit  of large Schmidt numbers \cite{StokesEinstein},
and  describes  the  important  influence of  boundaries  in  confined
suspensions  \cite{Nanopore_Fluctuations,BrownianBlobs}. If  one wants
to further  account for  inertial effects  and compressibility  of the
fluid, as crucial  for modeling the effect of  ultrasound on colloidal
particles  \cite{DirectForcing_Balboa}  or   the  acoustic  vibrations
produced by suspended \textit{\emph{particles \cite{Ohlinger2012}}} or
micro-organisms  \cite{Kirchner2014}, one  can use  a similar
model  but describe  the fluid  using \emph{compressible}  fluctuating
hydrodynamics \cite{DirectForcing_Balboa,CompressibleBlobs,ISIBM}.

In this work we  consider coupling compressible isothermal fluctuating
hydrodynamics to a suspended  nanocolloidal particle.  Unlike previous
phenomenological                                                models
\cite{DirectForcing_Balboa,CompressibleBlobs,ISIBM,BrownianBlobs,DiffusionJSTAT,StochasticImmersedBoundary,SIBM_Brownian,LB_IB_Points,LB_SoftMatter_Review,ForceCoupling_Fluctuations},
we obtain  our equations from  the underlying microscopic  dynamics by
using  the  Theory of  Coarse-Graining  (TCG)  as developed  by  Green
\cite{Green1952} and Zwanzig \cite{Zwanzig1961} (also see the textbook
\cite{Grabert1982}),   together    with   a   sequence    of   careful
approximations that preserve  the correct structure of  the exact (but
formal)  coarse-grained equations.  Our  derivation  is important  for
several reasons.  Firstly, our work provides  a microscopic foundation
for the types of models used in existing theoretical and computational
work
\cite{DirectForcing_Balboa,CompressibleBlobs,ISIBM,BrownianBlobs,DiffusionJSTAT,StochasticImmersedBoundary,SIBM_Brownian,LB_IB_Points,LB_SoftMatter_Review}.
Secondly, and  more importantly,  our derivation leads  to microscopic
Green-Kubo type formulas for the transport coefficients that appear in
the coarse-grained equations. This allows for these coefficients to be
estimated from molecular dynamics computations, thus fully taking into
account microscopic  effects that are  difficult if not  impossible to
include in purely continuum models.  Thirdly, our derivation will lead
us to first construct  a microscopically-justified fully discrete form
of   compressible  isothermal   fluctuating   hydrodynamics  that   is
second-order    accurate     while    also     maintaining    discrete
fluctuation-dissipation balance to second order.

This last contribution  is in itself a significant  extension of prior
work \cite{DiscreteLLNS_Espanol},  fully consistent with  the approach
to  nonlinear fluctuating  hydrodynamics proposed  in our  recent work
\cite{FluctDiff_FEM}.  Specifically, the  coarse-grained equations  we
derive here by following a  ``bottom-up'' approach can also be derived
by   a   ``top-down''   approach   in  which   one   starts   from   a
(phenomenological)  system of  formal stochastic  partial differential
equations and applies  a Petrov-Galerkin finite-element discretization
\cite{FluctDiff_FEM}.   Our  work  therefore  provides  a  direct  and
\emph{explicit}  link between  the microscopic  discrete dynamics  and
mesoscopic continuum fluctuating  hydrodynamics.  The physical insight
that   is   necessary   to  construct   phenomenological   fluctuating
hydrodynamics equations translates in this paper into physical insight
required when  constructing suitable  approximations or closures  of a
number  of  intractable  microscopic  expressions.  The  ``bottom-up''
procedure  clearly   reveals  all  of   the  required  terms   in  the
coarse-grained equations and provides  microscopic expressions for the
required coefficients.

At first  sight, it may seem  like the equations of  Smoluchowski that
underlie Brownian  dynamics have a well-known  microscopic derivation.
Indeed,  it is  not difficult  to construct  a text-book  TCG for  the
dynamic equation  describing the positions of  the colloidal particles
\cite{CoarseGraining_Pep}.   This leads  to the  well-known expression
for the hydrodynamic mobility (diffusion  tensor) as the time integral
of the correlation function of the velocities of the solute particles,
{conditional  on  the  particle's positions}.  It  should,
however, quickly be recognized  that this well-known expression, while
correct, is not useful in practice, for several reasons. Firstly, this
integral must be  computed anew for \emph{every}  configuration of the
suspended  particles.   Secondly, even  if  one  could  run a  new  MD
calculation  at every  step in  a BD  simulation, it  is important  to
realize that  these MD computations are  \emph{unfeasible} in practice
because they must  be very long on microscopic scales.   Namely, it is
well-known that  the slow viscous (diffusive)  dissipation of momentum
in  the  fluid makes  the  velocity  correlation functions  have  long
(power-law) hydrodynamic tails; it is the integral of these tails that
gives   the  hydrodynamic   correlations   (interactions)  among   the
particles, as well as finite-size effects on the diffusion coefficient
for confined particles  \cite{DiffusionRenormalization}. Therefore, to
correctly  capture  hydrodynamic  effects  the time  integral  in  the
Green-Kubo expression for the diffusion tensor must extend to at least
the time it takes for  momentum to diffuse throughout the \emph{whole}
system; while this time is typically  short compared to the time scale
at  which the  solute particles  move,  it is  very long  based on  MD
standards.

By contrast,  in the equations  derived here the  Green-Kubo integrals
can be computed  via \emph{feasible} (short) MD  simulations.  This is
because  all  of the  hydrodynamics,  such  as  the effects  of  sound
\cite{DirectForcing_Balboa}  or viscous  dissipation \cite{ISIBM}  are
captured by  explicitly resolving  the (fluctuating)  hydrodynamics of
the solvent using a grid of hydrodynamic cells, and only the remaining
\emph{local} and \emph{short-time} effects need  to be captured by the
microscopic simulations.  In the present work, we consider suspensions
that are  sufficiently dilute to  allow us  to neglect the  direct (as
opposed to hydrodynamic) interactions among the colloids and focus our
derivation  on  a  single  particle  immersed  in  a  viscous  liquid;
hydrodynamic  interactions  among  the particles  are  still  captured
because they are mediated by the explicitly resolved surrounding fluid
dynamics.  In  fact, we  believe that  in many  cases of  interest the
coarse-grained  diffusive   dynamics  can  effectively   be  simulated
by\emph{ a priori}  performing a small number of  short MD simulations
of    a    single    particle    in    a    small    (say    periodic)
domain.  \Pepmodified{Crucial to  the above  is the  fact that  in the
  present work the hydrodynamic cells  are assumed to be significantly
  larger than the nanoparticle itself.}

In the  next section we explain  in more detail the  basic assumptions
and thus limitations of our model. Briefly, our model assumes that the
solvent is a  simple isotropic single-component fluid. We  do not
explicitly  consider  energy transport  and  thus  limit our  work  to
isothermal suspensions.   We only consider dilute  suspensions of nano
particles. The extension to denser suspension leads to a significantly
more complicated  theory of  liquid mixtures that  is well  beyond the
scope  of this  work. The  limitation to  nanoscopic particles  is not
essential and the equations developed here can be used also for larger
particles such as micron-sized colloids;  however, in this case the MD
simulations required to obtain the  values of the Green-Kubo integrals
that  appear  in  the  coarse-grained  equations  would  again  become
unfeasible and  a different approach  is advised. We will  also assume
that  the particle  is effectively  spherical so  that describing  the
position of its  center of mass is sufficient without  requiring us to
also resolve its orientation. Our  theory assumes a separation of time
scales between  the positions of  the particles and  their velocities, 
and we do not include the velocities of the colloidal particles
in the description. More  precisely, it  requires that  the Schmidt  number of  the solute
particles  be very  large. This  is  not a  significant limitation  in
practice since the Schmidt number of even a single solvent molecule is
typically very large in liquids. In particular, our theory can be used
to describe  collective diffusion  of tagged solvent  particles (i.e.,
self-diffusion).

In Section  \ref{Sec:CG} we explain  the basic notation  and concepts,
and carefully select and define the coarse-grained (slow) variables in
terms  of the  microscopic degrees  of  freedom.  We  then proceed  to
carefully  examine  the  reversible   (non-dissipative)  part  of  the
dynamics.   In particular,  in Section  \ref{Sec:Exact} we  give exact
results that  are not  useful on  their own right  since they  lead to
equations that  are not closed explicitly  {. However}, by
making a series of approximations based  on a key ``linear for spiky''
approximation we  are able  to derive an  approximate closure  for the
reversible  dynamics   in  Section  \ref{Sec:LinSpiky}.    In  Section
\ref{Sec:Irrev} we  apply the  same approximation to  the irreversible
(dissipative)   part   of   the  dynamics,   together   with   another
{important} approximation in  which we replace constrained
Green-Kubo  expressions  with   unconstrained  equilibrium  Green-Kubo
averages.  The key results of  our calculations are then collected and
discussed  in  Section  \ref{Sec:FinalResults}.    We  first  give  an
approximate  but  {\em  closed}   form  for  the  coarse-grained  {\em
  discrete} dynamics, and then discuss  the relation of these discrete
equations  to  continuum  models in  Section  \ref{Sec:Continuum}.   A
comparison of our results to  phenomenological models and a discussion
of  their significance  and  range  of validity  is  given in  Section
\ref{Sec:Conclusions}.   A   number  of  technical   calculations  are
detailed in an extensive Appendix.

\section{\label{Sec:CG}Coarse-Graining}

In this section, we give the basic ingredients required to perform the
coarse-graining of  the microscopic dynamics for  our specific system.
We begin with a general overview  of the theory and then specialize to
the case of a nanoparticle suspended  in a simple liquid by explaining
the  details of  the microscopic  dynamics and  the definition  of the
coarse-grained variables.

\subsection{\label{Sec:ToCG}The Theory of Coarse-Graining}

In  this  section,   we  review  the  theory   of  Coarse-Graining  or
Non-Equilibrium   Statistical  Mechanics   as  established   by  Green
\cite{Green1952} and Zwanzig \cite{Zwanzig1961}.  The theory allows to
construct the dynamic equations for  the probability distribution of a
set  of coarse-grained  (CG) variables  that describe  the state  of a
system at a coarse {\em level of description}. The theory states that,
under the  assumption that the  CG variables are sufficiently  slow as
compared with the eliminated degrees  of freedom, the system follows a
diffusion process in the space of CG variables.  The resulting dynamic
equation for the probability distribution of the CG variables is given
by a Fokker-Planck equation (FPE), where both the drift and diffusion
terms are given in microscopic terms.

The coarse-grained  variables are  selected functions  $\hat{x}(z)$ in
phase space, i.e.  they depend on  the set of position and momenta $z$
of the molecules of the system. We follow the convention
that a  hatted symbol  like $\hat{x}(z)$ denotes  a function  in phase
space  that may  take  numerical  values $x$.   The  selection of  the
relevant variables $ \hat{x}(z)$ is  a crucial step in the description
of a non-equilibrium  system.  A crucial requirement is  that they are
{\em  slow} variables \cite{Hijon2010}.   When  this is  the case,  the
probability distribution of a set  of relevant variables $x$ obeys the
FPE
\begin{align}
\partial_t P(x,t)&=
-\frac{\partial}{\partial x}\!\cdot\! \left\{\left[ {\cal A}(x)
-{\cal D}(x)\!\cdot\!\frac{\partial {\cal H}}{\partial {x}}(x) \right]P(x,t) \right\}
\nonumber\\
&+k_BT\frac{\partial}{\partial x}\!\cdot\! \left\{{\cal D}(x)\!\cdot\!
\frac{\partial}{\partial {x}} P(x,t) \right\}
\label{ZFPE}
\end{align}
The different objects in this equation have a well-defined microscopic
definition. For example, the \textit{reversible drift} is
\begin{equation}
{\cal A}(x) = \langle L \hat{x}\rangle^x 
\label{Ax}
\end{equation}
where $L$ is the Liouville operator and the conditional expectation is defined by
\begin{equation}
\langle\ldots\rangle^{x}=\frac{1}{P^{\rm eq}(x)} \int
dz\rho^{\rm eq}(z)\delta(\hat{x}(z)-{x})\cdots 
\label{ca}
\end{equation}
where  $\rho^{\rm  eq}(z)$  stands  for the microscopic equilibrium  distribution  and
$\delta(  \hat{x}(z)-{x})$  is  actually  a  product  of  Dirac  delta
functions,  one for  every  function $  \hat{x}(z)$.  The  equilibrium
distribution of the relevant variables is
\begin{equation}
P^{\rm eq}(x)=\int dz\rho^{\rm eq}(z)\delta( \hat{x}(z)-{x}) 
\label{omega}
\end{equation}
and is closely  related to the \textit{bare free energy}  of the level
of description $x$ which is defined through
\begin{equation}
{\cal H}(x)\equiv -k_BT\ln P^{\rm eq}(x) 
\label{ent}
\end{equation}
Here  $k_B$ is  Boltzmann's constant  and $T$  the temperature  of the
equilibrium state. We will refer in  this work to the bare free energy
also as the coarse-grained Hamiltonian  because of the particular form
that ${\cal H}(x)$ acquires   at  the  hydrodynamic   level  of
description.  When non-isothermal situations are considered one rather
introduces  the  entropy  of  the   level  of  description  as  ${\cal
  S}(x)=k_B\ln  P^{\rm  eq}(x)$,  according to  Einstein  formula  for
fluctuations.

Finally,  the symmetric  and positive  semidefinite \cite{Grabert1982}
\textit{dissipative  matrix}   ${\cal  D}(x)$   is  the  matrix   of  transport
coefficients expressed in the form of Green-Kubo formulas,
\Pepmodified{\begin{equation}
{\cal D}(x)=\frac{1}{k_BT}\int_0^\infty\langle Q L  \hat{X}
\exp\{iQLt'\} Q L  \hat{X} \rangle^{x} dt' 
\label{M}
\end{equation}
}
The term  $Q L  \hat{x}$ is the  so called \textit{projected  current}.  The
projection  operator $Q$  is  defined  from its  action  on any  phase
function $\hat{B}(z)$ \cite{Zwanzig1961}
\begin{equation}
Q\hat{B}(z) = \hat{B}(z) - \langle \hat{B}\rangle^{ \hat{x}(z)} 
\label{qop}
\end{equation}
\Pepmodified{
The dynamic  operator $\exp\{iQLt'\}$  is usually named  the projected
dynamics,  which  is,  strictly   speaking  different  from  the  real
Hamiltonian  dynamics $\exp\{Lt'\}$.   The projected  dynamics can  be
usually approximated by  the real dynamics but, in order  to avoid the
so called \textit{plateau  problem} \cite{Hijon2010}, then the upper  infinite limit of
integration in  Eq.  (\ref{M}) has  to be  replaced by $\tau$,  a time
which is long  in front of the correlation time  of the integrand, but
short  in front  of the  time scale  of evolution  of the  macroscopic
variables \cite{Grabert1982,Kirkwood1946,Espanol1993,Hijon2010}, this is
\begin{equation}
{\cal D}(x)=\frac{1}{k_BT}\int_0^\tau\langle Q L  \hat{X}
\exp\{iLt'\} Q L \hat{X} \rangle^{x} dt' 
\label{M_L}
\end{equation}
In general, it is expected that different elements of the matrix may require different
values of $\tau$.}


The Ito stochastic differential  equation (SDE) that is mathematically
equivalent to the FPE (\ref{ZFPE}) is given by
\begin{equation}
\frac{dx}{dt} = {\cal A}(x) -{\cal D}(x)\!\cdot\!\frac{\partial {\cal H}}{\partial {x}}(x) +k_BT\frac{\partial
}{\partial {x}}\!\cdot\!{\cal D}(x) + \frac{d\tilde{x}}{dt}(x)
\label{sde}
\end{equation}
where  $\frac{d\tilde{x}}{dt}(x)=B(x)\frac{d{\cal   B}(t)}{dt}$  is  a
linear combination  of white  noises, formally  time derivatives  of a
collection of  independent Wiener processes (Brownian  motions) ${\cal
  B}(t)$,  where the  amplitudes  satisfy the  Fluctuation-Dissipation
Balance (FDB) condition
\begin{align}
B(x)^TB(x)&=2k_BT{\cal D}(x) 
\label{FD}
\end{align}
In summary, the three basic objects that determine the dynamics (either
in the  FPE (\ref{ZFPE}) or  the SDE (\ref{sde}) forms)  and that need  to be
computed in  the theory are  the bare  free energy ${\cal  H}(x)$, the
reversible  drift  ${\cal A}(x)$, and  the  dissipative matrix  ${\cal
  D}(x)$.

The reversible drift can also be written in the form \cite{Grabert1982}
\begin{align}
{\cal A}_{\mu}(x)
&=L_{\mu\nu}(x)\frac{\partial {\cal H}}{\partial x_\nu}(x)-k_BT\frac{\partial L_{\mu\nu}}{\partial x_\nu}(x)
\label{driftL}
\end{align}
where the skew-symmetric reversible matrix is defined as 
\begin{align}
 L_{\mu\nu}(x)&=\llangle\{X_\mu,X_\nu\}\rrangle^x
\label{Lskew}
\end{align}
{where $\{\cdot,\cdot\}$  is the Poisson bracket.   } Here
and  in what  follows,  Einstein convention  that  sums over  repeated
indices is  assumed.  Note that  the form of the  drift (\ref{driftL})
ensures  automatically  {the Gibbs-Boltzmann  distribution
  $P^{\rm  eq}(x)\propto  e^{-\beta{\cal  H}(x)}$ is  the  equilibrium
  solution of (\ref{ZFPE}),} \textit{even  for approximate forms} of the
  reversible matrix $L(x)$ and the  CG Hamiltonian ${\cal H}(x)$, and,
  thus, is the preferred form for  the reversible drift in the present
  work.

\subsection{\label{Sec:CGVars}Selection of Coarse-Grained Variables}

The  most  important  step  in  the   TCG  is  the  selection  of  the
\emph{relevant} (coarse-grained) \emph{variables}. This selection must
be  guided  by physical  intuition  and  the  presence or  absence  of
separation  of time  scales. The  key  guiding principle  is that  the
relevant  variables  must evolve  much more slowly  than \emph{all  }other
variables that cannot  be expressed entirely in terms  of the relevant
variables. This  allows us  to make a  Markovian approximation  of the
coarse-grained  dynamics,  which takes  the  form  of a  Fokker-Planck
equation for  the probability  distribution of relevant  variables, or
equivalently,   of  a   stochastic  differential   equation  for   the
instantaneous (fluctuating) relevant variables.

Ultimately,  one  is  often  only interested  in  the  positions  (and
possibly  orientations) of  the colloidal  particles, eliminating  the
solvent from  consideration entirely. This  is possible to do  via TCG
because  indeed in  liquids mass  diffusion is  very slow  compared to
momentum and heat  diffusion, and thus the positions  of the particles
are much  slower than the  hydrodynamic fields. Indeed,  following the
TCG using only the positions of  the particles leads to the well known
equations  of  Smoluchowski  or  Brownian  dynamics,  with  well-known
Green-Kubo  expressions for  the hydrodynamic  mobility (equivalently,
diffusion) matrix (see, for example, Section V in \cite{CoarseGraining_Pep}). 
As we explained above, this level of description is
not  sufficiently  detailed  to  allow  us to  describe  a  number  of
important  microscopic  effects that  occur  in  the vicinity  of  the
particle  surface. While  in the  present work  we do  not
  capture explicitly the slip at  the surface and the layering effects
  around  a  nanoparticle,  we  do  take  into  account  such  effects
  implicitly  through the  microscopic expressions  that enter  in the
  theory.   Furthermore, the  Green-Kubo formulas for the mobility  are not  useful in
practice  and  one  must  close  the equations  by  using  a  pairwise
approximation to the mobility matrix based on far-field expansions for
Stokes flow.

To go to  a more fundamental (microscopically more  informed) level of
description we  must include solvent  degrees of freedom as  well.  We
want to describe the solvent molecules at the hydrodynamic rather than
the microscopic level since it is  not reasonable to keep track of the
positions  and  momenta  of  every molecule  in  the  system.  At
macroscopic scales, a  fluid appears as a continuum  that is described
with  smooth fields  obeying the  well-known Navier-Stokes  equations.
The  ``field''  concept  is  tricky,  though, because  a  field  is  a
mathematical object  that has infinitely many  degrees of freedom,  while the
actual fluid  system has  a finite  number of  degrees of  freedom. Of
course, the fields are defined above a certain spatial resolution much
larger than  the typical size  and distances between molecules  of the
fluid. At  these macroscopic scales  the field  at one point  of space
effectively represents a very large number of molecules that move in a
coherent  manner.  When  one   descends  down  to  mesoscopic  scales,
molecules do not move that coherently, and one starts appreciating the
discrete nature of the fluid. In other words, the average behavior and
the actual behavior of the fluid  molecules start to differ, and it is
necessary to describe a fluid  system with hydrodynamic equations that
are intrinsically  stochastic. The  first phenomenological  theory for
such  fluctuating hydrodynamics  was proposed  by Landau  and Lifshitz,
who introduced  the concepts of  random stress  and heat fluxes,  to be
added to the usual Newtonian stress and Fourier heat flux \cite{Landau1959}.

From a  mathematical point of  view, the nonlinear  stochastic partial
differential  equations  (SPDEs)   of  fluctuating  hydrodynamics  are
ill-defined.  In other  words, a continuum limit of  sequences of more
refined  otherwise   reasonable  discrete  versions  of   the  partial
differential equation does  not exist. From a physical  point of view,
though,  this is  not  much of  a  problem because  we  know that  the
continuum  limit cannot  be  realized without  first encountering  the
atomistic nature  of matter.   For these reasons,  it is  necessary to
define  \emph{discrete} hydrodynamic  variables  by  averaging over  a
number of  nearby molecules, and  use these discrete variables  in the
TCG. In this  work, following the approach developed in  a sequence of
prior                                                            works
\cite{DiscreteDiffusion_Espanol,DiscreteLLNS_Espanol,FluctDiff_FEM},
we define discrete  hydrodynamic fields by placing  a fixed (Eulerian)
grid of hydrodynamic \emph{nodes} and associating to each node a fluid
density  and   momentum  averaged  over  a   hydrodynamic  \emph{cell}
associated to that node.  In  the present work we compute
  with  more  rigor  some  of  the  conditional  expectations  that  were
  plausibly approximated in \cite{DiscreteLLNS_Espanol}.  In order to
have  a   reasonable  hydrodynamics   description  we  need   to  have
hydrodynamic  cells  that  contain  many solvent  molecules;  here  we
consider simple  liquids for which hydrodynamic  cells containing many
molecules will also be much larger than the mean free path.

For  a  colloidal  particle  that  is much  larger  than  the  solvent
molecules,  the  hydrodynamic  flow  around the  nanoparticle  can  be
resolved  with  small  (compared  to the  size  of  the  nanoparticle)
hydrodynamic  cells that,  nevertheless,  still  contain many  solvent
molecules. In  this situation, the discrete  \emph{fluid mass} density
$\rho_{\mu}$, and  the discrete \emph{fluid} momentum  densities ${\bf
  g}_{\mu}$, where  $\mu$ indexes  the hydrodynamic nodes,  would only
include contributions from  the solvent particles. At such  a level of
description it  is necessary  to include both  the position  ${\bf R}$
\emph{and} the momentum ${\bf P}$ of the nanoparticle  in  the list of
relevant variables because even though  ${\bf P}$ is much faster
than  the  position,  it  evolves  on  the  same  time  scale  as  the
hydrodynamic momentum around the  particle. This level of
  description  has  been traditionally  used  for  the description  of
  Brownian  motion of  colloidal  particles  coupled with  fluctuating
  hydrodynamics \cite{Faxen_FluctuatingHydro,VACF_FluctHydro}. We
do not consider  this case here; for a phenomenological  model of this
type       we       refer        the       reader       to       Refs.
\cite{DirectForcing_Balboa,CompressibleBlobs,ISIBM}.  It  is important
to  note that  it  is inconsistent  to keep  the  velocities and  thus
inertial dynamics  of the  particles without  also accounting  for the
viscosity and inertia of the  surrounding fluid. This is because there
is  not a  separation of  time scales  between the  velocities of  the
particles and  the velocity of  the surrounding fluid; the  {\em only}
consistent coarse-grained {\em implicit-fluid} level of description is
that  of   Brownian  dynamics,   as  explained   in  detail   by  Roux
\cite{LangevinDynamics_Theory}.

Here we consider a nanoparticle that is not much larger than the fluid
molecules, so  that the  hydrodynamic cells are  much larger  than the
nanocolloidal   particle, i.e.,   we  have   a  ``subgrid''
  colloidal particle.  In  particular, the ``nanoparticle''
  particle  could  be  just  a tagged  fluid  molecule  when  modeling
  self-diffusion in  a liquid.   Since the  momentum of  the particle
evolves on the same time scale  as the solvent molecules with which it
collides, more precisely, since the fluctuations of the \emph{relative
}velocity  of  the colloid  are  fast  compared to  hydrodynamic  time
scales, we  define the hydrodynamic  mass and momentum  density fields
to\textit{  include the  nanoparticle contribution}.  In summary,  the
level of description that we consider in this work is characterized by
the  position of  the colloid  ${\bf  R}$, the  (total, i.e., including
the contribution from the nanoparticle) discrete  mass
density  $\rho_{\mu}$, and  the  (total)  discrete momentum  density
${\bf g}_{\mu}$, where $\mu$ indexes the hydrodynamic nodes.

We make  use of the  standard TCG of Zwanzig  where all the  terms (CG
free energy,  drift, and  diffusion matrix)  are given  in microscopic
terms \cite{Zwanzig1961,Grabert1982}. This allows one to obtain
the  general structure  of the  dynamics.   However in  order to  find
tractable results it  is crucial to make a number  of assumptions. All
the approximations  that we consider rely  on the fact that  the cells
used to  define the  hydrodynamic variables are  much larger  than the
typical  intermolecular  distances  in  such a  way  that  every  cell
contains many  molecules of the  fluid. In particular, we  assume that
the   microscopic  local   density  field   which  is   of  the   form
$\sum_{i}^{N}m_{i}\delta({\bf  r}-{\bf  q}_{i})$  gives,  once  inside
conditional expectations, the same result as the interpolated discrete
density    variables    (see    Eq.    (\ref{Approx0})    below    and
Fig. \ref{Fig:LFSA}).   This is  only plausible  if, again,  there are
many molecules  per cell and the  values of the discrete  variables in
neighboring cells are very similar.  While this is statement about the
flow regimes  for which the resulting  equations apply, it is  also an
statement  about  the size  of  the  fluctuation of  the  hydrodynamic
variables.  They  need to be  \textit{small}, otherwise, the  value in
neighbor  cells could  be very  different  just by  chance.  In  other
words, the number of molecules per  cell must be sufficiently large in
order for the relative fluctuations  to be sufficiently small.  In the
end, the  validity of the approximations  made and the utility  of the
final equations  we obtain can only  be judged by a computational
comparison to the true microscopic dynamics (molecular dynamics).

\subsection{\label{osec:var}Microscopic Dynamics}

\begin{figure}[t]
\begin{psfrags} 
\psfrag{r}{{${\bf r}_\mu$} }
\includegraphics{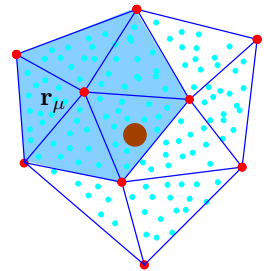}
\end{psfrags} 
\caption{Schematic representation of a nanoparticle (in brown) surrounded by molecules of a simple liquid solvent (in blue). Also shown is the triangulation that allows to define the discrete hydrodynamic variables at the nodes (in red). The shaded area around node $\mu$ located at ${\bf r}_\mu$ is the support of the finite element function $\psi_\mu({\bf r})$ and defines the hydrodynamic cell. }
\label{Fig:nano}
\end{figure}

In  the present  work  we consider  a simple  liquid  system of  $N+1$
particles described with  the position and momenta of  their center of
mass (see Fig.   \ref{Fig:nano} for a schematic  representation), in a
periodic box.  We distinguish particle $i=0$ as the nanoparticle which
has a  mass $m_0$,  typically larger  than the mass  $m$ of  a solvent
particle.  At the  {\em microscopic level} the system  is described by
the  set  $z$   of  all  positions  ${\bf  q}_i$   and  momenta  ${\bf
  p}_i=m_i{\bf  v}_i$   ($i=0,1,\cdots,N$)  of  the   particles.   The
microstate  of the  system evolves  according to  Hamilton's equations
with Hamiltonian given by
\begin{align}
\hat{H}(z) &= \frac{{\bf p}_0^2}{2m_0}+ 
\sum^N_{i=1} \frac{{\bf p}_i^2}{2m}+\hat{U}(q)
\nonumber\\
\hat{U}(q)&=\hat{U}^{\rm sol}(q)+\sum_{i=1}^N\Phi^{\rm int}(q_{0i})+\Phi^{\rm ext}({\bf q}_0)
\nonumber\\
\hat{U}^{\rm sol}(q)&= \frac{1}{2}\sum^N_{i,j=1}\phi(q_{ij})
\label{H}
\end{align}
We have assumed a pairwise  potential energy $\phi(q_{ij})$ between  liquid molecules $i,j$
separated a  distance  $q_{ij}$.  $\hat{U}^{\rm  sol}(q)$ is  the
potential energy  of the solvent  in the absence of  the nanoparticle,
$\Phi^{\rm int}(q)$  is the potential of interaction
of the  $i$-th solvent particle  with a nanoparticle a distance $q$ away,
and $\Phi^{\rm  ext}({\bf q}_0)$ is an external time-independent potential acting on
the nanoparticle.  The system
is assumed to have periodic boundary conditions.

Under the assumption that the Hamiltonian is mixing, the dynamics will
sample at  long times  the \textit{molecular  ensemble} \cite{Ray1999}
given by
\begin{align}
  \rho^{\rm eq}(z)=\frac{1}{\Omega(E_0,{\bf P}_0)}
\delta\left(\sum_{i=0}{\bf p}_i-{\bf P}_0\right)
\delta\left(H(z)-E_0\right)
\label{molens}
\end{align}
where ${\bf P}_0$ and $E_0$ are  the initial total momentum and energy
of the  system.  We will  assume that  in the thermodynamic  limit the
molecular ensemble can be approximated by the canonical ensemble
\begin{align}
\rho^{\rm eq}(z)&=\frac{1}{Z}  \exp\{-\beta \hat{H}(z)\},
\label{canens}
\end{align}
where $\beta=1/(k_B T)$, and we use the canonical ensemble in the theory for simplicity.


\subsection{Definition of Coarse-Grained Variables}


The first  step in  the Theory  of Coarse-Graining  is to  specify the
relevant  variables in  terms  of  the microscopic  state  $z$ of  the
system.   In the  present case,  we choose  as relevant  variables the
position of the nanoparticle 
\begin{align}
  \hat{\bf R}(z) &={\bf q}_0,
\end{align}
and  the  mass  and  momentum hydrodynamic  ``fields''.   As  we  will
consider fluctuations  in the hydrodynamic variables,  the latter need
to be  defined in  discrete terms  \cite{FluctDiff_FEM}.  This  is, we
want to look at the mass and momentum of collections of molecules that
are in a  given region of space.  To this end, we  seed physical space
with  a  set  of  $M$  \textit{nodes}, located  at  the  points  ${\bf
  r}_\mu$. Usually, the  nodes are arranged in a  regular lattice, but
this is not  necessary in what follows and  arbitrary simplicial grids
can be used (see Fig.  \ref{Fig:nano} for a schematic representation).

We  define the  mass and  momentum
densities of the node $\mu$ according to
\begin{align}
\hat{\rho}_\mu(z)&= \sum_{i=0}^Nm_i\delta_{\mu}({\bf q}_i)
\nonumber\\
\hat{\bf g}_\mu(z)&= \sum_{i=0}^N{\bf p}_i\delta_{\mu}({\bf q}_i)
\label{CGvariables}
\end{align}
{where the index $i=0$ labels the nanoparticle.}
The  basis  function   $\delta_{\mu}({\bf  r})$  is  a
function (with dimensions of inverse  of a volume) that is appreciably
different  from zero  only in  the  vicinity of  ${\bf r}_\mu$.   This
region is referred to as the hydrodynamic cell of node $\mu$.  We may
regard the basis function $\delta_\mu({\bf  r})$ as a ``discrete Dirac
delta function''. Its specific form is discussed below.
Note  that   both  the  mass   and  momentum  densities   contain  the
nanoparticle in their definition.  It  is convenient to introduce also
the hydrodynamic fields of the solvent
\begin{align}
\hat{\rho}^{\rm sol}_\mu(z)&= \sum_{i=1}^Nm_i\delta_{\mu}({\bf q}_i)
\nonumber\\
\hat{\bf g}^{\rm sol}_\mu(z)&= \sum_{i=1}^N{\bf p}_i\delta_{\mu}({\bf q}_i)
\label{rhogsol}
\end{align}
that  do  not  contain  in  its definition  the  contribution  of  the
nanoparticle \Pepmodified{(i.e. the particle $i=0$ is excluded in the sum)}.

We  may  express  the  discrete  hydrodynamic  variables
(\ref{CGvariables})  and   (\ref{rhogsol})  in  terms   of  the  usual
microscopic densities
\begin{align}
 \hat{\rho}_{\bf r}(z) &=\sum^{N}_{i=0}m_i\delta({\bf r}-{\bf q}_i),\quad
&\hat{\rho}^{\rm sol}_{\bf r}(z) =\sum^{N}_{i=1}m_i\delta({\bf r}-{\bf q}_i)
\nonumber\\
 \hat{\bf g}_{\bf r}(z)& =\sum^{N}_{i=0}{\bf p}_i\delta({\bf r}-{\bf q}_i),\quad
&\hat{\bf g}^{\rm sol}_{\bf r}(z) =\sum^{N}_{i=1}{\bf p}_i\delta({\bf r}-{\bf q}_i)
\label{omicdens}
\end{align}
as simple space integrals,
\begin{align}
\hat{\rho}_\mu(z)&= \int d{\bf r}\delta_\mu({\bf r})\hat{\rho}_{\bf r}(z),\quad
&
\hat{\rho}^{\rm sol}_\mu(z)= \int d{\bf r}\delta_\mu({\bf r})\hat{\rho}^{\rm sol}_{\bf r}(z)
\nonumber\\
\hat{\bf g}_\mu(z)&= \int d{\bf r}\delta_\mu({\bf r})\hat{\bf g}_{\bf r}(z),\quad
&\hat{\bf g}^{\rm sol}_\mu(z)= \int d{\bf r}\delta_\mu({\bf r})\hat{\bf g}^{\rm sol}_{\bf r}(z)
\label{omacmic1}
\end{align}

Note that the two sets of variables $\{\hat{\bf R},\hat{\rho},\hat{\bf
  g}\}$  and  $\{\hat{\bf   R},\hat{\rho}^{\rm  sol},\hat{\bf  g}^{\rm
  sol}\}$ are  not expressible in terms  of each other. While  we have
that the densities are related as
\begin{align}
\hat{\rho}^{\rm sol}_\mu(z)&=  \hat{\rho}_\mu(z)-m_0\delta_\mu(\hat{\bf R})
\label{rhos}
\end{align}
there is no way to express the momentum $\hat{\bf g}$ as a function of
${\bf R},\hat{\rho}^{\rm sol},\hat{\bf  g}^{\rm sol}$.  Therefore, the
dynamic  equations  to be  obtained  for  each  set of  variables  are
\textit{essentially different} and cannot  be obtained from each other
through a simple change of variables.  In other words, the two sets of
relevant variables lead to \textit{physically different} descriptions.
Since  the slowness  of  the hydrodynamic  variables  arises from  the
underlying  conservation  laws, and  only  the  {\em total}  mass  and
momentum fields  are conserved  quantities, the  appropriate variables
for    the    TCG    are     our    chosen    variables    $\{\hat{\bf
  R},\hat{\rho},\hat{\bf g}\}$.

\subsection{The basis functions}

\begin{figure}[t]
\begin{psfrags} 
\psfrag{f}{{$\psi_\mu({\bf r})$}} 
\psfrag{m}{{${\bf r}_\mu$} }
\psfrag{V}{ }
\includegraphics[scale=1]{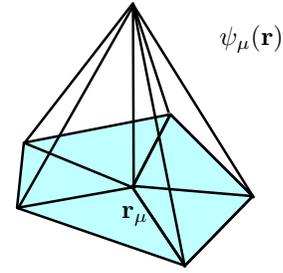}
\end{psfrags} 
\caption{The finite element basis function $\psi_\mu({\bf r})$ in two dimensions.}
\label{Fig:pyramid}
\end{figure}

The actual form of the  discrete Dirac delta function $\delta_\mu({\bf
  r})$  needs  to  be  specified.   One  possibility  is  to  use  the
characteristic function  (divided by  the volume of  the cell)  of the
Voronoi cell of node $\mu$. For $\hat{\rho}_\mu(z)$ this will give the
total mass (per unit volume) of the particles that happen to be within
the Voronoi cell $\mu$.  As  we discussed in Ref.  \cite{Espanol2009},
though, this selection is unsuited for the derivation of the equations
governing discrete  hydrodynamics from the Theory  of Coarse-Graining.
This is  because the  gradient of the  characteristic function  of the
Voronoi  cell  is singular  and  leads  to  ill-defined  Green-Kubo
expressions.    It  was   suggested   to  instead   use  the   Delaunay
triangulation associated  with the set  of nodes  as a grid  of finite
elements (FE), and  take the discrete delta function to  be the linear
FE basis  function $\psi_{\mu}({\bf  r})$ associated with  node $\mu$,
which  has the  characteristic shape  of a  tent in  one dimension,  a
pyramid in two  dimensions (as shown in  Fig.  \ref{Fig:pyramid}), and
more generally a $(d+1)$-dimensional simplex in $d$ dimensions.
Note that the use of a Voronoi/Delaunay tessellation is not required,
and any simplicial grid (i.e., a triangular grid in two dimensions or
a tetrahedral grid in three dimensions) whose vertices are the set of
hydrodynamic nodes can be used equally well (but for numerical
purposes the grid should be kept as close to uniform as possible).1

In recent  work \cite{FluctDiff_FEM,JaimeThesis}, we have  argued that
an even better selection (in terms  of numerical accuracy) is given by
a basis function $\delta_\mu({\bf r})$ that is a linear combination of
the (dimensionless)  finite element  linear basis  functions functions
$\psi_{\mu}({\bf r})$
\begin{align}
  \delta_\mu({\bf r})&=M^{\delta}_{\mu\nu}\psi_\nu({\bf r}),
\label{deltapsi}
\end{align}
The  crucial requirement  is  that these  basis
functions are mutually orthogonal
\begin{align}
  \norm\delta_\mu\psi_\nu\norm&=\delta_{\mu\nu}
\label{orthogon}
\end{align}
where we have introduced double bars to denote integration over space, this is
\begin{align}
\norm f\norm &\equiv  \int d{\bf r}f({\bf r})
\label{NotationInt}
\end{align}
for an arbitrary function $f({\bf r})$. Note that from (\ref{deltapsi}) and (\ref{orthogon})
it follows the explicit matrix form
\begin{align}
  M^\delta_{\mu\nu}&=\norm\delta_\mu\delta_\nu\norm
\end{align}
If we introduce the usual ``mass matrix'' of the finite element method
\begin{align}
  M^\psi_{\mu\nu}=\norm\psi_\mu\psi_\nu\norm
\label{Mpsi}
\end{align}
the  orthogonality  condition   implies  that  $M^\delta_{\mu\nu}$  in
(\ref{deltapsi}) is given by the inverse of $M^\psi_{\mu\nu}$, this is
\begin{align}
  M^\psi_{\mu\nu}M^\delta_{\nu\sigma}=\delta_{\mu\sigma}
\end{align}

The basis function  $\delta_\mu({\bf r})$ may be regarded as  a way of
discretizing a field $a({\bf  r})$ according to $a_\mu=\norm\delta_\mu
a\norm$.   The   basis  function  $\psi_{\mu}({\bf  r})$   permits  to
construct   interpolated   fields   out   of   the   discrete   fields
$\overline{a}({\bf   r})=\sum_\mu  a_\mu   \psi_\mu({\bf  r})$.    The
orthogonality condition (\ref{orthogon}) ensures that if we discretize
an interpolated field,  we recover the original  discrete values, i.e.
$\norm\delta_\mu\overline{a}\norm=a_\mu$.  This is the main motivation
to  use  the  slightly   more  involved  {basis}  function
$\delta_\mu({\bf  r})$ instead  of the  finite element  $\psi_\mu({\bf
  r})$ {for  the definition of  the CG variables}.  It turns
out  that  this  complication  pays   off,  as  the  resulting  finite
difference operators  are second order accurate  approximations of the
corresponding continuum differential operator, even in irregular grids
\cite{JaimeThesis}.

The  finite element  linear  basis  functions satisfy  a
partition of unity and give linear consistency,
\begin{align}
  \sum_\mu\psi_{\mu}({\bf r})&=1,
\quad\quad\quad\quad
  \sum_\mu{\bf r}_\mu\psi_{\mu}({\bf r})={\bf r}
\label{proppsi}
\end{align}
As  a   consequence  of  these  properties,  the
conjugate basis functions $\delta_\mu({\bf r})$ satisfy
\begin{align}
  \sum_\mu{\cal V}_\mu\delta_\mu({\bf r})&=1,
\quad\quad\quad\quad
  \sum_\mu{\cal V}_\mu{\bf r}_\mu\delta_\mu({\bf r})={\bf r}  
\label{propdelta}
\end{align}
where ${\cal V}_\mu$ is the volume of the hydrodynamic cell $\mu$
\begin{align}
{\cal V}_\mu\equiv\int d{\bf r}\psi_\mu({\bf r})
\label{vol}
\end{align}
Note that we have 
\begin{align}
\int d{\bf r}\delta_\mu({\bf r})&=1,
\quad\quad\quad\quad
\int d{\bf r}\;{\bf r}\delta_\mu({\bf r})={\bf r}_\mu
\label{voldel}
\end{align}
as can be proved by using  (\ref{proppsi})   and   the   orthogonality
(\ref{orthogon}).     These    properties   justify    to   call
$\delta_\mu({\bf  r})$  a  discrete  Dirac  delta  function.   

The partition of unity reflected in (\ref{propdelta}) implies
\begin{align}
  \sum_\mu{\cal V}_\mu\nabla\delta_\mu({\bf r})=0
\label{PartitionUnity}
\end{align}
which we will use often in proving that the resulting dynamic equations are conservative. In fact,  we define the total mass and total momentum of the system at
the CG level through,
\begin{align}
  M_T&\equiv\sum_\mu{\cal V}_\mu\hat{\rho}_\mu(z) = \sum_im_i
\nonumber\\
{\bf P}_T&\equiv\sum_\mu{\cal V}_\mu\hat{\bf g}_\mu(z) = \sum_i{\bf p}_i
\label{TotalInvariants}
\end{align}
which are, indeed,  the total mass and momentum.  These quantities are
conserved  by the  microscopic  dynamics \Pepmodified{and  need to  be
  conserved by the coarse-grained dynamics}.

It is convenient to  introduce also the following \textit{regularized}
Dirac delta function
\begin{align}
  \Delta({\bf r},{\bf r}')\equiv \delta_\mu({\bf r})\psi_\mu({\bf r}')=  
{\Delta({\bf r}',{\bf r})},
\label{Regularized}
\end{align}
which is closely related to what is called the discrete Delta function
or                interpolation               kernel                in
\incite{DirectForcing_Balboa,CompressibleBlobs,ISIBM,BrownianBlobs,StochasticImmersedBoundary,LB_IB_Points,LB_SoftMatter_Review}.
This function is  different from zero only for distances  of the order
of the size  of the hydrodynamic cells.  In the  limit of zero lattice
spacing  $\Delta({\bf  r},{\bf  r}')$   converges  in  weak  sense  to
$\delta({\bf r}-{\bf r}')$.  Therefore, $\Delta({\bf r},{\bf r}')$ can
be understood  as a Dirac delta  function regularized on the  scale of
the grid.  

The regularized Dirac delta  satisfies the exact identities
\begin{align}
  \int d{\bf r}'\Delta({\bf r},{\bf r}')\delta_\mu({\bf r}')&= \delta_\mu({\bf r})
\nonumber\\
  \int d{\bf r}'\Delta({\bf r},{\bf r}')\psi_\mu({\bf r}')&= \psi_\mu({\bf r})
\label{LFSAdelta}
\end{align}
One of the basic approximations that we will make in the present
work is the {\textit{smoothness approximation}}
\begin{align}
  \int d{\bf r}'A({\bf r}')\Delta({\bf r}',{\bf r})
&= \norm A\delta_\mu\norm\psi_\mu({\bf r})
\simeq A({\bf r})
\label{LFSA}
\end{align}
for a smooth function  $A({\bf r})$. {For smooth functions
  the  regularized  Dirac  delta  acts   like  a  Dirac  delta}.   The
approximation (\ref{LFSA})  is an  \textit{exact} identity  for linear
functions  $A({\bf r})=a+{\bf  r}\esc{\bf b}$.  Therefore, the  errors
committed  when  using  the   approximation  (\ref{LFSA})  for  smooth
functions    \textit{are   of    second   order    in   the    lattice
  spacing}. Sometimes, we will use the above identity in the form
\begin{align}
  \norm A \delta_\mu\norm \; \norm \psi_\mu B \norm \simeq
  \norm A B \norm 
\label{LFSAB}
\end{align}
for any two smooth functions $A({\bf r}),B({\bf r})$.

Finally, note that one property that is \textit{not} satisfied by
  the regularized Dirac delta function,  as opposed to the Dirac delta
  is the following symmetry
\begin{align}
\frac{\partial}{\partial {\bf r}}\Delta({\bf r},{\bf r}')=
-\frac{\partial}{\partial {\bf r}'}\Delta({\bf r},{\bf r}')
\end{align}
If the  regularized Dirac delta function  was translationally invariant,
i.e.  $\Delta({\bf r},{\bf r}')=\Delta({\bf  r}-{\bf r}')$, this would
be  obviously  true.  In  this  case,  we  would  have in addition to 
(\ref{LFSAdelta}) also the following relations,
\begin{align}
  \int d{\bf r}'\Delta({\bf r},{\bf r}')\boldsymbol{\nabla}'\delta_\mu({\bf r}')&= \boldsymbol{\nabla}\delta_\mu({\bf r})
\nonumber\\
  \int d{\bf r}'\Delta({\bf r},{\bf r}')\boldsymbol{\nabla}'\psi_\mu({\bf r}')&= \boldsymbol{\nabla}\psi_\mu({\bf r})
\label{LFSAGraddelta}
\end{align}
Even though these  identities are not fulfilled, we  will assume that
they  are reasonable  approximations, particularly  if both  sides are
multiplied with ``smooth discrete fields'', i.e.
\begin{align}
    \int d{\bf r}'\Delta({\bf r},{\bf r}')\boldsymbol{\nabla}'\overline{a}({\bf r}')
&\simeq \boldsymbol{\nabla}\overline{a}({\bf r})
\label{GradSmooth}
\end{align}
For a  sufficiently smooth  field $\overline{a}({\bf r})$,  the length
scale  of variation  of $\boldsymbol{\nabla}\overline{a}({\bf  r})$ is
much larger than the length scale of variation of $\Delta({\bf r},{\bf
  r}')$ and, therefore, $\Delta({\bf r},{\bf r}')$ acts as an ordinary
Dirac delta.

\subsection{Notation}
The notation  in the  present work is  unavoidably dense  because many
different    mathematical    objects     need    to    be    carefully
distinguished. Below we present a summary of the notation for the case
of the mass density variable alone.   Similar symbols are used for the
velocity  and momentum  density variables.  In general,  hatted symbol
like in
\begin{align}
    \hat{\rho}_\mu(z)&=\sum_{i=0}^Nm_i\delta_\mu({\bf q}_i),
\quad    \hat{\rho}_{\bf r}(z)=\sum_{i=0}^Nm_i\delta({\bf q}_i-{\bf r}),
\end{align}
denote phase functions.  The  numerical values taken by a
  phase  function  are   denoted  without  hat  as   in,  for  example,
  $\rho_\mu$.  The subscript is used here to distinguish the specific
node $\mu$  for discrete  variables such  as $\hat{\rho}_\mu$,  or the
specific point in space for  continuum fields such as $\hat{\rho}_{\bf
  r}$.  Overlined symbols like
\begin{align}
\overline{\rho}({\bf r})&=\psi_\mu({\bf r})\rho_\mu 
\label{orho}  
\end{align} 
denote  continuum fields  which are  interpolated  from  discrete ``fields''.   
Differential operators  act only  on the  symbol immediately  to their
left   unless  otherwise   indicated  by   parenthesis,  dot   denotes
contraction, and colon a double contraction.
\section{The reversible drift}
In this  section, we present  a number  of exact and  then approximate
results for the reversible part ${\cal  A}(x)$ of the dynamics and the
bare free energy ${\cal H}(x)$ for the present level of description.

The exact results presented in section \ref{Sec:Exact} are obtained by
integrating  the microscopic  momenta in  the microscopic  definitions
(\ref{Ax}) and  (\ref{omega}) for these quantities.   This integration
is possible because  we assume that the equilibrium  ensemble is given
by  the  canonical ensemble  (\ref{canens})  and  the resulting  space
integrals  involve relatively  simple Gaussian  integrals of  the kind
discussed  in  Appendix   \ref{App:MomentumIntegrals}.  The  molecular
ensemble  (\ref{molens}) can  also  be  used at  the  expense of  much
cumbersome  expressions.  We  assume that  in the  thermodynamic limit
both ensembles  are equivalent and  we opt  for the simpler  case.  In
Section \ref{Sec:LinSpiky}  we approximate the exact  results in order
to  obtain a  closed form  of the  reversible drift.   In the  present
section  we  simply  quote  the  exact results  and  redirect  to  the
appendices for the specific calculations.

\subsection{\label{Sec:Exact}The exact reversible drift}
We have obtained in  Eq.  (\ref{revLform}) of Appendix \ref{App:Exact}
the following exact form for the reversible drift ${\cal A}(x)$ in the
form  (\ref{driftL})  with  the  evidently  skew-symmetric  reversible
generator 
\begin{widetext}
\begin{equation}
    L = \left( \begin{array}{ccc}
        0&
        0&
        \delta_\mu({\bf R})
        \\
        \\
        0&
        0&
        \bra\hat{\rho}\delta_\nu\boldsymbol{\nabla}^\beta\delta_\mu \ket^{{\bf R}\rho{\bf g}}
        \\
        \\
        -\delta_\mu({\bf R}) 
        &-\bra\hat{\rho}\delta_\mu\boldsymbol{\nabla}^\alpha\delta_\nu \ket^{{\bf R}\rho{\bf g}}
        &\bra\hat{\bf g}^\alpha\delta_\nu\boldsymbol{\nabla}^\beta\delta_\mu\ket^{{\bf R}\rho{\bf g}}
        -\bra\hat{\bf g}^\beta \delta_\mu\boldsymbol{\nabla}^\alpha\delta_\nu\ket^{{\bf R}\rho{\bf g}}
 \end{array}\right)
\label{ExactDrift}
\end{equation}
\end{widetext}
The  double square  brackets  act on  arbitrary space-dependent  phase
functions $  \hat{f}_{\bf r}(z)$  and denote  the double  operation of
conditional averaging and space integration, this is
\begin{align}
  \bra \hat{f}\ket^{{\bf R}\rho{\bf g}}&\equiv \int d{\bf r}\llangle \hat{f}_{\bf r}\rrangle^{{\bf R}\rho{\bf g}}
\label{double}
\end{align}
where $\llangle  \hat{f}_{\bf r}\rrangle^{{\bf R}\rho{\bf g}}$  is the
conditional   expectation  (\ref{ca})   for  the   present  level   of
description.

The CG Hamiltonian ${\cal  H}({\bf  R},\rho,{\bf g})$ 
is shown in  Appendix \ref{App:Exact}, Eq. (\ref{CGHamiltonianExact})
to be given rigorously as 
\begin{align}
  {\cal H}({\bf R},\rho,{\bf g})&=
-k_BT\ln \llangle \frac{\exp\left\{-\frac{\beta}{2}{\bf g}_\mu\hat{M}^{-1}_{\mu\nu}{\bf g}_\nu\right\}}
{(2\pi/\beta)^{3M/2}\det \hat{M}^{3/2}}\rrangle^{{\bf R}\rho}
\nonumber\\
&+{\cal F}\left({\bf R},\rho\right)+\Phi^{\rm ext}({\bf R})
\label{CGHamiltonian}
\end{align}
In this expression the microscopic mass matrix is defined as
\begin{align}
 \hat{M}_{\mu\nu}(z)&\equiv\sum_{i=0}^Nm_i\delta_\mu({\bf q}_i)\delta_\nu({\bf q}_i)
\label{taghatMmunu}
\end{align}
This matrix depends on the  microscopic configuration of the particles and we
assume that for the typical configurations ${\bf R},\rho$ that condition the average
in (\ref{CGHamiltonian}) are such that give microscopic configurations for which
the inverse exists.

The fluid free energy is
the sum of two contributions
\begin{align}
 {\cal F}\left({\bf R},\rho\right)&={\cal F}^{\rm sol}\left(\rho^{\rm sol}\right)
+{\cal F}^{\rm int}({\bf R},\rho^{\rm sol})
\label{FreeEnergyDecomp}
\end{align}
where the discrete solvent density  $\rho^{\rm sol}_\mu$ is defined in
Eq. (\ref{rhos}).  The free energy of the solvent ${\cal F}^{\rm sol}$
and  the  free energy  of  interaction  ${\cal F}^{\rm  int}$  between
nanoparticle and  solvent are, respectively
\begin{align}
  {\cal F}^{\rm sol}(\rho_{\rm sol})\equiv&-k_BT\ln P^{\rm eq}_{\rm sol}(\rho_{\rm sol})
\nonumber\\
{\cal F}^{\rm int}({\bf R},\rho_{\rm sol})\equiv&-k_BT\ln \llangle
 \exp\left\{-\beta\sum_{i=1}^N\Phi^{\rm int}({\bf R}-{\bf q}_i)\right\} \rrangle^{\rho_{\rm sol}}
\label{S2:calFF}
\end{align}
where $P^{\rm eq}_{\rm sol}(\rho)$ is the equilibrium probability that
a  system  without  the  nanoparticle  has  a  particular  realization
$\rho_\mu$  for   the  mass  density.   The   conditional  expectation
$   \llangle  \cdots\rrangle^{\rho_{\rm   sol}}$  is   an
  equilibrium average  over solvent degrees of  freedom conditional to
  give the  realization $\rho_\mu$ for  the discrete density. 
  The  fact that the  free energy of the  system in
  Eq.   (\ref{FreeEnergyDecomp}) depends  on the  mass density  of the
  fluid  $\rho_\mu$ through  the combination  $\rho^{\rm sol}_\mu$  in
  (\ref{rhos}),  which is  the mass  density  of the  solvent in  cell
  $\mu$, is a non-trivial result.

\subsection{\label{Sec:LinSpiky}Approximate results for the reversible drift}
\begin{figure}[t]
\begin{psfrags} 
\psfrag{m}{{ $\mu$}} 
\psfrag{m1}{{$\mu+1$} }
\psfrag{m2}{{$\mu-1$} }
\psfrag{ri}{{${\bf q}_i$} }
\psfrag{rhom}{$\hat{\rho}_{\mu}$} 
\psfrag{rhom1}{$\hat{\rho}_{\mu+1}$} 
\psfrag{rhom2}{$\hat{\rho}_{\mu-1}$} 
\includegraphics[scale=0.8]{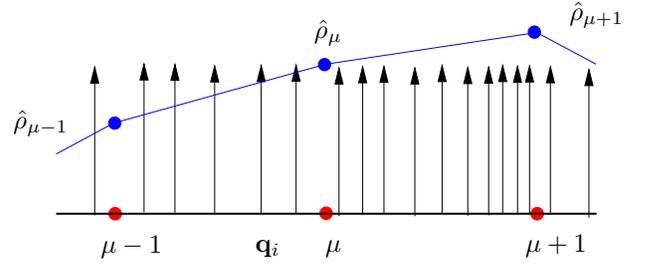}
\end{psfrags} 
\caption{The linear  for spiky approximation: The  microscopic density
  field  $\hat{\rho}_{\bf  r}(z)$,  which  is a  sum  of  Dirac  delta
  functions, each  located at the  particle's position ${\bf  q}_i$, is
  approximated   with    {the  linear   interpolation
  $\psi_\mu({\bf r})\hat{\rho}_\mu(z)$ } (blue line)  of the discrete  values of
  the density field $\rho_\mu$ at the nodes.}
\label{Fig:LFSA}
\end{figure}

The exact but formal results (\ref{ExactDrift}), (\ref{CGHamiltonian}) need
to   be  approximated   in  order   to  express   them  in   terms  of
\textit{explicit}   functions   of   the  relevant   variables   ${\bf
  R},\rho_\mu,{\bf   g}_\mu$.   These   results  involve   conditional
expectations  of  the   microscopic  density  fields  $\hat{\rho}_{\bf
  r}(z),\hat{\bf g}_{\bf r}(z)$.  The basic approximation that we will
consider when  computing conditional averages of  the microscopic mass
and momentum density  fields is that these fields  may be approximated
by linear interpolations of the CG densities, this is
\begin{align}
 \hat{\rho}_{\bf r}(z)&\simeq\psi_\mu({\bf r})\hat{\rho}_\mu(z)
\nonumber\\
 \hat{\bf g}_{\bf r}(z)&\simeq\psi_\mu({\bf r})\hat{\bf g}_\mu(z)
\label{Approx0}
\end{align}
A graphical representation of this approximation in 1D is shown in Fig
\ref{Fig:LFSA}.   Note  that   the  approximation  (\ref{Approx0})  is
equivalent to replacing the Dirac delta function $\delta({\bf r}-{\bf
  q}_i)$ in (\ref{omicdens}) with the regularized Dirac delta function
$\Delta({\bf r},{\bf q}_i)$ introduced in (\ref{Regularized}).

We  call this  approximation \textit{linear  for spiky  approximation}
because $  \hat{\rho}_{\bf r}(z)$,
as defined in Eq. (\ref{omicdens}), is  a sum of Dirac delta functions
while     $\psi_\mu({\bf     r})\hat{\rho}_{\mu}(z)$    defined     in
(\ref{Approx0}) is a piece-wise linear function of space.
The  approximation  assumes  that for  the  ``typically  encountered''
realization of  $\rho,{\bf g}$, the  above relation is  well satisfied
inside conditional expectations $\langle\cdots\rangle^{{\bf R}\rho{\bf
    g}}$.  It is  obvious that such an approximation  makes sense only
if the  conditioning values  $\rho_\mu,{\bf g}_\mu$ for  the densities
are  such that  they  correspond  to a  sufficiently  large number  of
particles in cell  $\mu$.  Eqs. (\ref{Approx0}) need  to be understood
in the weak  sense, this is, valid within  expressions involving space
integrals.   Note  that  if  we   multiply  both  sides  of  the  {\em
  approximate}  equations (\ref{Approx0})  with $\delta_\nu({\bf  r})$
and   integrate  over   space  we   get  an   \textit{exact}  identity
$\hat{\rho}_\mu(z)=\hat{\rho}_\mu(z)$ for all  microscopic states $z$;
this gives  us confidence in the  self-consistency of this
approximation.

As we demonstrate in the  Appendix, the linear for spiky approximation
allows us to replace hatted functions with overlined functions, and to
transform  the double  brackets $\bra\cdots\ket^{{\bf  R}\rho{\bf g}}$
into  simple space  averages $\norm\cdots\norm$.  This transforms  the
exact results  for the  reversible drift  into approximate  but closed
expressions, as we explain next.

\subsubsection{Approximate mass matrix}

The     microscopic    mass     matrix    $\hat{M}_{\mu\nu}(z)$     in
(\ref{taghatMmunu})  can   be  exactly  expressed  in   terms  of  the
microscopic    field   $\hat{\rho}_{\bf     r}(z)$    introduced    in
(\ref{omicdens}),
\begin{align}
 \hat{M}_{\mu\nu}(z)&=\norm\delta_\mu\delta_\nu\hat{\rho}(z)\norm
\label{Mrho}
\end{align}
Note that this matrix satisfies the following exact results
\begin{align}
{\cal V}_\mu \hat{M}_{\mu\nu}(z)&= \hat{\rho}_\nu(z),
&&
{\cal V}_\nu \hat{M}_{\mu\nu}(z)= \hat{\rho}_\mu(z)
\label{hydroconsistency}
\end{align}
where use has been made of the first equation (\ref{propdelta}).

Under  the linear  for spiky  approximation (\ref{Approx0}),  the mass
matrix in (\ref{Mrho}) becomes
\begin{align}
 \hat{M}_{\mu\nu}(z)&
\simeq\norm\delta_\mu\delta_\nu\psi_\sigma\norm\hat{\rho}_\sigma(z)
\label{Mapprox}
\end{align}
and therefore, in this approximation the matrix $ \hat{M}_{\mu\nu}(z)$
depends  on  the microstate  $z$  only  through the  \textit{discrete}
density     field    $\hat{\rho}_\sigma(z)$.      The    approximation
(\ref{Mapprox}) is  consistent in the sense that it fulfills the
exact properties  (\ref{hydroconsistency}).  Note that for  a function
of  relevant variables  $F(\hat{x}(z))$  the conditional  expectations
satisfies   $\llangle  F(\hat{x})\rrangle^x=F(x)$.    By  using   this
property, the conditional expectation  of the mass matrix (\ref{Mrho})
is
\begin{align}
\llangle \hat{M}_{\mu\nu}\rrangle^{{\bf R}\rho{\bf g}}
&\simeq\norm\delta_\mu\delta_\nu\psi_\sigma\norm\rho_\sigma
=\norm\overline{\rho}\delta_\mu\delta_\nu\norm \equiv  \overline{M}_{\mu\nu}(\rho)
\label{calM}
\end{align}
where the interpolated mass  density field $ \overline{\rho}({\bf r})$
is  defined in  (\ref{orho}) and  we have  introduced the  mass matrix
$\overline{M}_{\mu\nu}(\rho)$ (with dimensions of  mass over volume squared)
for notational convenience.

\subsubsection{Approximate reversible generator}

In Appendix  \ref{App:RevApprox}, Eq. (\ref{revLformApprox}),  we show
that under  the linear  for spiky approximations  (\ref{Approx0}) the
exact  reversible drift  {originating  from  the reversible  operator}
(\ref{ExactDrift}) becomes
\begin{widetext}
\begin{align}
&\left( \begin{array}{c}
\llangle L{\bf R}\rrangle^{{\bf R}\rho{\bf g}} \\
\\
\llangle L\rho_\mu\rrangle^{{\bf R}\rho{\bf g}} \\
\\
\llangle L{\bf g}^\alpha_\mu \rrangle^{{\bf R}\rho{\bf g}}
 \end{array}\right)
=\left( \begin{array}{ccc}
0&
0&
\delta_\mu({\bf R})
\\
\\
0&
0&
\norm\overline{\rho}\delta_\nu\boldsymbol{\nabla}^\beta\delta_\mu \norm
\\
\\
-\delta_\mu({\bf R}) 
&-\norm\overline{\rho}\delta_\mu\boldsymbol{\nabla}^\alpha\delta_\nu \norm
&\norm\overline{\bf g}^\alpha\delta_\nu\boldsymbol{\nabla}^\beta\delta_\mu\norm
-\norm\overline{\bf g}^\beta \delta_\mu\boldsymbol{\nabla}^\alpha\delta_\nu\norm
 \end{array}\right)
\left( \begin{array}{c}
\frac{\partial {\cal H}}{\partial {\bf R}} \\
\\
\frac{\partial {\cal H}}{\partial \rho_\nu} \\
\\
\frac{\partial {\cal H}}{\partial {\bf g}^\beta_\nu} 
 \end{array}\right)
-k_BT
\left( \begin{array}{c}
0\\
\\
0
 \\
\\
-\boldsymbol{\nabla}^\alpha\delta_\mu({\bf R})
 \end{array}\right)
\label{revLFSA}
\end{align}
\end{widetext}
The  interpolated density and velocity fields
are defined as
\begin{align}
  \overline{\rho}({\bf r})&={\rho}_\mu\psi_\mu({\bf r})
\nonumber\\
  \overline{\bf g}({\bf r})&={\bf g}_\mu\psi_\mu({\bf r})
\end{align}
and the  double bar notation introduced in (\ref{NotationInt}) describes
integration  over  all space.  The  stochastic  drift proportional  to
$k_BT$ emerging from  the divergence of the reversible  matrix is very
simple and, for the case of no suspended particles, indicates that the
reversible dynamics  follows a  Hamiltonian dynamics, i.e.,  the phase
space flow is incompressible.

\subsubsection{Approximate CG Hamiltonian}

In appendix \ref{App:Models}, see Eq. (\ref{CGHApprox}), we show
that under  the  linear for  spiky  approximation  (\ref{Mapprox}), the  CG
Hamiltonian (\ref{CGHamiltonian}) becomes
\begin{align}
  {\cal H}({\bf R},\rho,{\bf g})&= \frac{1}{2}{\bf g}_\mu\overline{M}^{-1}_{\mu\nu}{\bf g}_\nu
+{\cal F}\left({\bf R},\rho\right)+\Phi^{\rm ext}({\bf R})
\label{CGHamiltonianB}
\end{align}
The  CG Hamiltonian  is  the  free energy  of  the  selected level  of
description, but  we refer to  it as a  CG Hamiltonian because  of the
presence  of a  quadratic term  {in momenta}  that can  be
interpreted as a ``kinetic energy''  plus a ``potential energy'' given
by   the    intrinsic   fluid   free   energy    ${\cal   F}\left({\bf
    R},\rho\right)$.   This   free  energy  is  given   rigorously  by
(\ref{FreeEnergyDecomp}).

In     Appendix      \ref{App:Models},
Eq. (\ref{SimpleFreeEnergyModel})  we introduce an explicit  model for
the free energy (\ref{FreeEnergyDecomp}) 
\begin{align}
  {\cal F}({\bf R},\rho)&=\frac{c^2}{2\rho_{\rm eq}}\delta\rho_\mu M^\psi_{\mu\nu}\delta\rho_\nu
+\frac{m_0(c_0^2-c^2)}{\rho_{\rm eq}}\psi_\mu({\bf R})\rho_\mu
\label{OurModel}
\end{align}
where  $\delta\rho_\mu =  \rho_\mu  - \rho_{\rm  eq}$  is the  density
perturbation away  from the average  solvent density $\rho_{\rm  eq} =
M_T / {\cal V}_T$,  with ${\cal V}_T$ being the total system  volume. The motivation
behind  this model  is that  it  gives Gaussian  fluctuations for  the
solvent in the absence of any suspended nanoparticle, and describes in
a CG manner  the interaction between the nanoparticle  and the solvent
in  such  a way  that  gradients  of  density  produce forces  on  the
nanoparticle.  The  parameter $c_0$  with dimensions of  speed governs
the intensity of these forces. When the nanoparticle is simply
a tagged solvent particle, $c_0=c$.

The derivatives of the CG Hamiltonian (\ref{CGHamiltonianB})
are computed in Appendix \ref{App:Models}, Eq. (\ref{derHB0})
\begin{align}
  \frac{\partial {\cal H}}{\partial {\bf R}}&=
\frac{\partial{\cal F}}{\partial{\bf R}}
+  \frac{\partial \Phi^{\rm ext}}{\partial {\bf R}}
\nonumber\\
  \frac{\partial {\cal H}}{\partial \rho_\mu}&=
-\frac{1}{2}\norm\psi_\mu\overline{\bf v}\overline{\bf v}\norm
+   \frac{\partial {\cal F}}{\partial \rho_\mu}
\nonumber\\
  \frac{\partial {\cal H}}{\partial {\bf g}_\mu}&=M^\psi_{\mu\mu'}{\bf v}_{\mu'}
\label{DerAproxCGH}
\end{align}
where the discrete velocity is defined as
\begin{align}
{\bf v}_\mu &\equiv M^\delta_{\mu\nu}  \overline{M}_{\nu\nu'}^{-1}{\bf g}_{\nu'}
\label{DiscreteVel}
\end{align}
which is given  in terms of the density dependent  mass matrix and the
momentum density  field. The reason for introducing this somewhat involved
definition for the hydrodynamic velocity is justified by the resulting form of the discrete
hydrodynamic equations, resembling in form the structure of the continuum equations. Note that in
an ``incompressible''  limit in  which we
assume  that  the  density  fluctuations   are  very  small  and  then
$\rho_\mu=\rho_{\rm  eq}$, the above expression  simplifies to
${\bf v}_\mu =\rho_{\rm eq}^{-1}{\bf g}_\mu$ because of
\begin{align}
\overline{M}_{\mu\nu}=\left\Vert \bar{\rho}\delta_{\mu}\delta_{\nu}\right\Vert 
\simeq \rho_{\rm eq}\left\Vert \delta_{\mu}\delta_{\nu}\right\Vert 
=\rho_{\rm eq}M_{\mu\nu}^{\delta}.
\end{align}

Note that (\ref{DiscreteVel}) may be written as
\begin{align}
{\bf g}_{\mu}&\equiv   \overline{M}_{\mu\nu}M^\psi_{\nu\nu'}{\bf v}_{\nu'}
=\rho_\sigma\norm\psi_\sigma\delta_\mu\delta_\nu\norm 
M^\psi_{\nu\nu'}{\bf v}_{\nu'}
\nonumber\\
&=\norm\delta_\mu\psi_\sigma\psi_\nu\norm\rho_\sigma{\bf v}_\nu
=\norm\delta_\mu\overline{\rho}\;\overline{\bf v}\norm
\label{DiscreteVel1}
\end{align}
This  allows  to  write the interpolated momentum density field as
\begin{align}
  \overline{\bf g}({\bf r})&=\psi_\mu({\bf r})\norm 
\delta_\mu\overline{\rho}\overline{\bf v}\norm
  \label{gbar_approx}
\end{align}
If  we use  (\ref{LFSA}) under  an assumption  of sufficiently smooth
fields, which should apply in the  limit when the grid cells are large
and fluctuations are small, we obtain the {\em local} relationship
\begin{align}
  \overline{\bf g}({\bf r})\simeq \overline{\rho}({\bf r})\overline{\bf 
v}({\bf r})
  \label{local_g_v}
\end{align}
which  is  the familiar  continuum  definition  of velocity  from the
momentum and  mass densities. In general, however, (\ref{local_g_v}) does
not hold identically and we prefer to define $\overline{\bf v}({\bf r})$
as the interpolant based on the discrete velocities (\ref{DiscreteVel}).

\subsubsection{Approximate reversible drift}

We   may  perform   explicitly  the   matrix  multiplication   in  Eq.
(\ref{revLFSA}) with (\ref{DerAproxCGH}). This  leads to the following
approximate form for the reversible drift
\begin{align}
\llangle L{\bf R}\rrangle^{{\bf R}\rho{\bf g}} &= \overline{\bf v}({\bf R})
\nonumber\\
\llangle L\rho_\mu\rrangle^{{\bf R}\rho{\bf g}} 
&=\norm\overline{\rho}\;\overline{\bf v}\esc\boldsymbol{\nabla}\delta_\mu\norm
\nonumber\\
\llangle L{\bf g}^\alpha_\mu \rrangle^{{\bf R}\rho{\bf g}}&=
\norm \overline{\bf g}\;\overline{\bf v}\esc\boldsymbol{\nabla}\delta_\mu\norm
+k_BT\boldsymbol{\nabla}\delta_\mu({\bf R})
\nonumber\\
&-\delta_\mu({\bf R})\frac{\partial{\cal F}}{\partial{\bf R}}
-\norm \overline{\rho}\delta_\mu\boldsymbol{\nabla}\delta_\nu \norm \frac{\partial {\cal F}}{\partial \rho_\nu}+\delta_\mu({\bf R}){\bf F}^{\rm ext}
\nonumber\\
&+\frac{1}{2}\left(\norm \overline{\rho}\delta_\mu\boldsymbol{\nabla}\delta_\nu \norm
\norm\psi_{\nu}\overline{\bf v}^2\norm
-\norm \overline{\rho} \delta_\mu\boldsymbol{\nabla}\overline{\bf v}^2\norm\right)
\label{AproxDrift}
\end{align}
By conforming to the  structure (\ref{driftL}), the {reversible drift}
(\ref{AproxDrift})  preserves  the equilibrium  distribution  function
$e^{-\beta  {\cal H}}$.   The  total  mass (\ref{TotalInvariants})  is
conserved  by  the  above  equations,  as a  result  of  the  identity
(\ref{PartitionUnity}).   However,  total   momentum  is  not  exactly
conserved.  Since in the molecular ensemble (\ref{molens}) momentum is
conserved,  it  is important  to  conserve  momentum strictly  in  the
coarse-grained dynamics as well when ${\bf F}^{\rm ext}=0$, \Pepmodified{and we discuss this issue next}.

The rate of change of the total  momentum is given by
\begin{align}
  \frac{d{\bf P}_T}{dt}&=
-\frac{\partial{\cal F}}{\partial{\bf R}}
-\norm \overline{\rho}\boldsymbol{\nabla}\delta_\nu \norm \frac{\partial {\cal F}}{\partial \rho_\nu}
\nonumber\\
&+\frac{1}{2}\left(\norm \overline{\rho}\boldsymbol{\nabla}\delta_\nu \norm
\norm\psi_{\nu}\overline{\bf v}^2\norm
-\norm \overline{\rho}\boldsymbol{\nabla}\overline{\bf v}^2\norm\right)
\label{MomentumLeak}
\end{align}
which  does  not  necessarily   vanish.   The  violation  of  momentum
conservation    is    weak,     however.     First,    consider    the
{velocity}  terms  in   (\ref{MomentumLeak}).   Under  the
assumption of smooth fields, Eq.  (\ref{LFSAB}) applies and shows that
the  difference  of  two  terms  in  the  parenthesis  (last  term  in
(\ref{MomentumLeak}))    is    small    (second    order    in    grid
spacing). Therefore, we will neglect the last two term in the momentum
equation in (\ref{AproxDrift}).  Second,  consider the terms involving
the  free energy  in  (\ref{MomentumLeak}).
We have shown in Eqs.  (\ref{ExactTIofF}) and (\ref{r4})
in the Appendices that the translational invariance of the microscopic
Hamiltonian is reflected in the  following approximate property of the
free energy
\begin{align}
\frac{\partial{\cal F}}{\partial{\bf R}}
+\norm\overline{\rho}\boldsymbol{\nabla}\delta_\nu\norm\frac{\partial {\cal F}}{\partial\rho_\nu}
&=0
\label{TIofF}
\end{align}
relating the  gradient of  the free energy  to the  chemical potential
$\frac{\partial  {\cal F}}{\partial\rho_\mu}$.  This identity  implies
the  first  two  terms  in   (\ref{MomentumLeak})  cancel.  In  a  way
reminiscent   of  Noether's   theorem,  the   microscopic  translation
invariance (\ref{TIofF})  implies total  momentum conservation  in Eq.
(\ref{MomentumLeak}).

Unfortunately, the model for the free energy (\ref{OurModel}) does not
\textit{strictly} respects the property (\ref{TIofF}).  However, as we
explain  in Appendix  \ref{App:Models},  we can  restore the  property
(\ref{TIofF}) by  making the plausible approximation  that the density
field is sufficiently smooth
\begin{align}
\boldsymbol{\nabla}\overline{\rho}({\bf   R})\simeq
\norm\Delta\boldsymbol{\nabla}\overline{\rho}\norm   
\equiv \int d{\bf r}\Delta({\bf R},{\bf r})\boldsymbol{\nabla}\overline{\rho}({\bf r})
\label{rhoapp}
\end{align}
Recall  that the  reason why  (\ref{rhoapp}), which  is an  example of
(\ref{GradSmooth}), is not  an exact identity is due to  the fact that
the  regularized  Dirac  delta  is  not  translation  invariant,  i.e.
$\Delta({\bf r},{\bf  r'})\neq\Delta({\bf r}-{\bf  r'})$; this  is the
origin  of the  (small)  violation of  momentum  conservation.  If  we
nevertheless assume  that the  approximation (\ref{rhoapp})  is valid,
then   Eq.   (\ref{TIofF})  is   fulfilled   as   shown  in   Appendix
\ref{App:Models},  Eq.  (\ref{TIofFApp}) \Pepmodified{and we restore exact momentum
conservation}.   

In a  similar spirit,  the
terms involving the free energy  in the momentum equation are computed
in  Appendix \ref{App:Models},  in particular  (\ref{RhoGradMu}), with
the result
\begin{align}
-  \delta_\mu({\bf R})\frac{\partial{\cal F}}{\partial{\bf R}}
-\norm \overline{\rho}\delta_\mu\boldsymbol{\nabla}\delta_\nu \norm 
\frac{\partial {\cal F}}{\partial \rho_\nu}&=  -\norm\delta_\mu\boldsymbol{\nabla}P\norm
\label{form}
\end{align}
where we have introduced the ``pressure'' field
\begin{align}
P({\bf r})&=
\frac{c^2}{2\rho_{\rm eq}}\left(\overline{\rho}({\bf r})^2-{\rho}_{\rm eq}^2\right)
+m_0\frac{(c_0^2-c^2)}{\rho_{\rm eq}}\Delta({\bf R},{\bf r})\overline{\rho}({\bf r})
\label{TotalPressure0}
\end{align}
which consists  of two parts,  the first  being the equation  of state
corresponding  to  the Gaussian  model  for  the solvent  free  energy
density,  and  the  second   one  capturing  the  solvent-nanoparticle
interaction.  Note that this second contribution vanishes for a tagged
fluid molecule, when $c_0=c$.

Inserting  the result  (\ref{form}) in  (\ref{AproxDrift}) we  get the
final approximation of the reversible part of the momentum equation,
\begin{align}
\llangle L{\bf g}^\alpha_\mu \rrangle^{{\bf R}\rho{\bf g}}&=
\norm \overline{\bf g}\;\overline{\bf v}\esc\boldsymbol{\nabla}\delta_\mu\norm
+k_BT\boldsymbol{\nabla}\delta_\mu({\bf R})
\nonumber\\
&-\norm\delta_\mu\boldsymbol{\nabla}P\norm
+\delta_\mu({\bf R}){\bf F}^{\rm ext}
\label{gconserving}
\end{align}
This form exactly  conserves momentum, at the expense  of breaking the
structure   (\ref{driftL}).   As   a   consequence,  the   equilibrium
distribution  that   results  from   using  the   momentum  conserving
(\ref{gconserving})  instead of  (\ref{AproxDrift})  will be  slightly
different from  $\propto e^{-\beta{\cal H}}$.   \Pepmodified{Note that
  even if we has exactly $  e^{-\beta{\cal H}}$, the model of the free
  energy  (\ref{OurModel})  leads  to   the  a}  marginal  equilibrium
distribution of the particle  position \Pepmodified{that} is not given
by   the   Gibbs-Boltzmann  distribution   $\exp\left\{-\beta\Phi^{\rm
    ext}({\bf R})\right\}$ but rather by (\ref{PRnotBaro}).


\section{\label{Sec:Irrev}The irreversible part of the dynamics}

The  dissipative  matrix  (\ref{M}) involves  the  projected  currents
$\delta  L\hat{x}=L\hat{x}(z) -\langle  L\hat{x}\rangle^{\hat{x}(z)}$,
where       $\hat{x}(z)=\{\hat{\bf       R},\hat{\rho}_\mu(z),\hat{\bf
  g}_\mu(z)\}$ and $LX$ are the time derivatives of the
  relevant  variables. They  are  obtained by  applying the  Liouville
  operator  on the  position of  the nanoparticle,  mass and  momentum
  local densities.  In order to compute the time derivatives of the CG
  hydrodynamic  variables it  is  useful to  first  consider the  time
  derivatives  of   the  microscopic  local   fields  $\hat{\rho}_{\bf
    r}(z),\hat{\bf  g}_{\bf r}(z)$  defined in  (\ref{omicdens}) which
  are standard \cite{Grabert1982}. For pair-wise interactions they are
\begin{align}
 L \hat{\rho}_{\bf r}(z) =& -\boldsymbol{\nabla}\esc\hat{\bf g}_{\bf r}(z)
\nonumber\\
 L \hat{\bf g}_{\bf r}(z) =& -\boldsymbol{\nabla}\esc\hat{\boldsymbol{\sigma}}_{\bf r}
+{\bf F}^{\rm ext}({\bf q}_0)\delta({\bf q}_0-{\bf r})
\label{oildens}
\end{align}
where the stress tensor has the standard form
\begin{align}
\hat{\boldsymbol{\sigma}}_{\bf r}&=\sum_{i=0}^N{\bf p}_i{\bf v}_i\delta({\bf q}_i-{\bf r})
\nonumber\\
&+\frac{1}{2}\sum_{i,j=0}^N{\bf q}_{ij}{\bf F}_{ij}
\int_0^1d\epsilon\;\delta({\bf r}-{\bf q}_i+\epsilon{\bf q}_{ij})
\label{stresstensor}
\end{align}
Note that  the stress  tensor includes the  nanoparticle $i=0$  in its
definition.

The     time     derivatives     of     the     relevant     variables
$\hat{\rho}_\mu(z),\hat{\bf   g}_\mu(z)$   can    be   obtained   with
(\ref{omacmic1})  from   the  time  derivatives   of  $\hat{\rho}_{\bf
  r}(z),\hat{\bf g}_{\bf r}(z)$.  They are given by
\begin{align}
L{\bf R}&=\frac{{\bf p}_0}{m_0}
\nonumber\\
L\hat{\rho}_\mu(z)&=\sum_{i=0}^N{\bf p}_i\esc\boldsymbol{\nabla} \delta_\mu({\bf q}_i)
=\int d{\bf r}\boldsymbol{\nabla} \delta_\mu({\bf r}) \esc \hat{\bf g}_{\bf r}(z)
\nonumber\\
  L\hat{\bf g}_\mu(z)&=\int d{\bf r}\boldsymbol{\nabla}\delta_\mu({\bf r}) \esc \boldsymbol{\sigma}_{\bf r}(z)
+{\bf F}^{\rm ext}({\bf q}_0)\delta_\mu({\bf q}_0)
\label{LgSigmaB}
\end{align}
The
corresponding  reversible part  $\langle L\hat{x}\rangle^{\hat{x}(z)}$
that  is  subtracted   in  the  projected  current   \Pepmodified{has been}  computed  in
Eq. (\ref{AproxDrift}).  

We will discuss shortly the projected current corresponding
  to the  position of the  colloid, which  will be denoted  by $\delta
  L\hat{\bf R} \equiv  \delta \hat{\bf V}$.  By using  the linear for
spiky  approximation  (\ref{Approx0}),  we can  approximate  the  time
derivative of the density variable in (\ref{LgSigmaB}) as follows
\begin{align}
L\hat{\rho}_\mu(z)&\simeq\int d{\bf r} \; \psi_\nu({\bf r}) \boldsymbol{\nabla} \delta_\mu({\bf r}) \esc \hat{\bf g}_{\nu}(z)
\label{Lrhoapp2}
\end{align}
In this approximation, the time derivative of a relevant variable (the
density)  is  itself  given  in  terms of  a  relevant  variable  (the
momentum).  Therefore,  the corresponding projected  current vanishes,
i.e.  $\delta\rho_\mu(z)=0$,  resulting in  a great  simplification of
the  dissipative matrix.   From  Eq.  (\ref{LgSigmaB}),  the
projected current  corresponding to the  momentum may be  expressed in
the form
\begin{align}
 \delta L\hat{\bf g}_\mu(z)&=\int d{\bf r}\; \boldsymbol{\nabla}\delta_\mu({\bf r})  
\esc \delta\hat{\boldsymbol{\sigma}}_{\bf r}
\label{LgSigmaB2}
\end{align}
where the fluctuations of the stress tensor are
\begin{align}
\delta\hat{\boldsymbol{\sigma}}_{\bf r}&\equiv  \hat{\boldsymbol{\sigma}}_{\bf r}(z)-
\llangle\hat{\boldsymbol{\sigma}}_{\bf r}\rrangle^{\hat{\bf R}\hat{\rho}\hat{\bf g}}
\label{projsigma}\end{align}
The external force term in Eq. (\ref{LgSigmaB}) disappears from
the projected current (\ref{LgSigmaB2}) because  it is just a function
of ${\bf q}_0={\bf R}$ which is a relevant variable.

By using (\ref{LgSigmaB2}), we can write the dissipative matrix  ${\cal D}(x)$
as a collection of Green-Kubo integrals
\begin{widetext}
\begin{align}
&\frac{1}{k_BT}\int_0^\tau dt  
\left( \begin{array}{ccc}
\llangle \delta \hat{\bf V}^\beta(0)\delta \hat{\bf V}^\alpha(t)\rrangle^{\hat{\bf R}\hat{\rho}\hat{\bf g}}&
0&
\int d{\bf r}'\boldsymbol{\nabla}^{\beta'}\delta_\nu({\bf r}')  \llangle 
\delta\hat{\boldsymbol{\sigma}}^{\beta\beta'}_{{\bf r}'}(0)
\delta \hat{\bf V}^\alpha(t)\rrangle^{{\bf R}\rho{\bf g}}
\\
\\
0&0&0
\\
\\
\int d{\bf r}\boldsymbol{\nabla}^{\alpha'}\delta_\mu({\bf r}')  \llangle 
\delta \hat{\bf V}^\beta(0)\delta \hat{\boldsymbol{\sigma}}^{\alpha\alpha'}_{{\bf r}'}(t)
\rrangle^{{\bf R}\rho{\bf g}}
&
0
&
 \int d{\bf r}\int d{\bf r}'
\llangle 
\delta\hat{\boldsymbol{\sigma}}^{\beta\beta'}_{{\bf r}'}(0)
\delta\hat{\boldsymbol{\sigma}}^{\alpha\alpha'}_{{\bf r}}(t)
\rrangle^{\hat{\bf R}\hat{\rho}\hat{\bf g}}
(\boldsymbol{\nabla}^{\alpha'} \delta_\mu({\bf r})\boldsymbol{\nabla}^{\beta'} \delta_\nu({\bf r}')
\end{array}\right)
\label{MSigmaB}
\end{align}
\end{widetext}

In  general, the  {dissipative matrix}  depends on  the  values of  the
coarse-grained  variables  {${\bf R},\rho,{\bf  g}$}  that
condition the expectation values in (\ref{MSigmaB}).  {Consider, for
example, the colloid diffusion tensor defined as
\begin{align}
{\bf D}(x)&=
\llangle \delta \hat{\bf V}(0)\delta \hat{\bf V}(t)\rrangle^{\hat{\bf R}\hat{\rho}\hat{\bf g}}
\nonumber\\
&=
\int_0^\tau dt  \frac{\rho^{\rm eq}(z)\delta(\hat{x}(z)-x)}{P^{\rm eq}(x)}
\delta\hat{\bf V}(0) \delta\hat{\bf V}(t)
\label{D(x)}
\end{align}
}Indeed, even for
a  dilute nanocolloidal  suspensions,  had  we tried  to  jump to  the
Smoluchowski level (using  only the position of the  nanocolloids as a
slow variable) directly, the diffusion tensor
would depend strongly on the configuration because of the hydrodynamic
interactions (correlations)  between the  particles.  At our  level of
description, however, we can assume that, to a good approximation, the
dissipative matrix  does not  depend on the  configuration and  can be
approximated by its {\em equilibrium  average}, i.e., by replacing the
conditional expectations in (\ref{MSigmaB}) with equilibrium averages.
In this approximation,
\begin{align}
{\bf D}(x)\simeq {\bf D}^{\rm eq}\equiv\int dx' P^{\rm eq}(x'){\bf D}(x')
\label{eqapp}
\end{align}
By inserting (\ref{D(x)}) into  (\ref{eqapp}) and integrating over the
Dirac delta function gives
\begin{align}
{\bf D}^{\alpha\beta}(x) &\simeq \int_0^\tau dt \llangle\delta\hat{\bf V}^\beta(0)\delta\hat{\bf V}^\alpha(t)\rrangle_{\rm eq}
\label{D_x_approx}
\end{align}
where  the average  is now  an ordinary  equilibrium ensemble  average
rather than a constrained one.

Under the approximation in which the dissipative matrix is substituted
by  its   equilibrium  average,  the  non-diagonal   elements  of  the
dissipative  matrix  (\ref{MSigmaB}),  which  involve  a  third  order
tensor, will vanish because the  equilibrium ensemble is isotropic and
the  only  isotropic  third  order   tensor  is  the  null  one.   The
dissipative matrix becomes
\begin{widetext}
\begin{align}
{\cal D}(x) =  \left( \begin{array}{ccc}
\int_0^\tau dt \llangle\delta\hat{\bf V}^\beta(0)\delta\hat{\bf V}^\alpha(t)\rrangle_{\rm eq}&
0&
0\\
\\
0&0&0
\\
\\
0
&
0
&
 \int d{\bf r}\int d{\bf r}'
\boldsymbol{\eta}^{\alpha\alpha'\beta\beta'}_{{\bf  r}{\bf  r}'}
\boldsymbol{\nabla}^{\alpha'} \delta_\mu({\bf r})\boldsymbol{\nabla}^{\beta'} \delta_\nu({\bf r}')
\end{array}\right)
\label{MSigmaB2}
\end{align}
\end{widetext}
where we have introduced a  fourth order tensorial non-local viscosity
kernel
\begin{align}
\boldsymbol{\eta}^{\alpha\alpha'\beta\beta'}_{{\bf  r}{\bf  r}'} \equiv& \frac{1}{k_BT} \int_0^\tau dt\llangle 
\delta\hat{\boldsymbol{\sigma}}^{\beta\beta'}_{{\bf r}'}(0)
\delta\hat{\boldsymbol{\sigma}}^{\alpha\alpha'}_{{\bf r}}(t)
\rrangle^{\rm eq}
\label{Deta}
\end{align}

\subsection{Mass diffusion}

The projected  current corresponding  to the
position is given by
\begin{align}
\delta L\hat{\bf R}&=\hat{\bf V}-\hat{\bf v}^{\rm hydro}
\equiv \delta \hat{\bf V}
\label{deltaV}
\end{align}
where  we have  denoted by  $\hat{\bf V}=L\hat{\bf  R}=\frac{{\bf p}_0}{m_0}$  the
velocity of  the nanoparticle. The term 
$\hat{\bf v}^{\rm hydro}$ is the
reversible part of the evolution of ${\bf R}$, given in the first equation in (\ref{AproxDrift}),  evaluated at the microscopic value of
the phase functions, this is
\begin{align}
\hat{\bf v}^{\rm hydro}(z)=\llangle L\hat{\bf R}\rrangle^{\hat{\bf R}\hat{\rho}\hat{\bf g}}&=\psi_\mu(\hat{\bf R})M^{\delta}_{\mu\nu}\overline{M}^{-1}_{\nu\nu'}(\hat{\rho}(z))\hat{\bf g}_{\nu'}(z)
\label{FlucVel}
\end{align}

We  expect
that, being  an equilibrium average, which  is rotationally invariant,
the tensor ${\bf D}(x)$ given in (\ref{D_x_approx}) is, in fact, diagonal and of the form
\begin{align}
  {\bf D}^{\alpha\beta}&=D_0\delta^{\alpha\beta}
\end{align}
Here the scalar {\em bare diffusion coefficient} is given by
\begin{align}
  D_0&=\frac{1}{d}\int_0^\tau dt
\llangle\delta\hat{\bf V}(0)\esc\delta\hat{\bf V}(t)\rrangle_{\rm eq}
\label{Drenorm}
\end{align}
where $d$ is the dimensionality, and  $\delta\hat{\bf  V}$   is  defined   in  (\ref{deltaV})   with
(\ref{FlucVel}) as the fluctuation of the velocity of the nanoparticle
relative to the  surrounding flow velocity.  

Note  that the bare diffusion
coefficient is  different from  the macroscopic or {\em renormalized  diffusion coefficient},
\begin{align}
  D&=\frac{1}{d}\int_0^\tau dt\llangle 
\hat{\bf V}(0)\esc\hat{\bf V}(t)\rrangle^{\rm eq}
\label{D0}\end{align}
defined {\em  without} subtracting  the interpolated  fluid velocity.
We can  split the  renormalized diffusion  coefficient into  two parts
\cite{DiffusionJSTAT}, the  bare part which comes  from under-resolved
details of  the dynamics  occurring at length  and time  scales shorter
than  the ones  explicitly  represented by  the discrete  hydrodynamic
grid, and an  enhancement $\Delta D$ that comes from  the advection by
the  thermal  velocity  fluctuations  and  accounts  for  hydrodynamic
transport explicitly resolved by the discrete grid,
\begin{align}
  D&= D_0  + \Delta D = D_0  
\nonumber\\
&+ 
  \frac{1}{d}\int_0^\tau dt \llangle \Vav(0)\esc\Vav(t) \right.
\nonumber\\
&+ \left.
  \Vav(0)\esc\delta\hat{\bf V}(t) + \delta\hat{\bf V}(0)\esc\Vav(t)  \rrangle^{\rm eq} 
\label{D_split}
\end{align}
Observe that  $\Delta D$  contains a  lot of  hydrodynamic information
because of the  time lag in the time correlation  function; during the
time $t$ hydrodynamic information  (sound waves, viscous dissipation,
etc.)  propagates  around  the  particle  and  affects  its  diffusion
coefficient.

As we elaborate in more detail  in the Conclusions, the bare diffusion
coefficient (\ref{Drenorm})  depends on  the size of  the hydrodynamic
cells, i.e., on the resolution  at which hydrodynamics is represented.
By  contrast, the  renormalized  diffusion  coefficient (\ref{D0})  is
independent of  the resolution of  the grid. However, as  mentioned in
the  introduction,  $D$  is   not  really  computable  {in
  practice} in  MD, as opposed  to $D_0$,  since the upper  time limit
$\tau$   should  be   {\em  much}   larger  in   (\ref{D0})  than   in
(\ref{Drenorm}).

\subsection{Momentum Diffusion}

The range of the viscous kernel given in (\ref{Deta}) is that of the
correlation length  of the  stress tensor.  We  will assume  that this
range is much smaller than the size  of the cells, i.e.  in the length
scale in which $\boldsymbol{\eta}_{{\bf r}{\bf r}'}$ is different from
zero,  the function  $\boldsymbol{\nabla} \delta_\mu({\bf  r})$ hardly
changes.  Note  that the  stress tensor  (\ref{stresstensor}) contains
the contribution  of the  colloidal particle.  Therefore,  a condition
for this locality assumption is  that the colloidal particle itself is
much smaller  than the  grid size. If  this is the  case, then  we may
adopt a local approximation
\begin{align}
  \boldsymbol{\eta}_{{\bf r}{\bf r}'}\simeq
  \boldsymbol{\eta}\delta({\bf r}-{\bf r}')
\label{etadelta}
\end{align}
and therefore the viscous contribution to the dissipative matrix (\ref{MSigmaB2}) is
\begin{align}
& \int d{\bf r}\int d{\bf r}'
\boldsymbol{\eta}_{{\bf  r}{\bf  r}'}^{\alpha\alpha'\beta\beta'}
\boldsymbol{\nabla}^{\alpha'} \delta_\mu({\bf r})
\boldsymbol{\nabla}^{\beta'} \delta_\nu({\bf r}')
\nonumber\\
&\simeq
\boldsymbol{\eta}^{\alpha\alpha'\beta\beta'}
\norm\boldsymbol{\nabla}^{\alpha'}\delta_\mu
\boldsymbol{\nabla}^{\beta'} \delta_\nu\norm
\end{align}

The explicit microscopic expression for $\boldsymbol{\eta}$ in (\ref{etadelta}) is obtained
by integrating  the
viscosity kernel over  ${\bf  r},{\bf r}'$  to get
\begin{align}
  \int d{\bf r}\int d{\bf r}'  \boldsymbol{\eta}^{\alpha\alpha'\beta\beta'}_{{\bf r}{\bf r}'}
 \equiv& \frac{1}{k_BT} \int_0^\tau dt\llangle 
\delta\hat{\boldsymbol{\sigma}}^{\beta\beta'}(0)
\delta\hat{\boldsymbol{\sigma}}^{\alpha\alpha'}(t)
\rrangle^{\rm eq}
\label{eta1}
\end{align}
where the stress tensor of the whole system is, from (\ref{stresstensor})
\begin{align}
 \hat{\boldsymbol{\sigma}}^{\beta\beta'}&=\int d{\bf r}\;\hat{\boldsymbol{\sigma}}^{\beta\beta'}_{\bf r}
=
\sum_{i=0}^N{\bf p}_i^{\beta}{\bf v}_i^{\beta'}
+\frac{1}{2}\sum_{i,j=0}^N{\bf q}_{ij}^{\beta}{\bf F}_{ij}^{\beta'}
\label{wholestres}
\end{align}
By using (\ref{etadelta}) into (\ref{eta1}) gives
\begin{align}
\boldsymbol{\eta}^{\alpha\alpha'\beta\beta'} \equiv& \frac{1}{k_BT{\cal V}_T  } \int_0^\tau dt\llangle 
\delta\hat{\boldsymbol{\sigma}}^{\beta\beta'}(0)
\delta\hat{\boldsymbol{\sigma}}^{\alpha\alpha'}(t)
\rrangle^{\rm eq}  
\label{eta4}
\end{align}
where ${\cal V}_T$ is the volume of the system.  

The viscosity tensor,
being an equilibrium correlation, will  be isotropic. The general form
of the isotropic fourth order  tensor that accounts for the symmetries
of the stress tensor appearing in the Green-Kubo expression is
\begin{align}
   \boldsymbol{\eta}^{\alpha\alpha'\beta\beta'}
&\equiv
{\eta}\left(\delta^{\alpha\beta}\delta^{\alpha'\beta'}
+\delta^{\alpha\beta'}\delta^{\beta\alpha'}
-\frac{2}{\Pepmodified{d}}\delta^{\alpha\alpha'}\delta^{\beta\beta'}\right)
\nonumber\\
&+\zeta\delta^{\alpha\alpha'}\delta^{\beta\beta'}
\label{visctensorisotropic}
\end{align}
where $\eta,\zeta$  are shear and bulk  viscosities, respectively.  In
practice,  one  would  typically   neglect  the  contribution  of  the
nanoparticles to the  viscous stress and assume  that $\eta,\zeta$ are
the pure solvent equilibrium viscosities.

Finally, the dissipative matrix (\ref{MSigmaB2}) becomes
\begin{align}
{\cal D}(x)&\simeq  \left( \begin{array}{ccc}
\frac{D_0}{k_BT}\delta^{\alpha \beta}&
0&
0\\
\\
0&0&0
\\
\\
0
&
0
&
\boldsymbol{\eta}^{\alpha\alpha'\beta\beta'}
\norm\boldsymbol{\nabla}^{\alpha'} \delta_\mu\boldsymbol{\nabla}^{\beta'} \delta_\nu\norm
\end{array}\right)
\label{CalD}
\end{align}
Note the dissipative matrix is independent  of the state of the system
due to  its approximation with  its equilibrium average. As  a result,
the   stochastic    drift   term   $k_BT\partial_x\esc{\cal    D}(x)$   in
Eq. (\ref{sde}) should be taken as zero in this approximation.

\subsection{Noise terms }

In order to construct the Ito SDE (\ref{sde}) for the present level of
description,  we   need  to  specify  the   noise  terms  $\frac{d\tilde{\bf
  R}}{dt},\frac{d\tilde{\rho}_\mu}{dt},\frac{d\tilde{\bf g}_\mu}{dt}$. The  variance of the noise
is given  by the Fluctuation-Dissipation balance  (\ref{FD}) where the
matrix ${\cal D}(x)$ is given  by (\ref{CalD}).  From the structure of
this matrix we may infer that $\frac{d\tilde{\rho}_\mu}{dt}=0$ and
\begin{align}
  \llangle\frac{d\tilde{\bf R}}{dt}(t)\frac{d\tilde{\bf R}}{dt}(t')\rrangle&=2k_BTD_0\delta(t-t')
\nonumber\\
\llangle\frac{d\tilde{\bf g}_\mu^\alpha}{dt}(t)  \frac{d\tilde{\bf g}_\nu^\beta}{dt}(t')\rrangle &=2k_BT
\boldsymbol{\eta}^{\alpha\alpha'\beta\beta'}
\norm\boldsymbol{\nabla}^{\alpha'} \delta_\mu\boldsymbol{\nabla}^{\beta'} \delta_\nu\norm
\delta(t-t')
\label{FDTRg}
\end{align}
We need to produce next explicit linear combinations of white noise that
give rise to the above variances. While the velocity noise term is very simple
\begin{align}
\frac{d\tilde{\bf R}}{dt}(t)&=\sqrt{2k_BTD_0}\;\boldsymbol{\cal W}(t)
\label{RandomVel}
\end{align}
where $\boldsymbol{\cal W}(t)$ is a white noise, the explicit form of the 
random force $\frac{d\tilde{\bf g}_\mu}{dt}$ is not so obvious \Pepmodified{and will be considered next.}

The  noise term  in  the theory  of  CG  is just  a  modelling of  the
projected current appearing in the Green-Kubo expression (\ref{M}) as
a white noise. For this reason, it  is useful to look at the structure
of the projected current in Eq. (\ref{LgSigmaB2})
\begin{align}
\delta   L{\bf g}_\mu^{\alpha}&  
=M^\delta_{\mu\mu'}\int d{\bf r}\boldsymbol{\nabla}^\beta\psi_{\mu'}({\bf r})\delta\hat{\boldsymbol{\sigma}}^{\alpha\beta}_{\bf r}
\label{deltaLgmu}
\end{align}
We will model
$\delta\hat{\boldsymbol{\sigma}}^{\alpha\beta}_{\bf r}$ as a linear combination
of white noises of  the following form \cite{Espanol1998a}
\begin{align}
\delta\hat{\boldsymbol{\sigma}}^{\alpha\beta}_{\bf
  r}\simeq \boldsymbol{\Sigma} ^{\alpha\beta}_{\bf r} &= 
  \sqrt{2k_BT\eta}
\left[\W^{\alpha\beta}_{\bf r}(t)
-\delta^{\alpha\beta}
\frac{1}{d}\sum_\mu \W^{\mu\mu}_{\bf r}(t)\right]
\nonumber\\
&
+ \sqrt{\frac{k_BT\zeta}{d}}
\delta^{\alpha\beta}\sum_\mu \W^{\mu\mu}_{\bf r}(t)
\label{randoms0}
\end{align}
where the symmetric white-noise tensor $\W^{\mu\nu}_{\bf r}$ satisfies
\begin{align}
  \langle \W^{\mu\nu}_{\bf r}(t) \W^{\mu'\nu'}_{{\bf r}'}(t')\rangle 
&=
[\delta^{\mu\mu'}\delta^{\nu\nu'}+\delta^{\nu\mu'}\delta^{\mu\nu'}]
\nonumber\\
&\times
\delta({\bf r}-{\bf r}')\delta(t-t')
\label{eta}
\end{align}
It is straightforward to show that 
\begin{align}
\langle\delta{\M{\sigma}}^{\alpha\beta}_{\bf r}(t)  
\delta{\M{\sigma}}^{\mu\nu}_{{\bf r}'}(t')  \rangle 
&=2k_BT\delta({\bf r}-{\bf r}')\delta(t-t')\boldsymbol{\eta}^{\alpha\beta\mu\nu}
\label{corr1c}
\end{align}
and, therefore, the correlation of the  random stress is a white noise
in space and time, proportional to the  viscosity tensor.  Now we use the
following  expression for  the  piece-wise constant gradient of  the  finite element linear basis
functions \cite{Espanol2009}
\begin{align}
 \boldsymbol{{\nabla}}\psi_\nu({\bf r}) &=
\sum_{e_\nu}{\bf b}_{e_\nu}
 \theta_{e_\nu}({\bf r})
\label{ogradb}\end{align}
where $e_\nu$ labels each of the sub-elements of the node $\nu$, ${\bf
  b}_{e_\nu}$ is a constant vector within the sub-element $e_\nu$ that
is pointing  towards the node  $\nu$ and $\theta_{e_\nu}({\bf  r})$ is
the      characteristic      function     of      the      sub-element
$e_\nu$.

The projected current, can be written, therefore, as
\begin{align}
  \delta   L{\bf g}_\mu^{\alpha} &  =M^\delta_{\mu\nu}\int d{\bf r}
\sum_{e_\nu}{\bf b}^\beta_{e_\nu}
 \theta_{e_\nu}({\bf r})
\delta{\M{\sigma}}^{\alpha\beta}_{\bf r}(t)
\end{align}
By using  the model (\ref{randoms0})  for the projected  stress tensor
and equating the random  term $\frac{d\tilde{\bf g}_\mu}{dt}$ with the
projected current $  \delta L{\bf g}_\mu$ we have  the following explicit model
for the random forces
\begin{align}
\frac{d\tilde{\bf g}^\alpha_\mu}{dt}(t)
&=M^\delta_{\mu\nu}
\sum_{e_\nu}{\bf b}^\beta_{e_\nu}\tilde{\boldsymbol{\Sigma}}_{e_\nu}^{\alpha\beta}(t)
\label{dtildeg}
\end{align}
where the random stress tensor of the sub-element $e_\nu$ is given by
\begin{align}
\tilde{\boldsymbol{\Sigma}}_{e_\nu}^{\alpha\beta}(t) &=\sqrt{2k_BT\eta}
\left[\W^{\alpha\beta}_{e_\nu}(t)
-\delta^{\alpha\beta}
\frac{1}{d}\sum_\mu \W^{\mu\mu}_{e_\nu}(t)\right]
\nonumber\\
&+\sqrt{\frac{k_BT\zeta}{d}}
\delta^{\alpha\beta}\sum_\mu \W^{\mu\mu}_{e_\nu}(t)
\label{deltaLgmu5}
\end{align}
Here, we have introduced a symmetric matrix of white noise processes associated
to each sub-element $e_\nu$
\begin{align}
  \W^{\mu\nu}_{e}(t)&\equiv \int d{\bf r}\theta_e({\bf r})
\W^{\mu\nu}_{\bf r}(t)
\label{We}
\end{align}
These symmetric white-noise processes are independent among elements due to (\ref{eta})
\begin{align}
\llangle\W^{\mu\nu}_{e}(t)\W^{\mu'\nu'}_{e'}(t')\rrangle=\delta_{ee'}
[\delta^{\mu\mu'}\delta^{\nu\nu'}+\delta^{\nu\mu'}\delta^{\mu\nu'}]\delta(t-t')
\label{wienersubelement}
\end{align}

The noise term (\ref{dtildeg}) is a discrete 
divergence of  a discrete  random stress  tensor. The  discrete random
stress tensor $\tilde{\boldsymbol{\Sigma}}_{e_\nu}$  is an independent
stochastic process associated  to each sub-element.  It is  a matter of
calculation  to  check that  the  postulated  noise term  $d\tilde{\bf
  g}_\mu$  in  (\ref{dtildeg},\ref{deltaLgmu5})  with  the  white  noise  per
element $ \W^{\mu\nu}_{e}(t)$ satisfying (\ref{wienersubelement}),
gives precisely the FD balance in (\ref{FDTRg}).

Note that  the noise (\ref{dtildeg}) contains  the matrix
$M^\delta_{\mu\nu}$,  which  is  the  inverse  of  $M^\psi_{\mu\nu}$
defined in (\ref{Mpsi}).  The elements of $M^\psi_{\mu\nu}$ are proportional to the
volume  (area   in  2D)  of   the  overlaping  region   between  two
hydrodynamic cells which,  in turn, scales as the  typical volume of
the hydrodynamic  cells.  Therefore, the  stochastic  force
$\frac{d\tilde{\bf g}^\alpha_\mu}{dt}(t)$ scales with the inverse of
the  square root  of  the cell  size. Larger  cells  are subject  to
smaller  fluctuations,  in accordance  with  the  usual concepts  in
equilibrium statistical mechanics.

\section{\label{Sec:FinalResults}Final approximate dynamic equations }

We now have all the ingredients to construct the SDE (\ref{sde})
for the  chosen coarse-grained level of  description.  By collecting  the reversible
part (\ref{AproxDrift}) with (\ref{gconserving}) and irreversible part
of  the dynamics  {given by  ${\cal D}\esc\partial_x{\cal  H}$
  (where the dissipative matrix is  (\ref{CalD}) and the derivatives of
  the CG Hamiltonian are in (\ref{DerAproxCGH})}), the final SODEs for the
selected CG variables are
\begin{align}
\frac{d{\bf R}}{dt}&=  \overline{\bf v}({\bf R})
-\frac{D_0}{k_BT}\frac{\partial {\cal F}}{\partial{\bf R}}
+\frac{D_0}{k_BT}{\bf F}^{\rm ext}
+\frac{d\tilde{\bf R}}{dt}
\nonumber\\
\frac{d\rho_\mu}{dt}&=\norm \overline{\rho} \,\overline{\bf v} \cdot \boldsymbol{\nabla}\delta_\mu\norm
\nonumber\\
\frac{d{\bf g}_\mu}{dt}& = 
\norm \overline{\bf g}\,\overline{\bf v}\esc\boldsymbol{\nabla}\delta_\mu\norm
+k_BT\boldsymbol{\nabla}\delta_\mu({\bf R})
-\norm \delta_\mu\boldsymbol{\nabla} P \norm
+\delta_\mu({\bf R}){\bf F}^{\rm ext}
\nonumber\\
&+\eta \norm \delta_\mu \nabla^2\overline{{\bf v}}\norm+
\left(\frac{\eta}{3}+\zeta\right)\norm \delta_\mu \boldsymbol{\nabla}\left(\boldsymbol{\nabla}\esc\overline{{\bf v}}\right)\norm+\frac{d\tilde{\bf g}_\mu}{dt}
\label{SDEFin}
\end{align} 
\emph{These  equations are  the main  result of  this paper.}
Recall that the double bar denotes the spatial average defined in (\ref{NotationInt}), and
      the overlined  symbols denote  interpolated
  fields out of the discrete values as in, for example, $\overline{\bf
    v}({\bf r})={\bf v}_\nu\psi_\nu({\bf r})$, etc. The velocity ${\bf
    v}_\nu$   is  given   in  terms   of  $\rho_\mu,{\bf   g}_\mu$  in
  (\ref{DiscreteVel}).  The pressure equation of state is
    given  in  (\ref{TotalPressure0}) and  the  gradient  of the  free
    energy (\ref{OurModel}) is given by 
\begin{align}
\frac{\partial{\cal F}}{\partial{\bf R}}
&=\frac{m_0(c_0^2-c^2)}{\rho_{\rm eq}}\boldsymbol{\nabla}\overline{\rho}({\bf R})
\simeq
m_0\frac{(c_0^2-c^2)}{\rho_{\rm eq}}
\norm\Delta\boldsymbol{\nabla}\overline{\rho}\norm 
\label{DerFDerR}
\end{align}
see (\ref{rhoapp}) for the definition of the notation $\norm\Delta\boldsymbol{\nabla}\overline{\rho}\norm$.
The SDEs (\ref{SDEFin}) are closed and explicit in the  relevant variables.  

\subsection{\label{Sec:Meaning}  Physical  meaning  of  the  different
  terms in the dynamic equations}

The  first equation  in (\ref{SDEFin})  governs the  evolution of  the
position of  the nanoparticle.  The first  term $\overline{{\bf v}}({\bf
  R})$ is purely reversible and says that the nanoparticle follows the
interpolated velocity field of the  fluid.  This is a purely kinematic
effect due  to the fact  that the momentum  of the fluid  contains the
contribution due  to the nanoparticle. It  has nothing to do  with any
force that the  fluid may perform on the particle  which are described
by the second  contribution. This contribution is  proportional to the
bare mobility  $D_0/k_BT$, given  in terms  of the  \textit{bare} diffusion
coefficient $D_0$  introduced in (\ref{Drenorm}) through  a Green-Kubo
relation.  This term involves the  (minus) gradient of the free energy
${\cal F}({\bf R},\rho)$, which plays  the role of a \textit{potential
  of    mean   force}    for   the    nanoparticle
given explicitly in (\ref{DerFDerR}) .   As    seen   in
(\ref{ForceNanoparticle1}) the   force  due  to  the   fluid  on  the
nanoparticle  involves  the gradients  of  the  solvent density. The
presence  of  the  two  parameters  $c_0$, that  is  due  entirely  to
interactions of the nanoparticle with  the solvent particles, and $c$,
which is due to interactions  of the solvent particles with themselves
alone, indicates that $-\frac{\partial {\cal F}}{\partial {\bf R}}$ is
not  simply the  force that  the solvent  exerts on  the nanoparticle.
Note that  in the limit  when the  nanoparticle becomes just  a tagged
solvent particle,  which is realized for  $c_0\to c$, $-\frac{\partial
  {\cal  F}}{\partial {\bf  R}}$ given  in (\ref{DerFDerR})  vanishes.
The third  term {in the position  equation in (\ref{SDEFin})} is  due to
the  external  force  that  obviously   affects  the  motion  of  the
nanoparticle.   Finally, the  nanoparticle is  subject to  an explicit
noise term $\frac{d\tilde{\bf R}}{dt}$ whose  variance is given by the
fluctuation-dissipation  balance  relation (\ref{FDTRg}).   This  term
will produce Brownian motion of the nanoparticle, \textit{in addition}
to  the advection  by  the fluctuating  velocity field  $\overline{\bf
  v}({\bf R})$.   In order to  not ``double  count'' the noise  in the
Brownian motion of the  particle \cite{DiffusionJSTAT}, the diffusion
coefficient  that  governs the  amplitude  of  the random  noise  term
$\frac{d\tilde{\bf  R}}{dt}$ is  given in  terms of  the \textit{bare}
diffusion  coefficient  $D_0$  in  (\ref{Drenorm}),  and  not  by  the
renormalized diffusion coefficient $D$ defined in (\ref{D0}).

The  second equation  in (\ref{SDEFin})  gives the  evolution for  the
discrete  mass density  $\rho_\mu$  and  has the  form  of a  discrete
continuity equation.  This  evolution is purely reversible  due to the
fact that, very approximately, the time derivative of the mass density
is  given in  terms  of  the momentum  density,  which  is a  relevant
variable.  Therefore, the  projected current  vanishes and  so do  the
Green-Kubo coefficients,  i.e., there are no  Brenner diffusion terms,
as argued in \cite{BrennerModification_Ottinger}.

The  third equation  in (\ref{SDEFin})  governs the  discrete momentum
density ${\bf g}_\mu$.  It has the  structure of a discrete version of
the fluctuating  isothermal compressible Navier-Stokes  equations with
some modifications due to the interactions with the nanoparticle.  The
first term  in the momentum  equation is a convective  non-linear term
quadratic  in the  discrete momenta,  which corresponds  to the  usual
convective  term  in the  Navier-Stokes  equations.   The second  term
originates from  the stochastic drift $k_BT\partial_x\esc  L$ term and
can  be interpreted  as an  \textit{osmotic pressure}  term due  to the
presence of  the nanoparticle.  The  third term is reminiscent  of the
pressure  gradient term  in  the usual  Navier-Stokes equations.   The
pressure  equation of  state  is  given by  the  pressure  due to  the
Gaussian  model  for the  solvent,  plus  a pressure  correction  term
(proportional to the difference of the squares of the speeds of sound)
that   describes  the   interaction  between   the  solvent   and  the
nanoparticle.  Finally,  the term proportional to  ${\bf F}^{\rm ext}$
in  (\ref{SDEFin}) describes  the  effect that,  because the  discrete
momentum variable  contains the contribution due  to the nanoparticle,
any external force on the nanoparticle  will translate into a force on
the fluid  itself.  All  the terms  discussed so  far in  the momentum
equation are  purely reversible.   The only  irreversible terms  in the
momentum   equation   are   proportional  to   the   viscosities
$\eta,\zeta$ and correspond to the usual viscous terms involving
second space derivatives in the Navier-Stokes equations.  Finally, the
term  $d\tilde{\bf g}_\mu$  is the  random forces  with explicit  form
given in (\ref{deltaLgmu5})  and whose amplitudes are  dictated by the
fluctuation-dissipation balance in (\ref{FDTRg}).

Note that when $c_0=c$ a number
of terms  in the equations above  drop out and the  equations simplify
considerably.   This  happens,  for example,  when  the  distinguished
particle is simply  a tagged fluid molecule.  This may  also be a good
approximation  for neutrally  buoyant  particles that  do  not have  a
strong  chemical  interaction  with  the surrounding  fluid,  and  the
majority  of  prior work  in  the  literature  has  in fact  used  the
simplified   model   $c_0=c$,   with    the   notable   exception   of
\cite{CompressibleBlobs}.   

\subsection{Scope and general properties of the dynamic equations}

The validity  of the SODEs  (\ref{SDEFin}) is limited to  situations in
which the values $\rho_\mu,{\bf g}_\mu$  of the relevant variables are
such that  give a large  number of solvent particles  per hydrodynamic
cell and, at the  same time, give values that do  not differ very much
from one  cell to  its neighbors.  In  other words,  the interpolated
fields $\overline{\rho}({\bf r}),\overline{\bf g}({\bf r})$ need to be
smooth on the hydrodynamic cell  length scale. These assumptions imply
that  the validity  of the  equations is  restricted to  situations in
which  \textit{thermal fluctuations  are  small}. Correspondingly,  we
have assumed that the solvent  density fluctuations are Gaussian. This
precludes   the  study   of  other   interesting  phenomenology   like
liquid-vapor  phase  transitions,  for   example.  However,  it  is  a
sufficiently simple and physically  realistic model in many situations
of interest.   Concerning the nanoparticle,  it is assumed that  it is
smaller  than   the  hydrodynamic  cell   and  it  is,   therefore,  a
\textit{subgrid} nanoparticle.

The SODE (\ref{SDEFin})  conserve exactly the total mass  of the system
defined in (\ref{TotalInvariants}).  In the absence of external forces
acting on the nanoparticle, ${\bf  F}^{\rm ext}=0$, the total momentum
is also exactly conserved by the  equations. This is just a reflection
of  the  definition  (\ref{CGvariables})  of  the  discrete  mass  and
momentum ``fields'' in  terms of the basis functions  that satisfy the
partition of unity property (\ref{propdelta}).
Momentum conservation is a direct consequence of translational
invariance and we restored exact momentum conservation in our approximate
equations by restoring translational invariance of our free-energy model.

Discrete fluctuation-dissipation balance (DFDB)  is a crucial property
that  has been  carefully  maintained  in prior  work  that relied  on
phenomenological equations, see for  example \incite{SELM} or Appendix
B  of  \incite{ISIBM}.   A  key   component  of  DFDB  is  the  energy
conservation property that any work done by the external forces on the
suspended nanoparticle  must be converted exactly  into kinetic energy
of     the     fluid.      In     the     terminology     of     Refs.
\incite{SELM,ISIBM,BrownianBlobs}, this means that the linear operator
(matrix) used  to \emph{interpolate} the (discrete)  fluid velocity to
the particle, as represented by the term $\vb _{\mu}\psi_{\mu}\left(\V
  R\right)$ in the first equation of (\ref{SDEFin}), is the adjoint of
the linear  operator used  to \emph{spread} the  force applied  on the
particle    to   the    fluid,    as   represented    by   the    term
$\delta_{\mu}\left({\bf      R}\right)\V      F^{\text{ext}}\left({\bf
    R}\right)$ in  the last  equation in (\ref{SDEFin}).   This energy
conservation follows directly from the skew-symmetry of the reversible
operator (\ref{revLFSA}).

If the  reversible drift were  exactly in the form  $L\partial_x {\cal
  H}-k_B T \partial_x \esc L$,  it would automatically maintain DFDB,
this  is,  the equilibrium  distribution  function  would be  $\propto
e^{-\beta H}$.   However, the smoothness approximation  taken in order
to arrive  at the model (\ref{SDEFin})  imply that this is  true up to
small second order terms in the lattice spacing.  Our selection of the
model  for the  free energy  is not  exactly translational  invariant,
{i.e. it  does not satisfy (\ref{TIofF})  exactly}.  If it
were,  {as shown  in  Appendix  \ref{App:Models}}, then  we
would obtain that the marginal  distribution $P^{\rm eq}({\bf R})$ for
the position  of the  particle would  be given exactly by the  barometric law
(reduced Gibbs-Boltzmann distribution),
\begin{align}
  P^{\rm eq}({\bf R})& \sim \exp\left\{-\beta\Phi^{\rm ext}({\bf R})\right\}
  \label{barometric_law}
\end{align}   
However,  the violation  of  translation invariance  implies that  the
resulting  probability  distribution  is  given  by  (\ref{PRnotBaro})
instead, and the true barometric  distribution is obtained only in the
incompressible  limit  $c\rightarrow\infty$  or if  $c_0=c$  (e.g.,  a
tagged particle).

\newpage\subsection{\label{Sec:Continuum}The continuum equations}

We have obtained the SODEs (\ref{SDEFin}) from the Theory
of Coarse-Graining.  It can be  shown that  the same equations  can be
obtained     from     a    Petrov-Galerkin     discretization     (see
\cite{FluctDiff_FEM}  for   an  illustration  using  the   same  basis
functions  as  used  here)  of  the  following  system  of  stochastic
\textit{partial} differential equations (SPDEs)
\begin{align}
  \frac{d}{dt}{\bf R}&=\int d{\bf r}\Delta({\bf r},{\bf R}){\bf v}({\bf r})
\nonumber\\
&-\frac{D_0}{k_BT}
 \frac{m_0(c_0^2-c^2)}{\rho_{\rm eq}}
\int d{\bf r}\Delta({\bf R},{\bf r})\boldsymbol{\nabla}\rho({\bf r})
\nonumber\\
&+\frac{D_0}{k_BT}{\bf F}^{\rm ext}
+\frac{d\tilde{\bf R}}{dt}\nonumber\\
\partial_t\rho({\bf r},t)&=-\boldsymbol{\nabla}\esc{\bf g}
\nonumber\\
\partial_t{\bf g}({\bf r},t)&=-\boldsymbol{\nabla}\esc\left({\bf g}{\bf v}\right)
-k_BT\boldsymbol{\nabla}\Delta({\bf r},{\bf R})
\nonumber\\
&-\boldsymbol{\nabla}P({\bf r})+{\bf F}^{\rm ext}\Delta({\bf r},{\bf R})
\nonumber\\
&
+\eta \nabla^2{\bf v}+
\left(\frac{\eta}{3}+\zeta\right)\boldsymbol{\nabla}\left(\boldsymbol{\nabla}\esc
{{\bf v}}\right)
+\boldsymbol{\nabla}\esc{\boldsymbol{\Sigma} ^{\alpha\beta}_{\bf r}}
\label{SPDE}
\end{align}
where ${\bf v}={\bf g}/\rho$, and the pressure is given by 
\begin{align}
P({\bf r}) = \frac{c^2}{2\rho_{\rm eq}}\left({\rho}({\bf r})^2-\rho^2_{\rm eq}\right)
+\frac{m_0(c_0^2-c^2)}{\rho_{\rm eq}}
\Delta({\bf R},{\bf r}){\rho}({\bf r})
\label{PressureCont}
\end{align}
The random  velocity {$d\tilde{\bf  R}/dt$} is given  in (\ref{RandomVel}), and  the random stress tensor
$\boldsymbol{\Sigma} ^{\alpha\beta}_{\bf r}$  is given in (\ref{randoms0}).
The equations (\ref{SPDE}) are
very closely related to phenomenological  equations used in prior work
\cite{DirectForcing_Balboa,CompressibleBlobs,ISIBM},     with     some
differences that we further discuss in the Conclusions.

The  Petrov-Galerkin method  in  its most  pedestrian  form has  three
steps:  1) Multiply  the equations  (\ref{SPDE}) for  the hydrodynamic
fields with  the basis  functions $\delta_\mu({\bf r})$  and integrate
with respect to space. 2) Define the discrete variables $\rho_\mu=\int
d{\bf r}\delta_\mu({\bf  r})\rho({\bf r})$,  etc.  3)  Approximate the
fields in the  right hand side of the equations  (\ref{SPDE}) with the
linear    interpolations    $\overline{\rho}({\bf    r})=\psi_\mu({\bf
  r})\rho_\mu$,  etc.  This  procedure  applied  to (\ref{SPDE})  then
leads to  (\ref{SDEFin}).  As  an example, let  us consider  the first
term  in the  equation of  motion for  the particle,  representing the
advection by the  fluid velocity.  Replacing the  velocity with its
linear interpolant we get
\begin{align}
\int\vb \left({\bf r}\right)\Delta\left({\bf r},{\bf R}\right)d{\bf r} &{\rightarrow} \int\vb _{\mu}\psi_{\mu}\left({\bf r}\right)\Delta\left({\bf r},{\bf R}\right)d{\bf r}
\nonumber\\
&=\vb _{\mu}\int\psi_{\mu}\left({\bf r}\right)\Delta\left({\bf r},{\bf R}\right)d{\bf r}
\nonumber\\
&=\vb _{\mu}\psi_{\mu}\left({\bf R}\right)=\bar{\vb }\left({\bf R}\right),
\end{align}
where we used the property (\ref{LFSAdelta}). The right hand side is exactly our
discretization (derived here from the microscopic dynamics!) of the
term on the left hand side. The rest of the terms are discretized in a similar way.

\section{\label{Sec:Conclusions}Discussion and Conclusions}

By  performing a  systematic  coarse-graining procedure  based on  the
Zwanzing projection operator,  we have derived a  system of stochastic
\emph{ordinary} differential  equations (\ref{SDEFin})  describing the
dynamics of a nano-sized particle immersed  in a simple liquid.  A key
to the procedure was  the use of a dual set  of linear basis functions
familiar from  finite element methods  (FEM) as a way  to coarse-grain
the microscopic degrees of freedom. Another key ingredient was the use
of  a ``linear  for spiky'' weak approximation which  replaces microscopic
``fields'',  i.e.,  sums of  delta  functions  centered at  the  fluid
molecules, with a  linear interpolant in the FEM basis  set. These two
steps enabled us to obtain \emph{closed} approximations for all of the
terms in the reversible or non-dissipative dynamics, in a manner which
gives them a  clear physical interpretation and  preserves the correct
structure of the equations. Notably, the reversible dynamics preserves
a  discrete Gibbs-Boltzmann  distribution to  high accuracy.   For the
irreversible or dissipative dynamics,  we approximated the constrained
Green-Kubo  expressions for  the dissipation  coefficients with  their
equilibrium  averages,  and  assumed  a local  form  for  the  viscous
dissipation  suitable  when the  hydrodynamic  cells  contain a  large
number of fluid molecules.

The  coarse-grained  equations  we  derived  here can  be  seen  as  a
particular Petrov-Galerkin FEM discretization of a system of continuum
stochastic \emph{partial} differential equations (\ref{SPDE}) coupling
the familiar isothermal fluctuating Navier-Stokes (FNS) equations with
the  Brownian motion  of the  immersed particle.  These equations  are
similar in structure to phenomenological equations used in a number of
prior                                                            works
\cite{DirectForcing_Balboa,CompressibleBlobs,ISIBM,BrownianBlobs,DiffusionJSTAT,StochasticImmersedBoundary,SIBM_Brownian,LB_IB_Points,LB_SoftMatter_Review,ForceCoupling_Fluctuations},
and therefore  provide a justification  for those types of  models via
the Theory of Coarse Graining. This  is not just an academic exercise,
but one  that also  has some important  practical utility.  First, our
derivation provides Green-Kubo expressions for transport coefficients,
notably,  for   the  \emph{bare}   diffusion  coefficient   which  was
phenomenologically added in \cite{StokesEinstein}  as a way to account
for under-resolved microscopic details that  cannot be captured with a
hydrodynamic   approach.   Our   derivation   also  introduces   novel
terms  that come from the  microscopic interaction between
the  suspended particle  and the  liquid molecules;  these terms  give
additional  modeling  capability  to   account  for  more  microscopic
information  in  the   coarse-grained  description.  Another,  perhaps
unexpected, benefit  of the  microscopic derivation  was that  it lead
directly to a discrete form of the divergence of the stochastic stress
which  obeys the  fluctuation-dissipation  balance  relation. In  more
empirical approaches,  such a  structure has to  be either  guessed, or
constructed from  suitable discrete  stochastic fluxes  and a  pair of
discrete divergence and  gradient operators that are  skew adjoints of
one another \cite{LLNS_Staggered}.

The  coarse-graining procedure  carried out  here can  be viewed  as a
\emph{systematic}  derivation of  the  isothermal  FNS equations  from
molecular dynamics.   Formal derivations of these  equations have been
done  before  many  times,  see   for  example  early  work  including
non-linearities  in  \cite{Espanol1998a,DePablo2001},  however,  these
derivations   lead    either   to    linearized   equations    or   to
\emph{ill-defined} nonlinear  SPDEs exhibiting  an \textit{ultraviolet
  catastrophe}. As  we argued in more  detail in \cite{FluctDiff_FEM},
the proper  way to interpret  such \emph{formal} nonlinear SPDEs  is to
first \emph{discretize}  them by applying a  systematic discretization
procedure,  for example,  using  a  Petrov-Galerkin weak  formulation.
The justification for this prescription is
the  fact  that  here  we   obtain  \emph{exactly}  the  same  set  of
discretized  SPDEs  by  systematic   coarse  graining.  This  gives  a
direct\emph{ link}  between the  ``bottom-up'' approach of  going from
microscopic to mesoscopic equations,  and the ``top-down'' approach in
which one  starts from  continuum PDEs  and formally  adds white-noise
stochastic  forcing and  then applies  a standard  computational fluid
dynamics (CFD) method to the resulting equations.
Renormalization  techniques should  be applied  to the  discrete
equations  rather  than  the   continuum  ones  in   order  to
systematically   increase  the   coarse-graining   scale  from   the
mesoscopic   to   the  macroscopic in
order to  recover a continuum limit, where predictions  of physical
quantities like space-time correlations do not depend on the lattice
spacing $h$ as $h\to0$.

 The crucial link between
the  top-down  and  bottom-up   approaches  was  already  foreseen  in
\cite{DiscreteLLNS_Espanol},  and then  explicitly  demonstrated on  a
significantly  simpler   microscopic  model   in  the   Ph.D.   thesis
\cite{JaimeThesis}.  At the  same time,  the new  derivation
given   here  significantly   improves  on   the  earlier   derivation
\cite{DiscreteLLNS_Espanol},  which  did   not  consider  a  suspended
nanoparticle,  in three  key ways.  Firstly, here  we account  for the
presence of  a suspended particle.  Secondly, by  using a dual  set of
basis functions,  the resulting discretization is  second-order rather
than first-order accurate as the  earlier derivation based on a single
set of  basis functions \cite{JaimeThesis}.   Thirdly, in the  present work  the discrete
equations (\ref{SDEFin}) have a very precise relation to the continuum
equations  (\ref{SPDE}),  rather  than  simply  being  reminiscent  of
\emph{some}   ``sensible''   discretization.     We   have given   an
\emph{explicit} prescription  of how to  connect the two worlds  of MD
and CFD: use the same dual set of basis functions when coarse-graining
as  you do  when discretizing.   We believe  this prescription  can be
applied to a  variety of other problems, however, as  this work shows,
the bottom-up approach  requires a lot more work to  complete than the
top-down approach.

\subsection{Relation to phenomenological models}

The continuum equations (\ref{SPDE}) we  {proposed} here bear a strong
similarity, but  also some crucial differences,  with existing models.
In order  to explain the relation  to prior work more  clearly, let us
review a variety of existing  models starting from more ``refined'' to
more  ``coarse.'' A  more  formal mathematical  presentation of  these
levels of description is given  by Atzberger \cite{SELM}, here we give
a physical summary.  In many works an incompressible approximations is
made  in  order   to  eliminate  fast  sound  waves   from  the  model
\cite{ISIBM,StochasticImmersedBoundary}.  If one is interested only in
the  long-time  diffusive  dynamics  the fluid  inertia  can  also  be
eliminated       by       taking       an       overdamped       limit
\cite{BrownianBlobs,DiffusionJSTAT,SELM,ForceCoupling_Fluctuations}.
Here we focus  on the coupling between the  nanocolloidal particle and
the fluctuating fluid and not on the specifics of the fluid equations.

In a number of ``point particle'' \emph{frictional} coupling approaches,
as reviewed in detail in \cite{LB_SoftMatter_Review}, the colloid's
velocity is also included as a physical variable and a phenomenological
``friction'' force proportional to the velocity of the colloid relative
to the local fluid velocity is added to an inertial particle equation.
As we discussed in more detail in Section \ref{Sec:CGVars}, such
a level of description is not suitable under our assumption that the
colloid is smaller than the typical size of the hydrodynamic cells.
Instead, we include the momentum of colloid in the \emph{total} hydrodynamic
momentum field. We note that, although not usually presented in this
way, the same is actually true to a large extent for the model used
in \cite{LB_SoftMatter_Review} because what is called the mass of
the colloid is actually the \emph{excess} mass of the colloid over
the expelled fluid \cite{ISIBM,DirectForcing_Balboa}. This is because
the inertia of the expelled fluid is already included in the FNS equations,
which are assumed to apply everywhere including the volume occupied
by the particle. Therefore, part of the momentum of the particle is
in fact included in the ``fluid momentum''. For this reason, we
find it difficult to imagine how one can justify the frictional point
particle coupling model from a microscopic derivation.

In \cite{ISIBM,DirectForcing_Balboa}, an \emph{instantaneous} inertial
coupling is  proposed in which  the particle  is forced to  follow the
local    fluid    velocity;    the    main    difference    is    that
\cite{DirectForcing_Balboa}  considers  a  compressible  fluid,  while
\cite{ISIBM}  focuses  on  the   incompressible  limit.  As  shown  in
\cite{SELM_Reduction},        the        instantaneous        coupling
\cite{ISIBM,DirectForcing_Balboa}  can be  derived as  a limit  of the
frictional  coupling  when  the   friction  coefficient  becomes  very
large. In the language of the TCG, the (fast) momentum of the particle
is no longer included as a  relevant variable, instead, just as in our
description,  a total  momentum  field  is defined  \cite{ISIBM,SELM}.
This  total momentum  field follows  an equation  which has  a similar
structure to  the momentum equation  in (\ref{SPDE}), see  for example
Eq. (16) in \cite{ISIBM} in  the incompressible limit. It is important
to  note that,  even in  the limit  of an  incompressible liquid,  the
density  $\rho$  appearing  in (\ref{SPDE})  is  \emph{not}  constant,
rather, it includes the contribution  from the colloid. Therefore, for
a dense particle (e.g., gold nanocolloid) the discrete density will be
larger at  the nodes in  the vicinity  of the particle.  This ``excess
inertia''    is    explicitly    included     in    the    model    in
Refs. \cite{ISIBM,DirectForcing_Balboa} (see for example the left-hand
side of (13)  in \cite{ISIBM}) by ``spreading'' the  excess inertia to
the fluid  grid. In the equations  derived here the excess  inertia is
hidden in the definition of  the coarse-grained density to include the
contribution from the colloid, and the fact that the equation of state
is   only   applied    to   the   solvent   part    of   the   density
$\rho^{\text{sol}}$.

The terms in the particle equation related to the bare diffusion coefficient
$D_{0}$ do not appear in either \cite{DirectForcing_Balboa} or \cite{SELM,SELM_Reduction};
these terms are suggested in Appendix B.1 of \cite{ISIBM} but not
included in numerical simulations. A bare diffusion coefficient is
also introduced in the theoretical work \cite{DiffusionJSTAT} but
it is argued there that this term should somehow be small. Interestingly,
a renormalization of the diffusion coefficient similar in spirit to
$D_{0}$ is present in the frictional coupling formulation \cite{LB_SoftMatter_Review}
and can be expressed in the Einstein form $D_{0}=k_{B}T/\gamma$,
where $\gamma$ is the phenomenological friction coefficient,
see Eq. (290) in \cite{LB_SoftMatter_Review}. In the
limit of infinite friction, which is how \cite{SELM,SELM_Reduction}
derives the instantaneous coupling equations, $D_{0}\rightarrow0$.
However, this is ``throwing the baby out with the bathwater'' and
is not consistent with our microscopic derivation. The fact that $D_{0}>0$
is easy to appreciate: the sum $D=D_{0}+\D D$ is a physical parameter
that can be measured and is independent of the grid spacing (i.e.,
the coarse-graining length scale) while $\D D<D$ depends \emph{strongly}
on the grid spacing, as we explain in more detail shortly. In our
derivation, $D_{0}$ emerges naturally as does a Green-Kubo expression
for it, giving it a precise microscopic interpretation. The fact our
instantaneous coupling equations with bare diffusion cannot be consistently
derived from the frictional coupling formulation \cite{LB_SoftMatter_Review}
points to the lack of a microscopic foundation of that formulation,
and justifies once again the advantage of systematic bottom-up approaches
over phenomenological ones.

The stochastic thermal drift term $-k_{B}T\boldsymbol{\nabla}\Delta({\bf r},{\bf R})$
in the momentum equation is (wrongly) missing in \cite{DirectForcing_Balboa};
the term is also (rightfully) missing in the frictional coupling
\cite{LB_SoftMatter_Review} formulation. This term ought be there
for instantaneous coupling, as explained in Appendix B of \cite{ISIBM}
based on fluctuation-dissipation balance arguments, and derived by
taking the limit of infinite friction in the frictional coupling in
\cite{SELM_Reduction}. This osmotic pressure contribution from the
particle, spread to the fluid via the regularized delta function,
can be seen as coming from the eliminated fluctuations of the particle
velocity around the local fluid velocity \cite{SELM,SELM_Reduction}.
In the inertial coupling formulation, as explained in Appendix B of
\cite{ISIBM} and also in \cite{CompressibleBlobs}, this osmotic
pressure is split into two pieces just like the particle momentum
is split into two pieces, one piece attached to the fluid momentum
and another excess piece. When the two pieces are added together one
correctly recovers exactly the osmotic pressure term given in (\ref{SPDE}) \cite{CompressibleBlobs}.

The gradient of  the solvent pressure appears  in all phenomenological
models, and seems  very natural but it should be  recognized that this
is only an approximation.  Notably,  the approximation consists in the
assumption that this pressure is given by the equation of state of the
fluid  in  the   {\em  absence}  of  the   colloidal  particle.   This
approximation is exact for a labeled  or tagged particle of the fluid,
i.e., in case  of self-diffusion.  One can argue that  the same should
\emph{approximately} hold for colloidal  particles that have a similar
structure  to the  fluid,  notably,  that have  the  same density  and
compressibility    as    the    fluid.    Balboa    {\em    et    al.}
\cite{CompressibleBlobs}  have proposed  adding  an additional  excess
pressure  term  to account  for  a  different compressibility  of  the
colloid     relative     to     the     surrounding     fluid,     see
(\ref{FlorenRafaModel}).    These   terms    are   postulated   on   a
phenomenological basis. 

In this  work we derived equations containing
similar terms, however,  as already explained, our model  for the free
energy  differs from  that  used in  \cite{CompressibleBlobs} and  our
final      equations     (\ref{SPDE})      are     different      from
their,
which can be written in our notation as
\newcommand{\kt}{k_B T}
\newcommand{\bs}[1]{\boldsymbol{#1}}
\newcommand{\ps}[1]{\partial_{#1}}
\newcommand{\pare}[1]{\left( #1 \right) }
\newcommand{\corchete}[1]{\left[ #1 \right]}
\newcommand{\fr}[2]{\frac{#1}{#2}}
\newcommand{\wtil}[1]{\widetilde{#1}}
\newcommand{\mc}[1]{\mathcal{#1}}
\newcommand{\ang}[1]{\langle #1 \rangle}
\def\bna{\bs \nabla}
\def\vol{\mathcal{V}}              
\def\bq{\bs{R}}
\def\bv{{\vb}}              %
\def\br{{\bf r}}              %
\def\bg{{\bf g}}              %
\def\bF{{\bs F}}              %
\def\dd{\mathrm{d}}              %
\begin{eqnarray}
\label{eq:dqdt}
\frac{d}{dt}{\bf R} &=& \int \Delta(\bq-\br) \bv(\br) \; d \br , \\
\label{eq:drhodt}
\ps{t}\rho &=& - \bna \esc \bg, \\
\label{eq:dpdt}
\ps{t} \bg &=& -\bna \esc \pare{ \bg \bv } - (\kt)\bna \Delta(\bq-\br) \nonumber \\&&
-\bna P 
+ {\bf F}^{\rm ext} \Delta(\bq-\br) + \bna \esc \bs{\sigma}.
\label{FlorenRafaModel}
\end{eqnarray}
where
\begin{align}
P&=  c^2 (\rho(\br)-\rho_{\rm eq}) 
\nonumber\\
&+ \vol (c_0^2 -c^2) \Delta(\bq-\br) \int  \Delta(\bq-\br') (\rho(\br')-\rho_{\rm eq}) \; d \br'
\end{align}
where the regularized delta function $\Delta(\bq,\br)=\Delta(\bq-\br)$
is  given by  a Gaussian-like  isotropic kernel  $\Delta(r)$ of  width
comparable to  the hydrodynamic  radius of  the nanoparticle  and that
integrates  to  unity.   Here  the   fluid  velocity  is  defined  via
$\bv=\bg/\rho$,   and  $\bs{\sigma}$   denotes   the  viscous   stress
(deterministic and  fluctuating components)  with a form  identical to
ours.  The volume  associated to the particle can be  expressed in our
notation  as $\vol  \equiv m_0  /  \rho_{\rm eq}$  but is  interpreted
differently in \cite{CompressibleBlobs} to be a geometric rather than an
inertial quantity.  Note that we have  set here the excess mass of the
particle \cite{CompressibleBlobs}  over the  fluid $m_e=0$  because in
our  notation $\rho$  includes  the  total mass  of  the particle  (see
additional discussion in Section \ref{Sec:Conclusions}).

The similarities between our formulation and the formulation of Balboa
{\em et al.} are evident, but there are also some notable differences.
The equations  (\ref{FlorenRafaModel}) do  not include  bare diffusion
(i.e., $D_0=0$) and  therefore a number of terms are  missing from the
particle equation.  In the  model of Ref. \cite{CompressibleBlobs} the
parameter $c_0$  is used to  represent the  speed of sound  (i.e., the
isothermal compressibility)  inside the  colloidal particle,  while in
our model $c_0$ models the mean  force that the colloid experiences in
a density gradient, see  {(\ref{ForceNanoparticle1})}.  Balboa {\em et
  al.} justify  the ``particle compressibility''  pressure contribution
proportional  to   $c_0^2  -c^2$  starting  from   a  {\em  quadratic}
contribution to the free energy density of the form
\[
\mc{F}_{\rm comp} = \frac{\vol(c_0^2 -c^2)}{2\rho_{\rm eq}}
\left(\int  \Delta(\bq-\br') (\rho(\br')-\rho_0) \; d \br' \right)^2
\]
whereas the ``continuum'' analog of our {\em linear} model for the interaction free energy
found in the last term in (\ref{OurModel}) is
\[
\mc{F}_{\rm int} = \vol(c_0^2 -c^2) \left(\int  \Delta(\bq-\br') (\rho(\br')-\rho_0) \; d \br' \right)
\]

It  is  important  to  emphasize,  however,  that  our  discussion  of
similarities to  prior work above  only concerns the  formal continuum
formulation (\ref{SPDE}) and focuses on the structure of the equations
and  the  physics of  the  various  terms. Our  fully  \emph{discrete}
formulation (\ref{SDEFin}) is completely new  and is different in many
crucial ways  from existing  discretizations. The first  difference is
that  the  discretization   of  the  FNS  equations  is   based  on  a
second-order conservative  FEM method,  rather than the  more commonly
used     finite-difference    \cite{StochasticImmersedBoundary}     or
finite-volume approach  \cite{LLNS_Staggered}. A second  difference is
that  in all  prior  models we  are aware  of,  the regularized  delta
function is  used to represent the  particle itself, and its  width is
chosen to be on the order  of the hydrodynamic radius of the particle.
As  such,   the  regularized  delta   function  is  attached   to  the
\emph{particle},  i.e., it  is taken  to be  a smeared  delta function
kernel  such as  a Gaussian  kernel  or the  tensor-product kernel  of
Peskin \cite{IBM_PeskinReview}  centered at the particle  position. By
contrast,  in  the  formulation  (\ref{SPDE})  the  regularized  delta
function represents  the coarsening  of the  solvent dynamics  and its
width is assumed to be larger than the subgrid colloidal particle.  As
such,  our  discrete   delta  function  kernel  is   attached  to  the
\emph{grid}  and is  not centered  around the  particle; this  is most
obvious  in  (\ref{SDEFin}) where  it  is  clearly  seen that  in  the
equation for the  (fixed in space) grid node  $\mu$ the regularization
enters  via  the  basis  function $\delta_{\mu}$  associated  to  that
node. Note that in the immersed-boundary formulation of Peskin used in
a             number             of            prior             works
\cite{DirectForcing_Balboa,CompressibleBlobs,ISIBM,BrownianBlobs,StochasticImmersedBoundary,LB_SoftMatter_Review}
the \emph{width}  of the discrete delta  function is tied to  the grid
spacing just as it is for  our regularized delta function, however, in
those prior works the regularized  kernel is still centered around the
particle.  This was  proposed by  Peskin as  a very  effective way  to
maximize  translational invariance  of the  particle-grid interactions
\cite{IBM_PeskinReview}; we expect our formulation will not perform as
well  in  terms  of  translational   invariance  because  it  was  not
explicitly constructed with that goal in mind.

\subsection{Renormalization of the diffusion coefficient}

Let us consider, for a moment, a single freely-diffusing isolated
spherical nanoparticle suspended in a quiescent fluid. At large time
scales, the particle will perform a standard Brownian motion with
a renormalized diffusion coefficient $D$ given in (\ref{D0}) by
the familiar Green-Kubo integral of the particle's velocity autocorrelation
function. Since this quantity only involves the nanoparticle position,
it cannot depend on how we chose to perform the coarse graining of
the fluid. In particular, it must be a number independent of the typical
grid spacing $h$. For sufficiently large Schmidt numbers \cite{StokesEinstein}
we expect it to be well-predicted by the Stokes-Einstein formula (in
three dimensions) 
\[
D\approx D_{\text{SE}}=\frac{k_{B}T}{\alpha\eta R},
\]
where $R$ is the radius of the spherical particle and $\alpha$ is a coefficient
that depends on the boundary conditions applicable at the surface
of the sphere, equal to $6\pi$ for a stick surface and $4\pi$ for
a slip surface, or something in-between for more realistic models
\cite{StokesEinstein_BCs,StokesEinstein_MD}.

In the fluctuating hydrodynamic formalism presented here, the renormalized
diffusion coefficient $D$ is split into a bare part $D_{0}$ and
a renormalization $\D D$ defined by (\ref{D_split}). Let us try
to get a more quantitative understanding of the diffusion enhancement
$\Delta D$ for our specific discretization (approximation) of the
equations. First, we can replace the instantaneous interpolated fluid
velocity $\Vav$ by its approximation $\bar{\vb}({\bf R})=\vb_{\mu}\psi_{\mu}({\bf R})$.
Second, we can ignore the cross-correlation terms since $\hat{{\bf V}}$
evolves on a much faster time scale than the hydrodynamic fields and
can be assumed to be a white-noise process uncorrelated with $\bar{\vb}({\bf R})$.
This gives the approximation to the diffusion enhancement produced
by our discrete equations, 
\begin{align}
\Delta D({\bf R}) & =\frac{1}{d}\int_{0}^{\tau}dt\llangle\bar{\vb}({\bf R}(0))\esc\bar{\vb}({\bf R}(t))\rrangle^{{\rm eq}}\nonumber \\
 & =\frac{1}{d}\int_{0}^{\tau}dt\llangle\psi_{\mu}({\bf R}(0))\left(\vb_{\mu}(0)\esc\vb_{\mu^{\prime}}(t)\right)\psi_{\mu^{\prime}}({\bf R}(t))\rrangle^{{\rm eq}}.
\end{align}
In the overdamped limit of large Schmidt numbers (see \cite{StokesEinstein}
for corrections at moderate Schmidt numbers), the particle moves much
slower than the hydrodynamic correlations decay, and one can express
the diffusion enhancement in terms of the {\em equilibrium} correlation
of the fluid velocity (conditional on the particle being fixed at
a particular location),
\begin{align}
\Delta D({\bf R}) & =\frac{1}{d}\int_{0}^{\infty}dt \; \psi_{\mu}({\bf R})
\llangle\vb_{\mu}(0)\esc\vb_{\mu^{\prime}}(t) \rrangle_{\bf R}^{{\rm eq}}
\psi_{\mu^{\prime}}({\bf R}),\label{D_split_lgSc}
\end{align}
which can in principle be computed exactly by linearizing the fluid
equations around a quiescent state. Note that $\Delta D$ depends
on ${\bf R}$ explicitly; for confined systems this dependence is physical
but for translationally invariant system such dependence is a discretization
artifact that is hopefully small. Note that the immersed-boundary
discrete delta function used in Refs. \cite{DirectForcing_Balboa,CompressibleBlobs,ISIBM,BrownianBlobs,StochasticImmersedBoundary}
is specifically designed to obtain such translational invariance on
regular grids to a high accuracy \cite{IBM_PeskinReview}.

We can obtain  a physical estimate for the diffusion  due to advection
by the  thermal velocity  fluctuations by  assuming that  the discrete
velocities are consistent with  a Petrov-Galerkin procedure applied to
continuum  equations.  This  allows  us to  approximate  the  discrete
velocity  $\vb_{\mu}$  in  terms  of  a  continuum  fluctuating  field
$\vb({\bf  r},t)$  as   $\vb_{\mu}(t)=\int  d{\bf  r}\delta_{\mu}({\bf
  r})\vb({\bf r},t)$.   If we substitute this  in (\ref{D_split_lgSc})
we obtain the estimate
\begin{align}
\Delta D({\bf R}) & =\int d{\bf r}d{\bf r}^{\prime}\;\psi_{\mu}({\bf R})\delta_{\mu}({\bf r})
\nonumber\\
&\times\left(\frac{1}{d}\int_{0}^{\infty}dt\llangle\vb({\bf r},0)\esc\vb({\bf r}^{\prime},t)\rrangle^{{\rm eq}}\right)\delta_{\mu^{\prime}}({\bf r}^{\prime})\psi_{\mu^{\prime}}({\bf R})\nonumber \\
 & =\int d{\bf r}d{\bf r}^{\prime}\;\Delta({\bf r},{\bf R})
\nonumber\\
&\times\left(\frac{1}{d}\int_{0}^{\infty}dt\llangle\vb({\bf r},0)\esc\vb({\bf r}^{\prime},t)\rrangle^{{\rm eq}}\right)\Delta({\bf r}^{\prime},{\bf R}).\label{D_enh}
\end{align}
If one assumes that the evolution of $\vb({\bf r},t)$ can be described
by  a fluctuating  Stokes  equation  (i.e., linearized  incompressible
flow),  the time  integral can  easily be  expressed in  terms of  the
inverse Stokes operator  (i.e., the Green's function  for Stokes flow)
\cite{StokesEinstein}.  In  this case  the relation  (\ref{D_enh}) can
directly be matched with Eq. (10) in \incite{StokesEinstein} (see also
Eq. (288) in \cite{LB_SoftMatter_Review}), where the regularized delta
function      is      denoted      with      $\Delta({\bf      r},{\bf
  R})\rightarrow{\delta}_a({\bf r}-{\bf R})$.  In our notation Eq. (10)
in \incite{StokesEinstein} becomes,
\begin{align}
\Delta D({\bf R})&=\frac{k_{B}T}{\eta}\int d{\bf r}d{\bf r}^{\prime}\;\Delta({\bf r},{\bf R})
\nonumber\\
&\times
\left(\frac{1}{d}\mbox{Trace}\; \M G\left({\bf r},{\bf r}^{\prime}\right)\right)\Delta({\bf r}^{\prime},{\bf R}),\label{eq:chi_r_Stokes}
\end{align}
where $\M G$ is the Green's function for Stokes flow (Oseen tensor
for an unbounded domain at rest at infinity). This is nothing else
but an Einstein formula relating the diffusion coefficient with the
mobility of the particle, i.e., with the linear response of the particle
to a weak applied force.

A simple calculation based on the expression for $\M G$ in Fourier
space, or, equivalently, based on replacing the continuum Green's
function with its discrete equivalent, estimates that in three dimensions
\cite{StokesEinstein} 
\[
\D D=\frac{k_{B}T}{\alpha^{\prime}\eta h},
\]
where $h$ is the width of the regularized Delta function (i.e., the
grid spacing), and $\alpha^{\prime}$ is a coefficient that depends
on the geometric details of the grid. This suggests that 
\[
D_{0}=\frac{k_{B}T}{\eta}\left(\frac{1}{\alpha^{\prime}h}-\frac{1}{\alpha R}\right),
\]
which must be non-negative, i.e., it must be that $\alpha R<\alpha^{\prime}h$,
which is consistent with the assumption that the nanoparticle is smaller
than a typical grid cell. Observe that at large Schmidt numbers $\D D$
can be expressed purely in terms of geometric quantities and the equilibrium
(discrete) fluid correlation functions, and should therefore depend
mildly if at all on the details of the interaction between the particle
and the fluid. This suggests that it is the \emph{bare} diffusion coefficient
that must capture essentially \emph{all} of the microscopic details
such as slip versus no-slip on the particle surface or layering of
the fluid around the particle.

In  the  microscopic derivation  presented  here,  the bare  diffusion
coefficient $D_{0}$ is  to be computed using  (\ref{Drenorm}) from the
Green-Kubo integral  of the  autocorrelation function of  the particle
\emph{peculiar}  velocity, i.e.,  the velocity  relative to  the local
(interpolated)  fluid  velocity.   Only  a  combination  of  molecular
dynamics  and fluctuating  FEM calculations  can tell  us whether  the
effective diffusion  coefficient $D=D_{0}+\D  D$ indeed be  a constant
(approximately)  independent of  the grid  resolution.  In  prior work
based   on   phenomenological   fluctuating   hydrodynamics   theories
\cite{DiffusionJSTAT}, $D_{0}$ was treated  as an adjustable parameter
that is chosen  so as to give a \emph{desired}  (input) effective $D$,
since $\D  D$ follows from  the discretization of the  fluid equations
and cannot be  adjusted independently. This is similar to  how one can
treat  the fluid-particle  interaction strength  parameter {$c_0$}  as
fitting  parameters  used to  match  the  coarse-grained and  particle
dynamics  as  best as  possible,  instead  of  computing it  from  its
(approximate!)  microscopic definition (\ref{a_def}).

\subsection{Future Directions}

The work described here is purely theoretical and proposes a model
with the correct structure but leaves a number of terms to be approximated
and modeled. As such, the usefulness and accuracy of our equations
cannot be judged until a number of numerical studies are performed.

Firstly, a temporal discretization needs to be developed to go with
the spatial discretization (\ref{SDEFin}); the required tools are
readily available \cite{DFDB,MultiscaleIntegrators}. Secondly, it
is important to study the numerical aspects of the Petrov-Galerkin
FEM discretization developed here using existing numerical analysis
tools \cite{LLNS_S_k,LLNS_Staggered} and compare to existing discretizations.
Lastly, one needs to include immersed particles in the numerics as
well and study a number of standard test problems to evaluate the
performance of (\ref{SDEFin}) as a standalone method for simulating
dilute colloidal suspensions. Particular emphasis should be payed
to the violations of discrete fluctuation dissipation balance and
of translational invariance, both of which are in a formal sense second
order in the grid spacing, but may be significant in practice at scales
comparable to the grid spacing.

It is an important task for future work to perform molecular dynamics
(MD) simulations and compare the results to the coarse-grained description
proposed here. We expect that if the grid cells are too small we will
see unphysical artifacts, and if the grid cells are too large, the
MD simulations will become unfeasible. By confirming whether the correct
effective (renormalized) diffusion coefficient is obtained over a
reasonable range of grid spacings, we can access how good the approximations
made in our coarse graining theory are, and ultimately how useful
the proposed equations are in practice. One should begin these studies
with a single nanocolloidal particle in solvent in a periodic box
of varying sizes, perhaps starting in two dimensions, where there
are very strong (in fact, asymptotically dominant rather than decaying)
finite size effects \cite{DiffusionJSTAT}.

A key question  that we did not  fully address here is  how to compute
the various  coefficients that  appear in (\ref{SDEFin}).   We already
discussed  the  subtlety   of  this  issue  for   the  bare  diffusion
coefficient.   Even  more  freedom  exists  for  the  free  energy  of
interaction  between the  solute and  the solvent,  which needs  to be
modeled with  some number of adjustable  parameters.  These parameters
are  to  be  tuned  by matching  the  coarse-grained  and  microscopic
descriptions. How to do this matching in practice remains an important
open question. We  proposed a specific model with  a single adjustable
parameter  here but  it  remains  to be  seen  whether  this model  is
appropriate on a  case-by-case basis, and if not,  to make adjustments
to the equations by following the approach developed in this work.

It is important  to emphasize here that (\ref{SDEFin}) can  be used to
study  colloidal  suspensions of  more  than  one colloidal  particle,
however, the description  will only be accurate when  the colloids are
further than about  one grid cell apart. This is  because our modeling
of the bare  diffusion coefficient is based  on equilibrium Green-Kubo
expressions for a single particle. This  will fail to give an accurate
approximation when two particles come closer to each other than a grid
spacing; at  such short distances the  hydrodynamic correlations among
the diffusing particles will  \emph{not} be captured accurately.  This
is         no         different        from         prior         work
\cite{DirectForcing_Balboa,CompressibleBlobs,ISIBM,BrownianBlobs,DiffusionJSTAT,StochasticImmersedBoundary,SIBM_Brownian,LB_IB_Points,LB_SoftMatter_Review}
where the  hydrodynamic interactions are  only resolved up to  at most
the   Stokeslet  or   Rotne-Prager   level.   Furthermore,  when   the
nanocolloids  come close  to each  other we  expect that  their direct
interactions with  the solvent  molecules or with  each other  will be
affected and  other terms in  (\ref{SDEFin}) will need to  modified as
well.  Ultimately, as  the density  is  increased there  will be  many
nanoparticles per hydrodynamic cell and  in this case a coarse-grained
theory of  fluid mixtures should  emerge. Such a theory  could perhaps
provide  a   bridge  between   macroscopic  fluid   mixture  equations
\cite{FluctHydroNonEq_Book}  and dynamic  density functional  theories
with hydrodynamic effects \cite{DDFT_Hydro,DDFT_Hydro_Lowen,DDFT_Pep}.

Finally, the  present theory is isothermal  as the energy
  density of  the fluid is assumed  to be a fast  decaying variable as
  compared with mass and momentum variables. Of course, this precludes
  the  study   of  thermal  processes  that   arise  in  nanocolloidal
  suspensions in the presence of thermal gradients. The formulation of
  non-isothermal   models   is   the   subject   of   ongoing   work.
  
\acknowledgements We thank Florencio Balboa  for a critical reading of
the manuscript  and numerous helpful discussions.   Useful discussions
with  Rafael Delgado-Buscalioni  are also  greatly appreciated.   P.E.
acknowledges financial  support from  the Spanish Ministry  of Economy
and  Competitivity under  grant  FIS2013-47350-C5-3-R.   A. Donev  was
supported in part  by the Office of Science of  the U.S. Department of
Energy through Early Career award DE-SC0008271.

\appendix

\begin{widetext}
\section{\label{App:Exact} Derivation of exact results}

In this appendix we obtain a number of exact results for the equilibrium probability distributions, the free energies, and the reversible drift term. 
\subsection{Equilibrium distributions}
\label{osec:loc}
The equilibrium probability $P(x)$ in Eq. (\ref{omega}) in the present
level of description takes the form
\begin{align}
P^{\rm eq}({\bf R},\rho,{\bf g})&= 
\int dz \frac{1}{Z}\exp\{-\beta \hat{H}(z)\}
\delta({\bf R}-{\bf q}_0)\prod_\mu^M
\delta(\rho_\mu-\hat{\rho}_\mu(z))
\delta({\bf g}_\mu-\hat{{\bf g}}_\mu(z))
\label{Prg}
\end{align}
and the conditional expectation  $\langle\cdots\rangle^x$ in Eq. (\ref{ca})
takes the form
\begin{align}
\langle \cdots\rangle^{{\bf R}\rho{\bf g}}&= \frac{1}{P^{\rm eq}({\bf R},\rho,{\bf g})}
\int dz \frac{1}{Z}\exp\{-\beta \hat{H}(z)\}
\delta({\bf R}-{\bf q}_0)\prod_\mu^M
\delta(\rho_\mu-\hat{\rho}_\mu(z))
\delta({\bf g}_\mu-\hat{{\bf g}}_\mu(z))\cdots
\label{CondExpectRrhog}
\end{align}
It  is   convenient  to   introduce   the   marginal  equilibrium
probability $P^{\rm eq}({\bf R},\rho)$,
\begin{align}
  P^{\rm eq}\left({\bf R},\rho\right)&=
\int dz \frac{1}{Z}\exp\{-\beta \hat{H}(z)\}
\delta({\bf R}-{\bf q}_0)\prod_\mu^M
\delta(\rho_\mu-\hat{\rho}_\mu(z))
\label{PeqB}
\end{align}
and the corresponding conditional expectation 
 $\llangle\cdots\rrangle^{{\bf R}\rho}$
conditional on ${\bf R},\rho$, and not on ${\bf g}$.
Finally,   we  will   also   consider   the  equilibrium   probability
distribution of  the density  field, in  the absence  of nanoparticle,
which is defined as
\begin{align}
P^{\rm eq}_{\rm sol}(\rho)
\equiv&\int \prod_{i=1}^Nd{\bf q}_id{\bf p}_i\frac{1}{Z^{\rm sol}}
\exp\left\{-\beta \left(\sum_{i=1}^N\frac{{\bf p}_i^2}{2m}+\hat{U}^{\rm sol}(q)\right)\right\}
\prod_\mu^M\delta(\rho_\mu-\hat{\rho}^{\rm sol}_\mu(z))
\label{Prg25}
\end{align}
where $Z^{\rm sol}$ is the  normalization and the solvent mass density
$\hat{\rho}^{\rm  sol}_\mu(z)$ is  introduced in  Eq. (\ref{rhogsol}).
The  corresponding equilibrium  expectation  over  solvent degrees  of
freedom conditional to the solvent density is denoted by $\llangle\cdots\rrangle^\rho_{\rm sol}$.

The three  probabilities (\ref{Prg}), (\ref{PeqB}),  (\ref{Prg25}) are
related  to each  other.   By integrating  the  momentum variables  in
(\ref{Prg}), {occurring in the kinetic energy of the Hamiltonian and in
the  Dirac   delta  functions},   with  (\ref{III})  in   the  Appendix
\ref{App:MomentumIntegrals} we have
\begin{align}
P^{\rm eq}({\bf R},\rho,g) &= \frac{1}{Z'}
\int \prod_{i=0}^Nd{\bf q}_i\exp\{-\beta \hat{U}(q)\}
\delta({\bf R}-{\bf q}_0)
\prod_\mu^M\delta(\rho_\mu-\hat{\rho}_\mu(z))
\frac{\exp\left\{-\frac{\beta}{2}{\bf g}_\mu\hat{M}^{-1}_{\mu\nu}(z){\bf g}_\nu
\right\}
}{(2\pi/\beta)^{3M/2}\det \hat{M}(z)^{3/2}}
\label{Prg21}
\end{align}
where the  mass matrix is defined in (\ref{taghatMmunu}).

In a similar  way, by integrating the atomic  momenta in (\ref{PeqB}),
gives the following form for the marginal equilibrium probability
\begin{align}
P^{\rm eq}({\bf R},\rho)&=
\int \prod_{i=0}^Nd{\bf q}_i
\frac{1}{Q}\exp\left\{-\beta\hat{U}(q)\right\}
\delta({\bf R}-{\bf q}_0)
\prod_\mu^M\delta\left(\rho_\mu-\hat{\rho}_\mu(z)\right)
\label{136}
\end{align}
where all momentum variables have been integrated out and  $Q$ is the normalization.
We may write (\ref{Prg21}) in the following form
\begin{align}
  P^{\rm eq}({\bf R},\rho,g) &= 
  P^{\rm eq}\left({\bf R},\rho\right) 
\llangle \frac{\exp\left\{-\frac{\beta}{2}{\bf g}_\mu\hat{M}^{-1}_{\mu\nu}{\bf g}_\nu\right\}}{(2\pi/\beta)^{3M/2}\det \hat{M}^{3/2}}
\rrangle^{{\bf R}\rho}
\label{416}
\end{align}
that gives  the relation between  $ P^{\rm eq}({\bf R},\rho,g)$  and $
P^{\rm eq}\left({\bf  R},\rho\right)$. 

At the  same time, the marginal  $P^{\rm eq}\left({\bf R},\rho\right)$
in  Eq.  (\ref{PeqB})  can be  expressed in  terms of  the probability
distribution of the solvent density $P^{\rm eq}_{\rm sol}(\rho)$ in the absence
of nanoparticle. First, integrate momenta in (\ref{Prg25}) to get
\begin{align}
P^{\rm eq}_{\rm sol}(\rho)
\equiv&
 \int \prod_{i=1}^Nd{\bf q}_i\frac{1}{Q^{\rm sol}}
 \exp\left\{-\beta \hat{U}^{\rm sol}(z)\right\}
\prod_\mu^M\delta(\rho_\mu-\hat{\rho}^{\rm sol}_\mu(z))
\label{PeqSol}
\end{align}
where $Q^{\rm sol}$ is the normalization. Then, Eq. (\ref{136}) becomes
\begin{align}
P^{\rm eq}\left({\bf R},\rho\right)&= 
\int \prod_{i=1}^Nd{\bf q}_i
\frac{1}{Q}\exp\left\{-\beta\left(\hat{U}^{\rm sol}(q)+\sum_{i=1}^N\Phi^{\rm int}({\bf R}-{\bf q}_i)+\Phi^{\rm ext}({\bf R})\right)\right\}
\prod_\mu^M\delta\left(\rho_\mu-m_0\delta_\mu({\bf R})-\hat{\rho}^{\rm sol}_\mu(z)\right)
\nonumber\\
&=
\frac{1}{Q'}P^{\rm eq}_{\rm sol}\left(\rho-m_0\delta({\bf R})\right)
\llangle
 \exp\left\{-\beta\int d{\bf r}\Phi^{\rm int}({\bf R}-{\bf r})\hat{n}^{\rm sol}_{\bf r}\right\} 
\rrangle^{\rho-m_0\delta({\bf R})}_{\rm sol}
\exp\left\{-\beta\Phi^{\rm ext}({\bf R})\right\}
\label{PBsol}
\end{align}
where  in  the first  equality  we  have  integrated the  Dirac  delta
function $\delta({\bf  R}-{\bf q}_0)$  and in  the second  equality we
have used the definition  (\ref{PeqSol}). 

\subsection{Free energies}
The relationships (\ref{416}) and (\ref{PBsol}) between the probabilities reflects
into an  exact expression for the free energy of the system.
Let us introduce the  free  energy  of  the solvent  ${\cal  F}^{\rm
  sol}(\rho)$, the free energy of  the fluid ${\cal F}({\bf R},\rho)$,
and the  CG Hamiltonian  ${\cal H}({\bf R},\rho,g)$  (which is  also a
free energy) through the following expressions
\begin{align}
P^{\rm eq}_{\rm sol}(\rho)&\propto\exp\{-\beta{\cal F}^{\rm sol}(\rho)\}
\nonumber\\
P^{\rm eq}\left({\bf R},\rho\right)&\propto\exp\{-\beta{\cal F}\left({\bf R},\rho\right)-\beta\Phi^{\rm ext}({\bf R})\}
\nonumber\\
P^{\rm eq}({\bf R},\rho,g)&\propto
\exp\left\{-\beta{\cal H}({\bf R},\rho,g)\right\}
\label{Fsolvent}
\end{align}

We now  look at the relationships  between these free energies  (up to
irrelevant constants). 
Because of (\ref{416}) and (\ref{PBsol}), the CG Hamiltonian has the form
\begin{align}
  {\cal H}({\bf R},\rho,{\bf g})&=
-k_BT\ln \llangle \frac{\exp\left\{-\frac{\beta}{2}{\bf g}_\mu\hat{M}^{-1}_{\mu\nu}{\bf g}_\nu\right\}}
{(2\pi/\beta)^{3M/2}\det \hat{M}^{3/2}}\rrangle^{{\bf R}\rho}
+{\cal F}\left({\bf R},\rho\right)+\Phi^{\rm ext}({\bf R})
\label{CGHamiltonianExact}
\end{align}
and the fluid free energy is
\begin{align}
  {\cal F}\left({\bf R},\rho\right)={\cal F}^{\rm sol}\left(\rho-m_0\delta({\bf R})\right)+{\cal F}^{\rm int}({\bf R},\rho-m_0\delta({\bf R}))
\label{freedecomp}
\end{align}
where  the free  energy  of
interaction between nanoparticle and solvent is
\begin{align}
{\cal F}^{\rm int}({\bf R},\rho)\equiv&-k_BT\ln \llangle
 \exp\left\{-\beta\int d{\bf r}\Phi^{\rm int}({\bf R}-{\bf r})\hat{n}^{\rm sol}_{\bf r}\right\} \rrangle^{\rho}_{\rm sol}
\label{calFF}
\end{align}
The exact result (\ref{freedecomp}) that decomposes the free energy of
the system into a solvent and  an interaction part will be very useful
for modelling.   The fact that  the free energy depends  on $\rho_\mu$
only through the combination $\rho^{\rm sol}_\mu=\rho_\mu-m_0\delta_\mu({\bf R})$, which is
the mass density of the solvent, is a non-trivial result.

\subsection{The role of translation invariance on the free energy}

In this appendix we demonstrate the following exact identity involving derivatives
of the free energy
\begin{align}
\frac{\partial {\cal F}}{\partial {\bf R}}\left({\bf R},\rho\right) &=
k_BT
\frac{\partial}{\partial \rho_\nu}\bra\hat{\rho}\boldsymbol{\nabla}\delta_\nu\ket^{{\bf R}\rho}
-\bra\hat{\rho}\boldsymbol{\nabla}\delta_\nu\ket^{{\bf R}\rho}\frac{\partial{\cal F}}{\partial \rho_\nu}
\label{ExactTIofF}
\end{align}
This identity is a direct consequence of translation invariance of the
microscopic Hamiltonian.  It  is an important result  because it gives
an exact relationship between the  derivatives of the free energy with
respect to ${\bf R}$ and $\rho_\mu$. {This mathematical identity
gives a strong condition on the modelling of the free energy.}

The proof is as follows. In the integrals over  positions in (\ref{PeqB}) perform the
change of  variables ${\bf  q}_i={\bf q}_i'-{\bf a}$  which is  a pure
translation.  The  solvent potential  is translation  invariant and,
therefore,
\begin{align}
P^{\rm eq}\left({\bf R},\rho\right) &= 
\int \prod_{i=1}^Nd{\bf q}_i'
\frac{1}{Q} \exp\left\{-\beta\left( \hat{U}^{\rm sol}(q)+\sum_{i=1}^N\Phi^{\rm int}({\bf R}+{\bf a}-{\bf q}_i')
+\Phi^{\rm ext}({\bf R})\right)\right\}
\nonumber\\
&\times
\prod_\mu^M\delta\left(\rho_\mu-m_0\delta_\mu({\bf R})-\sum_{i=1}^Nm\delta_\mu({\bf q}_i'-{\bf a})\right)
\end{align}
Note  that  the right  hand  side  does  not  depend really  on  ${\bf
  a}$. Take the  derivative with respect to ${\bf a}$  and multiply by
$k_B T$ to obtain
\begin{align}
0&= \int \prod_{i=0}^Nd{\bf q}_i'
\frac{1}{Q} \exp\left\{-\beta\left( \hat{U}^{\rm sol}(q)
+\sum_{i=1}^N\Phi^{\rm int}({\bf R}+{\bf a}-{\bf q}_i')
+\Phi^{\rm ext}({\bf R})\right)\right\}
\int d{\bf r}{\bf F}^{\rm int}({\bf R}+{\bf a}-{\bf r})\hat{n}^{\rm sol}_{\bf r}(z)
\nonumber\\
&\times\prod_\mu^M\delta\left(\rho_\mu-m_0\delta_\mu({\bf R})
-\sum_{i=1}^Nm\delta_\mu({\bf q}_i'-{\bf a})\right)
\nonumber\\
&+k_BT\int \prod_{i=0}^Nd{\bf q}_i'
\frac{1}{Q} \exp\left\{-\beta\left( \hat{U}^{\rm sol}(q)
+\sum_{i=1}^N\Phi^{\rm int}({\bf R}+{\bf a}-{\bf q}_i')
+\Phi^{\rm ext}({\bf R})
\right)\right\}
\nonumber\\
&\times\sum_\nu\frac{\partial}{\partial {\rho_\nu}}\int d{\bf r}\boldsymbol{\nabla}\delta_\nu({\bf r}-{\bf a})
\hat{\rho}^{\rm sol}_{\bf r}(z)\prod_\mu^M\delta\left(\rho_\mu-m_0\delta_\mu({\bf R})
-\sum_{i=1}^Nm\delta_\mu({\bf q}_i'-{\bf a})\right)
\end{align}
{Here, ${\bf F}^{\rm int}({\bf R}-{\bf r})$ is the force that a solvent
particle located  at ${\bf r}$  exerts on the nanoparticle  located at
${\bf  R}$. } Evaluate  this expression  at ${\bf  a}=0$ and  divide by
$P^{\rm eq}\left({\bf R},\rho\right)$ to obtain
\begin{align}
0&= 
\int d{\bf r}{\bf F}^{\rm int}({\bf R}-{\bf r})
\llangle\hat{n}^{\rm sol}_{\bf r}\rrangle^{{\bf R}\rho}
+k_BT\int d{\bf r}\sum_\nu\boldsymbol{\nabla}\delta_\nu({\bf r})
\frac{1}{P^{\rm eq}\left({\bf R},\rho\right)}\frac{\partial}{\partial {\rho_\nu}}
P^{\rm eq}\left({\bf R},\rho\right)\llangle\hat{\rho}^{\rm sol}_{\bf r}\rrangle^{{\bf R}\rho}
\end{align}
Therefore, translation invariance implies the following exact result
\begin{align}
\int d{\bf r}{\bf F}^{\rm int}({\bf R}-{\bf r})\llangle\hat{n}^{\rm sol}_{\bf r}\rrangle^{{\bf R}\rho}
&=- k_BT\frac{\partial}{\partial {\rho_\nu}}\bra\hat{\rho}^{\rm sol}\boldsymbol{\nabla}\delta_\nu\ket^{{\bf R}\rho}
+\bra\hat{\rho}^{\rm sol}\boldsymbol{\nabla}\delta_\nu\ket^{{\bf R}\rho}
\frac{\partial}{\partial {\rho_\nu}}{\cal F}\left({\bf R},\rho\right)
\label{TI}
\end{align}
where we have used the notation (\ref{double}). 
This result is important because
it relates the actual force on the nanoparticle due to the solvent with the
derivatives of the free energy.

Now,   the   gradient  of   the   free   energy  ${\cal   F}\left({\bf
    R},\rho\right)$ introduced in Eq. (\ref{Fsolvent}) satisfies
\begin{align}
\frac{\partial}{\partial{\bf R}}  {\cal F}\left({\bf R},\rho\right)
+\frac{\partial}{\partial{\bf R}}  \Phi^{\rm ext}\left({\bf R}\right)
&=- \frac{1}{P^{\rm eq}\left({\bf R},\rho\right)}
\int \prod_{i=1}^Nd{\bf q}_i
\frac{1}{Q}\exp\left\{-\beta\left( \hat{U}^{\rm sol}(q)+\sum_{i=1}^N\Phi^{\rm int}({\bf R}-{\bf q}_i)-\beta\Phi^{\rm ext}({\bf R})\right)\right\}
\nonumber\\
&\times\int d{\bf r}{\bf F}^{\rm int}({\bf R}-{\bf r})\hat{n}^{\rm sol}_{\bf r}(z)
\prod_\mu^M\delta\left(\rho_\mu-m_0\delta_\mu({\bf R})-\sum_{i=1}^Nm_i\delta_\mu({\bf q}_i)\right)
\nonumber\\
&-k_BT \frac{1}{P^{\rm eq}\left({\bf R},\rho\right)}
\int \prod_{i=1}^Nd{\bf q}_i
\frac{1}{Q}\exp\left\{-\beta\left( \hat{U}^{\rm sol}(q)+\sum_{i=1}^N\Phi^{\rm int}({\bf R}-{\bf q}_i)-\beta\Phi^{\rm ext}({\bf R})\right)\right\}
\nonumber\\
&\times
\sum_\nu\frac{\partial}{\partial\rho_\nu}\left(-m_0\boldsymbol{\nabla}\delta_\nu({\bf R})\right)\prod_\mu^M\delta\left(\rho_\mu-m_0\delta_\mu({\bf R})-\sum_{i=1}^Nm_i\delta_\mu({\bf q}_i)\right)
\end{align}
\end{widetext}
Hence we have the exact result
\begin{align}
{\bf F}({\bf R}) &= \int d{\bf r}{\bf F}^{\rm int}({\bf R}-{\bf r})\llangle\hat{n}^{\rm sol}_{\bf r}\rrangle^{{\bf R}\rho}
\nonumber\\
&= -\frac{\partial {\cal F}}{\partial{\bf R}}\left({\bf R},\rho\right)
- m_0\boldsymbol{\nabla}\delta_\nu({\bf R})\frac{\partial{\cal F}}{\partial\rho_\nu}\left({\bf R},\rho\right)
\label{exactRFB}
\end{align}
This  result shows  that  the force  on the  nanoparticle  due to  the
solvent   is  not   simply  the   gradient  of   the  free   energy  $
-\frac{\partial {\cal  F}}{\partial{\bf R}}\left({\bf R},\rho\right)$,
but  also depends  on the  chemical potential  of the  fluid near  the
particle.   This is  because  the  density includes  the  mass of  the
colloid  in our  formulation.   The two  exact  result (\ref{TI})  and
(\ref{exactRFB})    combine    to     give    the    exact    relation
(\ref{ExactTIofF}).

\subsection{The exact form of the reversible drift}

We  now  consider   the  form  of  the   drift  given  by
(\ref{driftL}).
The Poisson brackets entering the elements of the reversible matrix are computed as follows
\begin{align}
\left\{\hat{\bf R},\hat{\bf R}\right\}&=0
\nonumber\\
\left\{\hat{\bf R},\hat{\rho}_\nu\right\}&=0
\nonumber\\
\left\{\hat{\bf R},\hat{\bf g}_\nu\right\}&=\delta_\mu({\bf r}_0)
\nonumber\\
\left\{\hat{\rho}_\mu,\hat{\bf R}\right\}&=0
\nonumber\\
\left\{ \hat{\bf g}_\mu,\hat{\bf R}\right\}&=-\delta_\mu({\bf r}_0)
\nonumber\\
\left\{ \hat{\rho}_\mu,\hat{\rho}_\nu\right\}&=\sum_{i}
\frac{\partial \hat{\rho}_\mu}{\partial {\bf r}_i}
\frac{\partial \hat{\rho}_\nu}{\partial {\bf p}_i}
-
\frac{\partial \hat{\rho}_\mu}{\partial {\bf p}_i}
\frac{\partial \hat{\rho}_\nu}{\partial {\bf r}_i} 
=0
\end{align}
\begin{align}
\left\{ \hat{\rho}_\mu, \hat{\bf g}_\nu\right\}
&=\sum_im_i\delta_\nu({\bf r}_i)\boldsymbol{\nabla}\delta_\mu({\bf r}_i)
\nonumber\\
&=\int d{\bf r}\hat{\rho}_{\bf r}(z)\delta_\nu({\bf r})\boldsymbol{\nabla}\delta_\mu({\bf r})
\nonumber\\
\left\{\hat{\bf g}_\mu,\hat{\rho}_\nu\right\}
&=-\sum_im_i\delta_\mu({\bf r}_i)\boldsymbol{\nabla}\delta_\nu({\bf r}_i)
\nonumber\\
&=\int d{\bf r}\hat{\rho}_{\bf r}(z)\delta_\mu({\bf r})\boldsymbol{\nabla}\delta_\nu({\bf r})
\nonumber\\
\left\{\hat{\bf g}^\alpha_\mu, \hat{\bf g}^\beta_\nu\right\}
&=\sum_i\left({\bf p}^\alpha_i\delta_\nu({\bf r}_i)\boldsymbol{\nabla}^\beta\delta_\mu({\bf r}_i)
-{\bf p}^\beta_i\delta_\mu({\bf r}_i)\boldsymbol{\nabla}\delta_\nu({\bf r}_i)\right)
\nonumber\\
&=\int d{\bf r}\left[\hat{\bf g}^\alpha_{\bf r}
\delta_\nu({\bf r})\boldsymbol{\nabla}^\beta\delta_\mu({\bf r})
-\hat{\bf g}^\beta_{\bf r}\delta_\mu({\bf r})\boldsymbol{\nabla}\delta_\nu({\bf r})\right]
\end{align}
The conditional averages are
\begin{align}
 \llangle\left\{\hat{\bf R}, \hat{\bf g}_\nu\right\}\rrangle^{{\bf R}\rho^B{\bf g}^B}
&= \delta_\mu({\bf R})
\nonumber\\
 \llangle\left\{ \hat{\rho}_\mu, \hat{\bf g}_\nu\right\}\rrangle^{{\bf R}\rho^B{\bf g}^B}
&=\bra\hat{\rho}\delta_\nu\boldsymbol{\nabla}\delta_\mu\ket^{{\bf R}\rho{\bf g}}
\nonumber\\
 \llangle
\left\{\hat{\bf g}^\alpha_\mu,\hat{\bf g}^\beta_\nu\right\}\rrangle^{{\bf R}\rho^B{\bf g}^B}
&=
{\bra\hat{\bf g}^\alpha\delta_\nu\boldsymbol{\nabla}^\beta\delta_\mu\ket^{{\bf R}\rho{\bf g}}
-\bra\hat{\bf g}^\beta \delta_\mu\boldsymbol{\nabla}^\alpha\delta_\nu\ket^{{\bf R}\rho{\bf g}}}
\end{align}
\begin{widetext}
where the double bracket notation is introduced in (\ref{double}). Therefore, the 
reversible part of the dynamics takes the form
\begin{align}
&\left( \begin{array}{c}
\llangle L\hat{\bf R}\rrangle^{{\bf R}\rho{\bf g}} \\
\\
\llangle L\hat{\rho}_\mu\rrangle^{{\bf R}\rho{\bf g}} \\
\\
\llangle L\hat{\bf g}^\alpha_\mu \rrangle^{{\bf R}\rho{\bf g}}
 \end{array}\right)
=\left( \begin{array}{ccc}
0&
0&
\delta_\mu({\bf R})
\\
\\
0&
0&
\bra\hat{\rho}\delta_\nu\boldsymbol{\nabla}^\beta\delta_\mu \ket^{{\bf R}\rho{\bf g}}
\\
\\
-\delta_\mu({\bf R}) 
&-\bra\hat{\rho}\delta_\mu\boldsymbol{\nabla}^\alpha\delta_\nu \ket^{{\bf R}\rho{\bf g}}
&\bra\hat{\bf g}^\alpha\delta_\nu\boldsymbol{\nabla}^\beta\delta_\mu\ket^{{\bf R}\rho{\bf g}}
-\bra\hat{\bf g}^\beta \delta_\mu\boldsymbol{\nabla}^\alpha\delta_\nu\ket^{{\bf R}\rho{\bf g}}
 \end{array}\right)
\left( \begin{array}{c}
\frac{\partial {\cal H}}{\partial {\bf R}} \\
\\
\frac{\partial {\cal H}}{\partial \rho_\nu} \\
\\
\frac{\partial {\cal H}}{\partial {\bf g}^\beta_\nu} 
 \end{array}\right)
\nonumber\\
\nonumber\\
&-k_BT
\left( \begin{array}{c}
0\\
\\
\frac{\partial}{\partial {\bf g}^\alpha_\nu}
\bra\hat{\rho}\delta_\nu\boldsymbol{\nabla}^\alpha\delta_\mu\ket^{{\bf R}\rho{\bf g}}
 \\
\\
-\boldsymbol{\nabla}^\alpha\delta_\mu({\bf R})
-\frac{\partial}{\partial \rho_\nu} 
\bra\hat{\rho}\delta_\mu\boldsymbol{\nabla}^\alpha\delta_\nu\ket^{{\bf R}\rho{\bf g}}
+\frac{\partial}{\partial {\bf g}^\beta_\nu}
\left(\bra\hat{\bf g}^\alpha\delta_\nu\boldsymbol{\nabla}^\beta\delta_\mu\ket^{{\bf R}\rho{\bf g}}
-\bra\hat{\bf g}^\beta \delta_\mu\boldsymbol{\nabla}^\alpha\delta_\nu\ket^{{\bf R}\rho{\bf g}}\right)
 \end{array}\right)
\label{revLform}
\end{align}
where no approximations have been taken so far.
\newpage\end{widetext}

\color{black}
\section{\label{App:RevApprox}Approximate form for the reversible drift}

We will now use  the approximations (\ref{Approx0}) (\ref{Mapprox}) in
order to  compute all  the different  terms that  appear in  the exact
equations (\ref{ExactDrift}).
Consider  first the term
\begin{align}
\bra\hat{\rho}\delta_\mu\boldsymbol{\nabla}^\alpha\delta_\nu\ket^{{\bf R}\rho{\bf g}}
&=
\int d{\bf r}\llangle \hat{\rho}_{\bf r}\rrangle^{{\bf R}\rho{\bf g}}
\delta_{\mu}({\bf r})\boldsymbol{\nabla}\delta_\mu({\bf r})
\end{align}
By using the linear for spiky approximation (\ref{Approx0}) this becomes
\begin{align}
\bra\hat{\rho}\delta_\mu\boldsymbol{\nabla}^\alpha\delta_\nu\ket^{{\bf R}\rho{\bf g}}
&
\simeq
\int d{\bf r}\llangle \psi_\sigma({\bf r})\hat{\rho}_{\sigma}\rrangle^{{\bf R}\rho{\bf g}}
\delta_{\mu}({\bf r})\boldsymbol{\nabla}\delta_\mu({\bf r})
\end{align}
Note that the conditional expectation of the discrete density field is just
the conditioning value, this is $\llangle \hat{\rho}_{\sigma}\rrangle^{{\bf R}\rho{\bf g}}=\rho_\sigma$. Therefore,
\begin{align}
\bra\hat{\rho}\delta_\mu\boldsymbol{\nabla}^\alpha\delta_\nu\ket^{{\bf R}\rho{\bf g}}
=\int d{\bf r}\psi_\sigma({\bf r})\rho_\sigma\delta_\mu\boldsymbol{\nabla}\delta_\mu
=\norm\overline{\rho}\delta_\mu\boldsymbol{\nabla}\delta_\mu\norm
\label{r1}\end{align}
where  we  have  used  the  definition  of  the  interpolated  density
field. 
{By using the LFSA (\ref{Approx0}) for the momentum field, the other
required term becomes
\begin{align}
  \bra\hat{\bf g}^\alpha\delta_\nu\boldsymbol{\nabla}^\beta\delta_\mu\ket^{{\bf R}\rho{\bf g}}
&\simeq\norm\overline{\bf g}^\alpha\delta_\nu\boldsymbol{\nabla}^\beta\delta_\mu\norm
\end{align}
}
We see that, formally, the linear for spiky approximation approximate hatted
functions with overlined functions, and allows to transform the double
brackets  $\bra\cdots\ket^{{\bf  R}\rho{\bf  g}}$  into  simple  space
averages $\norm\cdots\norm$.

Consider now the derivative of these terms that are required in (\ref{revLform})
\begin{align}
\frac{\partial}{\partial {\bf g}^\alpha_\nu}
\norm \overline{\rho}\delta_\nu\boldsymbol{\nabla}^\alpha\delta_\mu\norm&=0
\end{align}
This vanishes because the mass density and momentum density variables are independent.
The next term is of the form
\begin{align}
\frac{\partial}{\partial \rho_\nu} 
\bra\hat{\rho}\delta_\mu\boldsymbol{\nabla}^\alpha\delta_\nu\ket^{{\bf R}\rho{\bf g}}
&\simeq\frac{\partial}{\partial \rho_\nu} 
\norm \overline{\rho}\delta_\mu\boldsymbol{\nabla}^\alpha\delta_\nu\norm
\nonumber\\
&=\norm \psi_\nu\delta_\mu\boldsymbol{\nabla}^\alpha\delta_\nu\norm=0
\label{r2}
\end{align}
where the term vanishes due to (\ref{void}).
Finally, we need to compute the following derivative
\begin{align}
&\frac{\partial}{\partial {\bf g}^\beta_\nu}
\left(\bra\hat{\bf g}^\alpha\delta_\nu\boldsymbol{\nabla}^\beta\delta_\mu\ket^{{\bf R}\rho{\bf g}}
-\bra\hat{\bf g}^\beta \delta_\mu\boldsymbol{\nabla}^\alpha\delta_\nu\ket^{{\bf R}\rho{\bf g}}\right)
\nonumber\\
&\simeq\frac{\partial}{\partial {\bf g}^\beta_\nu}
\left({\bf g}^\alpha_\sigma\norm \psi_\sigma\delta_\nu\boldsymbol{\nabla}^\beta\delta_\mu\norm
-{\bf g}_\sigma^\beta\norm \psi_\sigma \delta_\mu\boldsymbol{\nabla}^\alpha\delta_\nu\norm\right)\nonumber\\
&=
\norm \psi_\nu\delta_\nu\boldsymbol{\nabla}^\alpha\delta_\mu\norm
-3\norm \psi_\nu \delta_\mu\boldsymbol{\nabla}^\alpha\delta_\nu\norm
=0\label{g1}
\end{align}
where the terms vanish due to (\ref{void}).
\begin{widetext}
In summary, the form of the reversible dynamics under the linear for spiky approximation is
\begin{align}
&\left( \begin{array}{c}
\llangle L{\bf R}\rrangle^{{\bf R}\rho{\bf g}} \\
\\
\llangle L\rho_\mu\rrangle^{{\bf R}\rho{\bf g}} \\
\\
\llangle L{\bf g}^\alpha_\mu \rrangle^{{\bf R}\rho{\bf g}}
 \end{array}\right)
=\left( \begin{array}{ccc}
0&
0&
\delta_\mu({\bf R})
\\
\\
0&
0&
\norm\overline{\rho}\delta_\nu\boldsymbol{\nabla}^\beta\delta_\mu \norm
\\
\\
-\delta_\mu({\bf R}) 
&-\norm\overline{\rho}\delta_\mu\boldsymbol{\nabla}^\alpha\delta_\nu \norm
&\norm\overline{\bf g}^\alpha\delta_\nu\boldsymbol{\nabla}^\beta\delta_\mu\norm
-\norm\overline{\bf g}^\beta \delta_\mu\boldsymbol{\nabla}^\alpha\delta_\nu\norm
 \end{array}\right)
\left( \begin{array}{c}
\frac{\partial {\cal H}}{\partial {\bf R}} \\
\\
\frac{\partial {\cal H}}{\partial \rho_\nu} \\
\\
\frac{\partial {\cal H}}{\partial {\bf g}^\beta_\nu} 
 \end{array}\right)
-k_BT
\left( \begin{array}{c}
0\\
\\
0
 \\
\\
-\boldsymbol{\nabla}^\alpha\delta_\mu({\bf R})
 \end{array}\right)
\label{revLformApprox}
\end{align}
\end{widetext}
Let  us now  apply the  linear for  spiky approximation  to the  exact
translation  invariance identity  (\ref{ExactTIofF}).  By  multiplying
(\ref{r1}) and (\ref{r2}) with the  volume ${\cal V}_\mu$ and sum over
$\mu$, by using (\ref{propdelta}), we obtain
\begin{align}
\bra\hat{\rho}\boldsymbol{\nabla}^\alpha\delta_\nu\ket^{{\bf R}\rho{\bf g}}
&\simeq\norm\overline{\rho}\boldsymbol{\nabla}\delta_\mu\norm
\nonumber\\
\frac{\partial}{\partial \rho_\nu} 
\bra\hat{\rho}\boldsymbol{\nabla}^\alpha\delta_\nu\ket^{{\bf R}\rho{\bf g}}
&\simeq0
\label{r4}
\end{align}
By  using  these  approximations  in  the  exact  translation  property
(\ref{ExactTIofF}) we obtain the approximation
\begin{align}
\frac{\partial{\cal F}}{\partial{\bf R}}
+\norm\overline{\rho}\boldsymbol{\nabla}\delta_\nu\norm\frac{\partial {\cal F}}{\partial\rho_\nu}
&=0
\label{TIofFApp}
\end{align}

\section{\label{App:Models}Approximate model for the 
CG Hamiltonian}

\subsubsection{Modelling the kinetic  part  of the CG Hamiltonian}
The kinetic part of exact CG Hamiltonian in (\ref{CGHamiltonianExact})
can be approximated under the LFSA (\ref{Mapprox}) in the form
\begin{align}
&-k_BT\ln \llangle \frac{\exp\left\{-\frac{\beta}{2}{\bf g}_\mu\hat{M}^{-1}_{\mu\nu}{\bf g}_\nu\right\}}
{(2\pi/\beta)^{3M/2}\det \hat{M}^{3/2}}\rrangle^{{\bf R}\rho}
\nonumber\\
&\simeq
\frac{1}{2}{\bf g}_\mu\overline{M}^{-1}_{\mu\nu}{\bf g}_\nu
+\frac{3k_BT}{2}\ln{\det \overline{M}}
\end{align}
up to  irrelevant constant terms. The  order of magnitude of  the term
proportional  to $k_BT$  can be  estimate by  assuming a  sufficiently
smooth density field for which we may approximate
\begin{align}
  {\cal M}_{\mu\nu}=(\delta_\mu\delta_\nu\psi_\sigma)\rho_\sigma\simeq\delta_{\mu\nu}\frac{\rho_\mu}{{\cal V}_\mu}
\end{align}
leading to a  diagonal matrix. This approximation  still satisfies the
exact  requirement (\ref{hydroconsistency}).  The  log det  term of  a
diagonal matrix is simple
\begin{align}
\frac{3k_BT}{2}  \ln\det{\cal M}=\frac{3k_BT}{2}{\rm tr}\ln{\cal M}\simeq\frac{3k_BT}{2}\sum_\mu\ln \rho_\mu
\end{align}
We observe that this term \textit{is not extensive}, this is, does not scale 
as the number of particles per node. On the other hand, the kinetic energy
\begin{align}
 \frac{1}{2}{\bf g}_\mu\overline{M}^{-1}_{\mu\nu}{\bf g}_\nu\simeq\sum_\mu{\cal V}_\mu
\frac{{\bf g}_\mu^2}{\rho_\mu}
\end{align}
scales with  the number  of particles  per node  because,
typically ${\bf g}_\mu\sim \rho_\mu{\bf v}_\mu $ and
$\rho_\mu\sim m \frac{N_\mu}{{\cal V}_\mu}$, giving
\begin{align}
   \frac{1}{2}{\bf g}_\mu\overline{M}^{-1}_{\mu\nu}{\bf g}_\nu
\sim \sum_\mu \frac{N_\mu m }{2}{\bf v}_\mu^2
\end{align}
which  is  an  extensive  quantity,  proportional  to  the  number  of
particles per node.  As we assume that the typical number of particles
per  node  is  large,  we may  neglect  the  term  $\frac{3k_BT}{2}\ln
\det{\cal M}$ in front of the kinetic energy term.

 From now on we will neglect this term and the CG Hamiltonian
has the form
\begin{align}
  {\cal H}({\bf R},\rho,{\bf g})&= \frac{1}{2}{\bf g}_\mu\overline{M}^{-1}_{\mu\nu}{\bf g}_\nu
+{\cal F}\left({\bf R},\rho\right)+\Phi^{\rm ext}({\bf R})
\label{CGHApprox}
\end{align}
We will need the derivatives of the
CG Hamiltonian that are given by 
\begin{align}
  \frac{\partial {\cal H}}{\partial {\bf R}}&=
  \frac{\partial {\cal F}}{\partial {\bf R}}
+  \frac{\partial \Phi^{\rm ext}}{\partial {\bf R}}
\nonumber\\
  \frac{\partial {\cal H}}{\partial \rho_\mu}&=
\frac{1}{2}{\bf g}_{\mu'}   \frac{\partial \overline{M}^{-1}_{\mu'\nu'}}{\partial \rho_\mu}
{\bf g}_{\nu'}
+   \frac{\partial {\cal F}}{\partial \rho_\mu}
\nonumber\\
  \frac{\partial {\cal H}}{\partial {\bf g}_\mu}&=
\overline{M}^{-1}_{\mu\mu'}{\bf g}_{\mu'}
\label{derHB}
\end{align} 
We now use the result
\begin{align}
   \frac{\partial \overline{M}^{-1}_{\mu'\nu'}}{\partial \rho_\mu}&=
-\overline{M}^{-1}_{\mu'\mu''}   \frac{\partial \overline{M}_{\mu''\nu''}}{\partial \rho_\mu}
\overline{M}^{-1}_{\nu''\nu'} 
\nonumber\\
&=
-\overline{M}^{-1}_{\mu'\mu''}   \norm\delta_{\mu''}\psi_\mu\delta_{\nu''}\norm
 \overline{M}^{-1}_{\nu''\nu'} 
\end{align}
where we have used the explicit form of the matrix in (\ref{calM}).
With the discrete velocity
(\ref{DiscreteVel}) the derivatives (\ref{derHB}) become, finally
\begin{align}
  \frac{\partial {\cal H}}{\partial {\bf R}}&=
  \frac{\partial {\cal F}}{\partial {\bf R}}+  \frac{\partial \Phi^{\rm ext}}{\partial {\bf R}}
\nonumber\\
  \frac{\partial {\cal H}}{\partial \rho_\mu}&=
-\frac{1}{2}\norm\psi_\mu\overline{\bf v}\overline{\bf v}\norm
+   \frac{\partial {\cal F}}{\partial \rho_\mu}
\nonumber\\
  \frac{\partial {\cal H}}{\partial {\bf g}_\mu}&=M^\psi_{\mu\mu'}{\bf v}_{\mu'}
\label{derHB0}
\end{align}

\subsubsection{Modelling the solvent part  of the free energy}
The free energy of the  solvent ${\cal F}^{\rm sol}(\rho)$ is obtained
from  the   first  equation  (\ref{Fsolvent})  from   the  probability
(\ref{PeqSol}).    The  explicit   calculation  of   $P^{\rm  eq}_{\rm
  sol}(\rho)$ is in general impossible  due to the high dimensionality
of the integrals in phase space.  Therefore, we are forced to consider
specific approximate  models for this probability  distribution.

  In
  accordance with the assumption  that each hydrodynamic cell contains
  many fluid  molecules, we will  assume that the  probability $P^{\rm
    eq}_{\rm sol}(\rho)$  is a Gaussian.    The Gaussian probability has the form
\begin{align}
  P^{\rm eq}_{\rm sol}(\rho)&=\frac{1}{N}\exp\left\{-\frac{1}{2}\delta\rho_\mu C^{-1}_{\mu\nu}\delta\rho_\nu\right\}
\end{align}
where   $N$  is   the  normalization,  $\delta\rho_\mu=\rho_\mu-\rho_{\rm eq}$    are    the
fluctuations  with respect  to the  homogeneous density $\rho_{\rm eq}$, and
the matrix of covariances is given by
\begin{align}
  C_{\mu\nu}=\llangle \delta\rho_\mu\delta\rho_\nu\rrangle_{\rm eq}=
\int d{\bf r}\int d{\bf r}\delta_{\mu}({\bf r})\delta_{\nu}({\bf r}')\llangle \delta\rho_{\bf r}\delta\rho_{{\bf r}'}\rrangle_{\rm eq}
\end{align}
We estimate the form of this  matrix as follows.  We assume that the
correlation  of density  fluctuations fluctuations  decay in  a length
scale  much smaller  than the  size of  the cell  and, therefore,  the
correlation can  be approximated  as proportional  to the  Dirac delta
function, according to a standard result 
\begin{align}
  \llangle\delta\rho_{\bf r}\delta\rho_{{\bf r}'}\rrangle_{\rm eq}=
\frac{k_B T \rho_{\rm eq}}{ c^2} \delta({\bf r}-{\bf r}')
\end{align}
where $c$ is the isothermal speed of sound.  The resulting free energy
is quadratic in  the density and will be  termed \textit{Gaussian free
  energy}. It has the explicit form
\begin{align}
  {\cal F}^{\rm sol}(\rho)=\frac{c^2}{2\rho_{\rm eq}}\delta\rho_\mu M^\psi_{\mu\nu}\delta\rho_\nu
\label{FGauss}
\end{align}

This free energy \textit{function} can be obtained from a local free energy
\textit{functional} of the form  (square  brackets   denote  a
functional, while rounded parenthesis denote a function)
\begin{align}
 {\cal F}^{\rm sol}[\rho]&=\int d{\bf r}f^{\rm sol}(\rho({\bf r}))
\label{Flocal}
\end{align}
where $f^{\rm sol}(\rho)$  is the thermodynamic free  energy density of
the solvent which, for the Gaussian model is
\begin{align}
  f(\rho)=\frac{c^2}{2\rho_{\rm eq}}\left(\rho-\rho_{\rm eq}\right)^2
\label{fGauss}
\end{align}
This functional  is perhaps the  simplest model familiar  from Density
Functional Theory.   The model neglects molecular  correlations, which
is appropriate  for the coarse  description in which  the hydrodynamic
cells are  much larger than  molecular correlation lengths.   The free
energy    (\ref{FGauss})    is    obtained   from    the    functional
(\ref{Flocal})-(\ref{fGauss})   by   using  the   interpolated   field
$\overline{\rho}({\bf  r})=\rho_\sigma\psi_\sigma({\bf   r})$  in  the
functional, as advocated in Ref.  \cite{FluctDiff_FEM}. 

Once we have a free energy density, we may compute the pressure of the
Gaussian model from the well-known thermodynamic relation
\begin{align}
  P^{\rm sol}(\rho)=& \rho \frac{df^{\rm sol}}{d\rho}(\rho)-f^{\rm sol}(\rho)
\label{Pressdef}
\end{align}
The pressure (\ref{Pressdef}) that corresponds to (\ref{fGauss}) 
is given by the quadratic equation of state (EOS)
\begin{align}
  P^{\rm sol}(\rho)&=\frac{c^2}{2\rho_{\rm eq}}\left(\rho^2-\rho^2_{\rm eq}\right)
\label{Pressure}
\end{align}
Observe that for small deviations from equilibrium we obtain the expected
linear EOS $P^{\rm sol}(\rho)=c^2 \left(\rho-\rho_{\rm eq}\right)$.

\subsubsection{Modelling the interaction part  of the free energy}
The  interaction part  of the  free energy  has the  exact microscopic
expression given in (\ref{S2:calFF}). We  may obtain a simple model for
this function if
we consider the linear  for spiky approximation (\ref{Approx0}).  Note
that  the  microscopic potential  energy  of  interaction between  the
nanoparticle and  the solvent molecules  can be expressed in  terms of
the microscopic solvent mass density as
\begin{align}
  \sum_{i=1}^N\Phi^{\rm int}({\bf R}-{\bf q}_i)= \frac{1}{m}\int d{\bf r}\Phi^{\rm int}({\bf R}-{\bf r})\hat{\rho}_{\bf r}^{\rm sol}(z)
\label{Phin}
\end{align}
Within   the  linear  for  spiky   approximation,  we  will
approximate the spiky field $  \hat{\rho}_{\bf r}^{\rm sol}(z)$ with a
linear interpolation
\begin{align}
  \hat{\rho}_{\bf r}^{\rm sol}(z)\simeq \psi_\mu({\bf r})\hat{\rho}^{\rm sol}_\mu(z)
\label{naprox}
\end{align}
by using the approximation (\ref{naprox}) into (\ref{Phin}) we obtain
\begin{align}
  \sum_{i=1}^N\Phi^{\rm int}({\bf R}-{\bf q}_i)&\simeq
\frac{1}{m}\Phi^{\rm int}_\mu({\bf R})\hat{\rho}^{\rm sol}_\mu(z)
\label{aproxphi0}
\end{align}
where the nodal potential $\Phi^{\rm int}_\mu({\bf R})$ is defined according
to
\begin{align}
\Phi^{\rm int}_\mu({\bf R})\equiv&
\int d{\bf r}\Phi^{\rm int}({\bf R}-{\bf r})\psi_\mu({\bf r})
\label{PhimuR}
\end{align}

We consider situations in  which the nanoparticle is much smaller
than the hydrodynamic cells and the range of the interaction potential
$\Phi^{\rm  int}({\bf  R}-{\bf r})$  is  also  much smaller  than  the
support  of  $\psi_\mu({\bf  r})$.    Therefore,  we  may  approximate
$\psi_\mu({\bf  r}) \simeq\psi_\mu({\bf  R})$  in Eq.  (\ref{PhimuR}),
leading to
\begin{align}
\Phi^{\rm int}_\mu({\bf R})&\simeq a\psi_\mu({\bf R})
\label{PhimuRLoc}
\end{align}
where the constant $a$ is the volume integral of the  interaction potential
\begin{align}
a = \int d{\bf r}\Phi^{\rm int}({\bf r})\equiv \frac{m m_0 c_0^2}{\rho_{\rm eq}} 
  \label{a_def}  
\end{align}
and we have introduced the ``particle speed of
sound''  $c_0$ whose  physical  interpretation is  that  it gives  the
strength  of the  interaction  of the  nanoparticle  with the  solvent
particles.  Under  these approximations, the microscopic  potential of
interaction  between the  nanoparticle  and the  solvent molecules  is
approximated by
\begin{align}
  \sum_{i=1}^N\Phi^{\rm int}({\bf R}-{\bf q}_i)&\simeq \frac{m_0c^2_0}{\rho_{\rm eq}}
\psi_\mu({\bf R})\hat{\rho}^{\rm sol}_\mu(z)
\label{aproxphi}
\end{align}

Note  that this  approximation breaks  translation invariance,  because
while  the left  hand side  of (\ref{aproxphi})  is invariant  under a
translation of all the particles, the  right hand side is not.  In the
approximation   (\ref{aproxphi}),   the   potential  energy   of   the
nanoparticle  depends  on the  microscopic  configuration  $z$ of  the
solvent  particles only  through the  discrete solvent  number density
$\hat{n}^{\rm   sol}_{\mu}(z)$.    Therefore,  by   substituting   the
approximation  (\ref{aproxphi})  for  the potential  energy  into  the
interaction part  of the  free energy  (\ref{S2:calFF}) we  obtain the
explicit model
\begin{align}
{\cal F}^{\rm int}({\bf R},\rho^{\rm sol})&\simeq \frac{m_0c^2_0}{\rho_{\rm eq}}
\psi_\mu({\bf R})\rho^{\rm sol}_\mu 
\label{CalFintApprox}
\end{align}
By collecting (\ref{FGauss}) and (\ref{CalFintApprox}) the free energy (\ref{freedecomp}) becomes 
\begin{align}
  {\cal F}({\bf R},\rho)&=\frac{c^2}{2\rho_{\rm eq}}\delta\rho^{\rm sol}_\mu M^\psi_{\mu\nu}\delta\rho^{\rm sol}_\nu
+\frac{m_0c_0^2}{\rho_{\rm eq}}
\psi_\mu({\bf R})
\delta\rho^{\rm sol}_\mu
\end{align}
In terms of the total mass density,  the free energy is
\begin{align}
  {\cal F}({\bf R},\rho)&=\frac{c^2}{2\rho_{\rm eq}}\delta\rho_\mu M^\psi_{\mu\nu}\delta\rho_\nu
+\frac{m_0(c_0^2-c^2)}{\rho_{\rm eq}}\psi_\mu({\bf R})\delta\rho_\mu
\nonumber\\
&+\epsilon({\bf R})
\label{FModelI}
\end{align}
where the last term is a density-independent term
\begin{align}
\epsilon({\bf R})&\equiv  \frac{m^2_0}{\rho_{\rm eq}}\left[\frac{c^2}{2}-c_0^2\right]
\delta_\mu({\bf R})\psi_\mu({\bf R})
\end{align}
The derivatives of the model (\ref{FModelI}) are
\begin{align}
\frac{\partial}{\partial{\bf R}}  {\cal F}({\bf R},\rho)
&=\frac{m_0(c_0^2-c^2)}{\rho_{\rm eq}}\rho_\mu\boldsymbol{\nabla}\psi_\mu({\bf R})
+\frac{\partial\epsilon}{\partial{\bf R}}({\bf R})
\nonumber\\
\frac{\partial}{\partial\rho_\mu}{\cal F}({\bf R},\rho)&=
\frac{c^2}{\rho_{\rm eq}}M^\psi_{\mu\nu}\delta \rho_\nu+
\frac{m_0(c_0^2-c^2)}{\rho_{\rm eq}}\psi_\mu({\bf R})
\label{DerFreeEnergy}
\end{align}
From Eq. (\ref{exactRFB}) the force on the nanoparticle due to the surrounding solvent 
is given by
\begin{align}
{\bf F}({\bf R})
&= -\frac{\partial {\cal F}}{\partial{\bf R}}\left({\bf R},\rho\right)
- m_0\boldsymbol{\nabla}\delta_\nu({\bf R})\frac{\partial{\cal F}}{\partial\rho_\nu}\left({\bf R},\rho\right)
\nonumber\\
&=-\frac{m_0c_0^2}{\rho_{\rm eq}}\rho^{\rm sol}_\mu\boldsymbol{\nabla}\psi_\mu({\bf R})
\label{ForceNanoparticle1}
\end{align}
This form of the force is consistent with the approximation (\ref{aproxphi}).

We now  consider the translation invariance  (\ref{TIofF}) property of
the free energy.  This property is a strong  guiding principle for the modelling
of the free energy.  In order to  see if this important
property is satisfied, we compute
\begin{align}
&    \frac{\partial}{\partial{\bf R}}{\cal F}({\bf R},\rho)+
\norm\overline{\rho}\boldsymbol{\nabla}\delta_\nu\norm\frac{\partial {\cal F}}{\partial\rho_\nu}\nonumber\\
&=
\frac{m_0(c_0^2-c^2)}{\rho_{\rm eq}}\rho_\mu
\left(\boldsymbol{\nabla}\psi_\mu({\bf R})
+\norm\psi_\mu\boldsymbol{\nabla}\delta_\nu\norm\psi_\nu({\bf R})\right)
\nonumber\\
&+\frac{c^2}{\rho_{\rm eq}}\underbrace{\norm\overline{\rho}\boldsymbol{\nabla}\overline{\rho}\norm}_{=0}
+\frac{\partial\epsilon}{\partial{\bf R}}({\bf R})
\nonumber\\
&=
\frac{m_0(c_0^2-c^2)}{\rho_{\rm eq}}
\left(\boldsymbol{\nabla}\overline{\rho}({\bf R})
-\norm \Delta\boldsymbol{\nabla}\overline{\rho}\norm\right)
+\frac{\partial\epsilon}{\partial{\bf R}}({\bf R})
\label{TI?}
\end{align}
where    we   used    that   in    a   periodic    domain   $    \norm
\overline{\rho}\boldsymbol{\nabla}\overline{\rho}\norm =0$  (as can be
seen from  integration by parts).  Also, $\norm\Delta\boldsymbol{\nabla}\overline{\rho}\norm$ is a 
compact notation for
\begin{align}
  \norm\Delta\boldsymbol{\nabla}\overline{\rho}\norm\equiv
\int d{\bf r}\Delta({\bf R},{\bf r})\boldsymbol{\nabla}\overline{\rho}({\bf r})
\end{align}
and it  depends on  the position  ${\bf R}$  of the  nanoparticle.  We
observe that, in general, (\ref{TI?})  does not vanish. However, note
that  for  sufficiently  smooth   density  fields  {Eq  (\ref{rhoapp})
  applies,  and} the  first  term  is small.   In  particular, in  the
incompressible limit  in which  the density  is constant,  it vanishes
identically.  This  strongly suggests that for  modelling purposes, it
is convenient  to set  $\epsilon({\bf R})=0$  and correct
  the free energy model developed so  far  in order to better respect
translational invariance.

In conclusion, in the present work we will use the following model for
the free energy of a fluid made of a solvent interacting with a single
nanoparticle
\begin{align}
  {\cal F}({\bf R},\rho)&=\frac{c^2}{2\rho_{\rm eq}}\delta\rho_\mu M^\psi_{\mu\nu}\delta\rho_\nu
+\frac{m_0(c_0^2-c^2)}{\rho_{\rm eq}}\psi_\mu({\bf R})\rho_\mu
\label{SimpleFreeEnergyModel}
\end{align}
Because this  free energy  gives the  probability $P({\bf
    R},\rho)$, and we  expect that for the case  that the nanoparticle
  is  identical  to a  tagged  solvent  particle this  probability  is
  Gaussian, we  conclude that the  limit of the  nanoparticle becoming
  just another solvent particle is realized for $c_0=c$.

The   derivatives   of   the   free   energy   model
(\ref{SimpleFreeEnergyModel}) are
\begin{align}
\frac{\partial}{\partial{\bf R}}  {\cal F}({\bf R},\rho)
&=\frac{m_0(c_0^2-c^2)}{\rho_{\rm eq}}\rho_\mu\boldsymbol{\nabla}\psi_\mu({\bf R})
\nonumber\\
\frac{\partial}{\partial\rho_\mu}{\cal F}({\bf R},\rho)&=
\frac{c^2}{\rho_{\rm eq}}M^\psi_{\mu\nu}\delta \rho_\nu+
\frac{m_0(c_0^2-c^2)}{\rho_{\rm eq}}\psi_\mu({\bf R})
\label{DerFreeEnergyOK}
\end{align}
With these derivatives, we  now compute the  term (\ref{form})  entering the
momentum equation
{
\begin{align}
&-\delta_{\mu}({\bf R})\frac{\partial{\cal F}}{\partial{\bf R}}
-\norm \overline{\rho}\delta_\mu\boldsymbol{\nabla}\delta_\nu \norm
\frac{\partial {\cal F}}{\partial \rho_\nu}
\nonumber\\
&= 
-\frac{c^2}{\rho_{\rm eq}}
\norm \delta_\mu\overline{\rho}\boldsymbol{\nabla}\overline{\rho} \norm
-
\frac{m_0(c_0^2-c^2)}{\rho_{\rm eq}}
\left(\delta_\mu({\bf R}) \boldsymbol{\nabla}\overline{\rho}({\bf R})
+\norm \overline{\rho}\delta_\mu\boldsymbol{\nabla}\Delta\norm\right) 
\nonumber\\
&=-\norm \delta_\mu\boldsymbol{\nabla}\overline{P} \norm
-\frac{m_0(c_0^2-c^2)}{\rho_{\rm eq}}
\left(\delta_\mu({\bf R}) \boldsymbol{\nabla}\overline{\rho}({\bf R})
-\norm \delta_\mu \Delta \boldsymbol{\nabla}\overline{\rho}\norm\right) 
\label{RhoGradMu00}
\end{align}

}where we  have  introduced  the following total pressure equation of state
\begin{align}
\overline{P}({\bf r})&\equiv\frac{c^2}{2\rho_{\rm eq}}\left(\overline{\rho}({\bf r})^2-\rho^2_{\rm eq}\right)
+\frac{m_0(c_0^2-c^2)}{\rho_{\rm eq}}
\Delta({\bf R},{\bf r})\overline{\rho}({\bf r})
\label{TotalPressure}
\end{align}
Note that in the limit when  the nanoparticle is just a tagged solvent
particle we  have $c_0=c$  and the last  contribution to  the pressure
vanishes,  giving   simply  the   pressure  of  the   Gaussian  model.
{The last term in  (\ref{RhoGradMu00}) is arguably small and
  will be neglected. Indeed, for smooth density fields
  \begin{align}
    \norm\delta_\mu\Delta\boldsymbol{\nabla}\overline{\rho}\norm\simeq
    \norm\delta_\mu\Delta\norm\boldsymbol{\nabla}\overline{\rho}({\bf R})=
    \delta_\mu({\bf R})\boldsymbol{\nabla}\overline{\rho}({\bf R})
\label{consist}  \end{align}
and we have, finally
\begin{align}
-\delta_{\mu}({\bf R})\frac{\partial{\cal F}}{\partial{\bf R}}
-\norm \overline{\rho}\delta_\mu\boldsymbol{\nabla}\delta_\nu \norm
\frac{\partial {\cal F}}{\partial \rho_\nu}
&=-\norm \delta_\mu\boldsymbol{\nabla}\overline{P} \norm
\label{RhoGradMu}
\end{align}

}

\subsection{Translation invariance and the barometric law}
In this appendix we examine  the marginal probability $P^{\rm eq}({\bf
  R})$  of  finding the  nanoparticle  at  position ${\bf  R}$.   This
probability is, by definition,
\begin{align}
P^{\rm eq}({\bf R})&=  \int d\rho P^{\rm eq}({\bf R},\rho)
\nonumber\\
&=\int d\rho
\exp\{-\beta\left({\cal F}({\bf R},\rho)+\Phi^{\rm ext}({\bf R})\right) \}
\end{align}
Take its gradient
\begin{align}
\frac{\partial}{\partial {\bf R}}P^{\rm eq}({\bf R})
&= -\beta\int d\rho \frac{\partial  {\cal F}}{\partial {\bf R}}
\exp\{-\beta\left({\cal F}({\bf R},\rho)+\Phi^{\rm ext}({\bf R})\right) \}
\nonumber\\
&-\beta\frac{\partial  \Phi^{\rm ext}}{\partial {\bf R}}P^{\rm eq}({\bf R})
\end{align}
and use the approximate translation invariance  of the free energy (\ref{TIofF})
\begin{align}
&\frac{\partial}{\partial {\bf R}}P^{\rm eq}({\bf R})
+\beta\frac{\partial  \Phi^{\rm ext}}{\partial {\bf R}}P^{\rm eq}({\bf R})
\nonumber\\
&= \beta\int d\rho 
\norm\overline{\rho}\boldsymbol{\nabla}\delta_\mu\norm
\frac{\partial {\cal F}}{\partial\rho_\mu}\exp\{-\beta\left({\cal F}({\bf R},\rho)+\Phi^{\rm ext}({\bf R})\right) \}
\nonumber\\
&  = \int d\rho 
\exp\{-\beta\left({\cal F}({\bf R},\rho)+\Phi^{\rm ext}({\bf R})\right) \}
\frac{\partial }{\partial\rho_\mu}\norm\overline{\rho}\boldsymbol{\nabla}\delta_\mu\norm
\nonumber\\
& = \int d\rho 
\exp\{-\beta{\cal F} \}
\norm\psi_\mu\boldsymbol{\nabla}\delta_\mu\norm=0
\end{align}
Therefore,  \textit{the (approximate)  translation invariance  property
  (\ref{TIofF}) \Pepmodified{rigorously} implies the well-known barometric law }
\begin{align}
  P^{\rm eq}({\bf R})&=\frac{1}{Q}\exp\left\{-\beta\Phi^{\rm ext}({\bf R})\right\}
\label{barometric}
\end{align}
where  $Q$  is  the  normalization  factor.   In the absence of
an external field the probability to find the particle at a particular
point ${\bf R}$ should be constant.

Should the free energy  model respect \textit{exactly} the translation
invariance  property  (\ref{TIofF}),  then the  marginal  distribution
function  would be  \textit{rigorously}  given by  the barometric  law
(\ref{barometric}).        However,      the       Gaussian      model
(\ref{SimpleFreeEnergyModel})   for   the    free   energy   satisfies
(\ref{TIofF})  only approximately,  up  to second  order  terms. As  a
consequence,   the  marginal   distribution   $P^{\rm  eq}({\bf   R})$
corresponding to the model (\ref{SimpleFreeEnergyModel}) does not give
\textit{exactly} the barometric law (\ref{barometric}) but rather
\begin{align}
  P^{\rm eq}({\bf R})&=\int d\rho \exp\{-\beta{\cal F}({\bf R},\rho)-\beta \Phi^{\rm ext}({\bf R})\}\nonumber\\
&\propto
\exp\left\{-\beta\Phi^{\rm ext}({\bf R})+\beta\frac{m_0^2(c_0^2-c^2)^2}{2\rho_{\rm eq}c^2}
\delta_\mu({\bf R})\psi_\mu({\bf R})
\right\}
\label{PRnotBaro}
\end{align}
as can  be seen by  explicitly performing the Gaussian  integral. When
$\Phi^{\rm ext}=0$, the nanoparticle  is not homogeneously distributed
in space but, rather, ``sees'' the underlying grid,
unless it is a tagged fluid particle in which case $c_0=c$.

\section{Derivatives of  the basis functions}
\label{App:ApproxBasis}

In this work we assume periodic boundary conditions. Therefore
any integration by parts give no surface terms. For example
\begin{align}
  \norm A\boldsymbol{\nabla}B\norm= -\norm B\boldsymbol{\nabla}A\norm
\end{align}
for arbitrary functions $A({\bf r}),B({\bf r})$.

We  consider   some  identities   that  involve  gradients   of  basis
functions. For example, note that
\begin{align}
  \norm\delta_\mu\psi_\nu\boldsymbol{\nabla}\delta_\nu\norm&=
  \norm\delta_\mu\delta_\nu\boldsymbol{\nabla}\psi_\nu\norm
=-  \norm\psi_\nu\boldsymbol{\nabla}\delta_\mu\delta_\nu\norm
\nonumber\\
&=-  \norm\psi_\nu\delta_\mu\boldsymbol{\nabla}\delta_\nu\norm
-  \norm\psi_\nu\delta_\nu\boldsymbol{\nabla}\delta_\mu\norm
\end{align}
where in the second identity we have performed an integration by parts. Therefore,
\begin{align}
\norm\psi_\nu\delta_\nu\boldsymbol{\nabla}\delta_\mu\norm&= -2 \norm\delta_\mu\psi_\nu\boldsymbol{\nabla}\delta_\nu\norm
\end{align}
By multiplying both sides of this equation with ${\cal V}_\mu$ and summing over $\mu$, we obtain
\begin{align}
\norm\delta_\nu\nabla\psi_\nu\norm=0  
\label{dnp}
\end{align}
Another identity is obtained by introducing 
the following vector defined at each node $\mu$
\begin{align}
  {\bf a}_\mu&\equiv \norm\delta_\mu\psi_\nu\boldsymbol{\nabla}\delta_\nu\norm
\end{align}
Because of (\ref{dnp}), this vector satisfies
\begin{align}
{\cal V}_\mu  {\bf a}_\mu=0
\end{align}
If the  mesh of nodes is  regular {in such a  way that for
  all  nodes $\mu$  we have  ${\bf a}_\mu={\bf  a}_0$, then  the above
  equation  implies ${\bf  a}_0=0$}. Therefore,  in regular  grids we
have the identities
\begin{align}
\norm\psi_\nu\delta_\nu\boldsymbol{\nabla}\delta_\mu\norm&=0
\nonumber\\
\norm\delta_\mu\psi_\nu\boldsymbol{\nabla}\delta_\nu\norm  &=0
\label{void}
\end{align}
It is expected that in non-regular meshes these quantities are also zero or very small.

\section{Momentum integrals} 
\label{App:MomentumIntegrals}
In this appendix we quote the results for the following momentum
integrals 
\begin{align}
I_0({\bf g})\equiv& \int \prod^N_{i=0}d{\bf p}_i\exp\left\{-\beta\sum_{i=0}^N\frac{{\bf p}_i^2}{2m_i}\right\}
\nonumber\\
&\times\prod_\mu^M\delta\left(\sum_{i=0}^N{\bf p}_i\delta_\mu({\bf q}_i)-{\bf g}_\mu\right)
\nonumber\\
&=\frac{\prod^N_{i=0}(2\pi m_ik_BT)^{3/2}}{(2\pi k_BT)^{3M/2}\det \hat{M}^{3/2}}
\exp\left\{-\frac{\beta}{2}
{\bf g}_\mu
\hat{M}_{\mu\nu}^{-1}
{\bf g}_\nu
\right\}
\end{align}
\begin{align}
{\bf I}_i^{(1)}({\bf g})\equiv& \int  \prod^N_{i=0}d{\bf p}_i\exp\left\{-\beta\sum_{i=0}^N\frac{{\bf p}_i^2}{2m_i}\right\}
\nonumber\\
&\times\prod_\mu^M\delta\left(\sum_{i=0}^N{\bf p}_i\delta_\mu({\bf q}_i)-{\bf g}_\mu\right)
{\bf p}_i
\nonumber\\
&=I_0({\bf g})m_i\delta_\mu({\bf q}_i)\hat{M}_{\mu\nu}^{-1}{\bf g}_\nu
\label{III}
\end{align}
where the configuration dependent mass matrix is defined as
\begin{align}
 \hat{M}_{\mu\nu}(z)&\equiv\sum_{i=0}^Nm_i\delta_\mu({\bf q}_i)\delta_\nu({\bf q}_i)
\label{hydrohatMmunu1}
\end{align}
The above integrals are relatively easy to compute 
by using the Fourier representation of the Dirac delta function.


\begin{thebibliography}{10}

\bibitem{Einstein1905}
A~Einstein.
\newblock {\"{U}ber die von der molekularkinetischen Theorie der W\"{a}rme
  geforderte Bewegung von in ruhenden Fl\"{u}ssigkeiten suspendierten
  Teilchen}.
\newblock {\em Ann. Phys.}, 19:549, 1905.

\bibitem{Faxen_FluctuatingHydro}
D.~Bedeaux and P.~Mazur.
\newblock {Brownian motion and fluctuating hydrodynamics}.
\newblock {\em Physica}, 76(2):247--258, 1974.

\bibitem{VACF_FluctHydro}
E.~H. Hauge and A.~Martin-Lof.
\newblock {Fluctuating hydrodynamics and Brownian motion}.
\newblock {\em J. Stat. Phys.}, 7(3):259--281, 1973.

\bibitem{BrownianCompressibility_Zwanzig}
R.~Zwanzig and M.~Bixon.
\newblock {Compressibility effects in the hydrodynamic theory of Brownian
  motion}.
\newblock {\em J. Fluid Mech.}, 69:21--25, 1975.

\bibitem{DiffusionRenormalization_I}
D.~Bedeaux and P.~Mazur.
\newblock {Renormalization of the diffusion coefficient in a fluctuating fluid
  I}.
\newblock {\em Physica}, 73:431--458, 1974.

\bibitem{DiffusionRenormalization_II}
P.~Mazur and D.~Bedeaux.
\newblock {Renormalization of the diffusion coefficient in a fluctuating fluid
  II}.
\newblock {\em Physica}, 75:79--99, 1974.

\bibitem{DiffusionRenormalization_III}
D.~Bedeaux and P.~Mazur.
\newblock {Renormalization of the diffusion coefficient in a fluctuating fluid
  III. Diffusion of a Brownian particle with finite size}.
\newblock {\em Physica A Statistical Mechanics and its Applications},
  80:189--202, 1975.

\bibitem{VACF_Langevin}
E.~J. Hinch.
\newblock {Application of the Langevin equation to fluid suspensions}.
\newblock {\em J. Fluid Mech.}, 72(03):499--511, 1975.

\bibitem{LangevinDynamics_Theory}
J.~N. Roux.
\newblock {Brownian particles at different times scales: a new derivation of
  the Smoluchowski equation}.
\newblock {\em Phys. A}, 188:526--552, 1992.

\bibitem{ConfinedDiffusion_2D}
Johannes Bleibel, A~Dom{\'\i}nguez, F~G{\"u}nther, J~Harting, and M~Oettel.
\newblock Hydrodynamic interactions induce anomalous diffusion under partial
  confinement.
\newblock {\em Soft matter}, 10(17):2945--2948, 2014.

\bibitem{DiffusionJSTAT}
Aleksandar Donev, Thomas~G Fai, and Eric Vanden-Eijnden.
\newblock {A reversible mesoscopic model of diffusion in liquids: from giant
  fluctuations to Fick's law}.
\newblock {\em J. Stat. Mech. Theory Exp.}, 2014(4):P04004, April 2014.

\bibitem{ActiveSuspensions}
Donald~L Koch and Ganesh Subramanian.
\newblock Collective hydrodynamics of swimming microorganisms: Living fluids.
\newblock {\em Annual Review of Fluid Mechanics}, 43:637--659, 2011.

\bibitem{Nanomotors_Kapral}
Raymond Kapral.
\newblock Perspective: Nanomotors without moving parts that propel themselves
  in solution.
\newblock {\em J. Chem. Phys.}, 138:020901, 2013.

\bibitem{HYDROPRO_Globular}
Jos{\'e} Garc{\'\i}a de~la Torre, Mar{\'\i}a~L Huertas, and Beatriz Carrasco.
\newblock Calculation of hydrodynamic properties of globular proteins from
  their atomic-level structure.
\newblock {\em Biophysical Journal}, 78(2):719--730, 2000.

\bibitem{RotationalBD_Torre}
Miguel~X Fernandes and Jos{\'e}~Garc{\'\i}a de~la Torre.
\newblock Brownian dynamics simulation of rigid particles of arbitrary shape in
  external fields.
\newblock {\em Biophysical journal}, 83(6):3039--3048, 2002.

\bibitem{Nanofluids_Review}
L.~Wang and M.~Quintard.
\newblock Nanofluids of the future.
\newblock {\em Advances in Transport Phenomena}, pages 179--243, 2009.

\bibitem{Nanofluidics_Review}
L.~Bocquet and E.~Charlaix.
\newblock {Nanofluidics, from bulk to interfaces}.
\newblock {\em Chemical Society Reviews}, 39(3):1073--1095, 2010.

\bibitem{NanoparticlesAtInterface}
Y.~Lin, H.~Skaff, T.~Emrick, A.~D. Dinsmore, and T.~P. Russell.
\newblock Nanoparticle assembly and transport at liquid-liquid interfaces.
\newblock {\em Science}, 299(5604):226, 2003.

\bibitem{BrownianDynamics_OrderNlogN}
A.~Sierou and J.~F. Brady.
\newblock {Accelerated Stokesian Dynamics simulations}.
\newblock {\em J. Fluid Mech.}, 448:115--146, 2001.

\bibitem{BrownianDynamics_OrderN}
J.~P. Hernandez-Ortiz, J.~J. de~Pablo, and M.~D. Graham.
\newblock {Fast Computation of Many-Particle Hydrodynamic and Electrostatic
  Interactions in a Confined Geometry}.
\newblock {\em Phys. Rev. Lett.}, 98(14):140602, 2007.

\bibitem{ExtraDiffusion_Vailati}
D.~Brogioli and A.~Vailati.
\newblock {Diffusive mass transfer by nonequilibrium fluctuations: Fick's law
  revisited}.
\newblock {\em Phys. Rev. E}, 63(1):12105, 2000.

\bibitem{DiffusionRenormalization}
A.~Donev, A.~L. Garcia, Anton de~la Fuente, and J.~B. Bell.
\newblock {Enhancement of Diffusive Transport by Nonequilibrium Thermal
  Fluctuations}.
\newblock {\em J. of Statistical Mechanics: Theory and Experiment},
  2011:P06014, 2011.

\bibitem{Nanopore_Fluctuations}
F.~Detcheverry and L.~Bocquet.
\newblock Thermal fluctuations in nanofluidic transport.
\newblock {\em Phys. Rev. Lett.}, 109:024501, 2012.

\bibitem{FluctHydroNonEq_Book}
J.~M.~O.~De Zarate and J.~V. Sengers.
\newblock {\em {Hydrodynamic fluctuations in fluids and fluid mixtures}}.
\newblock Elsevier Science Ltd, 2006.

\bibitem{LongRangeCorrelations_MD}
J.~R. Dorfman, T.~R. Kirkpatrick, and J.~V. Sengers.
\newblock {Generic long-range correlations in molecular fluids}.
\newblock {\em Annual Review of Physical Chemistry}, 45(1):213--239, 1994.

\bibitem{GiantFluctuations_Universal}
D.~Brogioli, A.~Vailati, and M.~Giglio.
\newblock Universal behavior of nonequilibrium fluctuations in free diffusion
  processes.
\newblock {\em Phys. Rev. E}, 61:R1--R4, 2000.

\bibitem{FractalDiffusion_Microgravity}
A.~Vailati, R.~Cerbino, S.~Mazzoni, C.~J. Takacs, D.~S. Cannell, and M.~Giglio.
\newblock {Fractal fronts of diffusion in microgravity}.
\newblock {\em Nature Communications}, 2:290, 2011.

\bibitem{SIBM_Brownian}
P.~J. Atzberger.
\newblock {A note on the correspondence of an immersed boundary method
  incorporating thermal fluctuations with Stokesian-Brownian dynamics}.
\newblock {\em Physica D: Nonlinear Phenomena}, 226(2):144--150, 2007.

\bibitem{BrownianBlobs}
S.~Delong, F.~Balboa Usabiaga, R.~Delgado-Buscalioni, B.~E. Griffith, and
  A.~Donev.
\newblock {Brownian Dynamics without Green's Functions}.
\newblock {\em J. Chem. Phys.}, 140(13):134110, 2014.
\newblock Software available at
  \url{https://github.com/stochasticHydroTools/FIB}.

\bibitem{DDFT_Hydro}
A.~Donev and E.~Vanden-Eijnden.
\newblock {Dynamic Density Functional Theory with hydrodynamic interactions and
  fluctuations}.
\newblock {\em J. Chem. Phys.}, 140(23):234115, 2014.

\bibitem{DDFT_Hydro_Lowen}
M~Rex and H~L{\"o}wen.
\newblock Dynamical density functional theory for colloidal dispersions
  including hydrodynamic interactions.
\newblock {\em The European Physical Journal E}, 28(2):139--146, 2009.

\bibitem{DDFT_Pep}
Pep Espa{\~n}ol and Hartmut L{\"o}wen.
\newblock Derivation of dynamical density functional theory using the
  projection operator technique.
\newblock {\em J. Chem. Phys.}, 131:244101, 2009.

\bibitem{StokesEinstein}
F.~Balboa Usabiaga, X.~Xie, R.~Delgado-Buscalioni, and A.~Donev.
\newblock {The Stokes-Einstein Relation at Moderate Schmidt Number}.
\newblock {\em J. Chem. Phys.}, 139(21):214113, 2013.

\bibitem{DirectForcing_Balboa}
F.~Balboa Usabiaga, I.~Pagonabarraga, and R.~Delgado-Buscalioni.
\newblock {Inertial coupling for point particle fluctuating hydrodynamics}.
\newblock {\em J. Comp. Phys.}, 235:701--722, 2013.

\bibitem{Ohlinger2012}
Alexander Ohlinger, Andras Deak, Andrey~a. Lutich, and Jochen Feldmann.
\newblock {Optically trapped gold nanoparticle enables listening at the
  microscale}.
\newblock {\em Phys. Rev. Lett.}, 108(1):1--5, 2012.

\bibitem{Kirchner2014}
Silke~R. Kirchner, Spas Nedev, Sol Carretero-Palacios, Andreas Mader, Madeleine
  Opitz, Theobald Lohm\"{u}ller, and Jochen Feldmann.
\newblock {Direct optical monitoring of flow generated by bacterial flagellar
  rotation}.
\newblock {\em Appl. Phys. Lett.}, 104(9):0--5, 2014.

\bibitem{CompressibleBlobs}
F.~Balboa~Usabiaga and R.~Delgado-Buscalioni.
\newblock {A minimal model for acoustic forces on Brownian particles}.
\newblock {\em Phys. Rev. E}, 88:063304, 2013.

\bibitem{ISIBM}
F.~Balboa Usabiaga, R.~Delgado-Buscalioni, B.~E. Griffith, and A.~Donev.
\newblock {Inertial Coupling Method for particles in an incompressible
  fluctuating fluid}.
\newblock {\em Comput. Methods Appl. Mech. Engrg.}, 269:139--172, 2014.
\newblock Code available at \url{https://code.google.com/p/fluam}.

\bibitem{StochasticImmersedBoundary}
P.~J. Atzberger, P.~R. Kramer, and C.~S. Peskin.
\newblock {A stochastic immersed boundary method for fluid-structure dynamics
  at microscopic length scales}.
\newblock {\em J. Comp. Phys.}, 224:1255--1292, 2007.

\bibitem{LB_IB_Points}
R.~W. Nash, R.~Adhikari, and M.~E. Cates.
\newblock Singular forces and pointlike colloids in lattice boltzmann
  hydrodynamics.
\newblock {\em Physical Review E}, 77(2):026709, 2008.

\bibitem{LB_SoftMatter_Review}
B.~D{\"u}nweg and A.J.C. Ladd.
\newblock {Lattice Boltzmann simulations of soft matter systems}.
\newblock {\em Adv. Comp. Sim. for Soft Matter Sciences III}, pages 89--166,
  2009.

\bibitem{ForceCoupling_Fluctuations}
Eric~E. Keaveny.
\newblock Fluctuating force-coupling method for simulations of colloidal
  suspensions.
\newblock {\em J. Comp. Phys.}, 269(0):61 -- 79, 2014.

\bibitem{Green1952}
M.S. Green.
\newblock {Markoff Random Processes and the Statistical Mechanics of
  Time-Dependent Phenomena}.
\newblock {\em J. Chem. Phys.}, 20:1281, 1952.

\bibitem{Zwanzig1961}
R~Zwanzig.
\newblock {Memory effects in irreversible thermodynamics}.
\newblock {\em Phys. Rev.}, 124:983, 1961.

\bibitem{Grabert1982}
H.~Grabert.
\newblock {\em Projection Operator Techniques in Nonequilibrium Statistical
  Mechanics}.
\newblock Springer Verlag, Berlin, 1982.

\bibitem{DiscreteLLNS_Espanol}
P.~Espa{\~n}ol, J.G. Anero, and I.~Z{\'u}{\~n}iga.
\newblock {Microscopic derivation of discrete hydrodynamics}.
\newblock {\em J. Chem. Phys.}, 131:244117, 2009.

\bibitem{FluctDiff_FEM}
J.A. de~la Torre, P.~Espa\ nol, and A.~Donev.
\newblock {Finite element discretization of non-linear diffusion equations with
  thermal fluctuations}.
\newblock {\em J. Chem. Phys.}, 142(9):094115, 2015.

\bibitem{CoarseGraining_Pep}
P.~Espa{\~n}ol.
\newblock Statistical mechanics of coarse-graining.
\newblock {\em Novel Methods in Soft Matter Simulations}, pages 69--115, 2004.

\bibitem{Hijon2010}
C.~Hij\'{o}n, E.~Vanden-Eijnden, R.~Delgado-Buscalioni, and P.~Espa\~{n}ol.
\newblock {Mori-Zwanzig formalism as a practical computational tool.}
\newblock {\em Faraday Discuss.}, 144(1):301--22; discussion 323--45, 467--81,
  January 2010.

\bibitem{Kirkwood1946}
J.G Kirkwood.
\newblock {The Statistical Mechanical Theory of Transport Processes. I. General
  Theory}.
\newblock {\em J. Chem. Phys.}, 14:180, 1946.

\bibitem{Espanol1993}
P.~Espa\~{n}ol and H.C. \"{O}ttinger.
\newblock {On the interpretation of random forces derived by projection
  operators}.
\newblock {\em Zeitschrift f\"{u}r Phys. B Condens. Matter}, 90(3):377--385,
  1993.

\bibitem{Landau1959}
L.~D. Landau and E.~M. Lifshitz.
\newblock {\em Fluid Mechanics (First Edition)}.
\newblock Pergamon Press, 1959.

\bibitem{DiscreteDiffusion_Espanol}
P.~Espa{\~n}ol and I.~Z{\'u}{\~n}iga.
\newblock On the definition of discrete hydrodynamic variables.
\newblock {\em J. Chem. Phys}, 131:164106, 2009.

\bibitem{Ray1999}
J~R Ray and H~Zhang.
\newblock {Correct microcanonical ensemble in molecular dynamics.}
\newblock {\em Phys. Rev. E. Stat. Phys. Plasmas. Fluids. Relat. Interdiscip.
  Topics}, 59(5 Pt A):4781--4785, 1999.

\bibitem{Espanol2009}
P.~Espa\~{n}ol and I.~Z\'{u}\~{n}iga.
\newblock {On the definition of discrete hydrodynamic variables.}
\newblock {\em J. Chem. Phys.}, 131(16):164106, October 2009.

\bibitem{JaimeThesis}
Jaime~Arturo de~la Torre~Rodriguez.
\newblock {\em Top-down and Bottom-up Approaches to Discrete Diffusion Models}.
\newblock PhD thesis, Universidad Nacional de Educacion a Distancia (UNED),
  2015.

\bibitem{Espanol1998a}
P.~Espa\~{n}ol.
\newblock {Stochastic differential equations for non-linear hydrodynamics}.
\newblock {\em Phys. A Stat. Mech. its Appl.}, 248(1):77--96, 1998.

\bibitem{BrennerModification_Ottinger}
H.~C. \"Ottinger, H.~Struchtrup, and M.~Liu.
\newblock Inconsistency of a dissipative contribution to the mass flux in
  hydrodynamics.
\newblock {\em Phys. Rev. E}, 80:056303, 2009.

\bibitem{SELM}
P.~J. Atzberger.
\newblock {Stochastic Eulerian-Lagrangian Methods for Fluid-Structure
  Interactions with Thermal Fluctuations}.
\newblock {\em J. Comp. Phys.}, 230:2821--2837, 2011.

\bibitem{LLNS_Staggered}
F.~Balboa Usabiaga, J.~B. Bell, R.~Delgado-Buscalioni, A.~Donev, T.~G. Fai,
  B.~E. Griffith, and C.~S. Peskin.
\newblock {Staggered Schemes for Fluctuating Hydrodynamics}.
\newblock {\em SIAM J. Multiscale Modeling and Simulation}, 10(4):1369--1408,
  2012.

\bibitem{DePablo2001}
Juan~Jos\'{e} {De Pablo} and Hans~Christian \"{O}ttinger.
\newblock {Atomistic approach to general equation for the nonequilibrium
  reversible-irreversible coupling}.
\newblock {\em J. Nonnewton. Fluid Mech.}, 96(1-2):137--162, 2001.

\bibitem{SELM_Reduction}
G.~Tabak and P.J. Atzberger.
\newblock Systematic stochastic reduction of inertial fluid-structure
  interactions subject to thermal fluctuations.
\newblock {\em arXiv preprint arXiv:1211.3798}, 2013.

\bibitem{IBM_PeskinReview}
C.S. Peskin.
\newblock {The immersed boundary method}.
\newblock {\em Acta Numerica}, 11:479--517, 2002.

\bibitem{StokesEinstein_BCs}
J.~T. Hynes, R.~Kapral, and M.~Weinberg.
\newblock {Molecular theory of translational diffusion: Microscopic
  generalization of the normal velocity boundary condition}.
\newblock {\em J. Chem. Phys.}, 70(3):1456, February 1979.

\bibitem{StokesEinstein_MD}
J.R. Schmidt and J.L. Skinner.
\newblock Hydrodynamic boundary conditions, the stokes--einstein law, and
  long-time tails in the brownian limit.
\newblock {\em J. Chem. Phys.}, 119:8062, 2003.

\bibitem{DFDB}
S.~Delong, B.~E. Griffith, E.~Vanden-Eijnden, and A.~Donev.
\newblock {Temporal Integrators for Fluctuating Hydrodynamics}.
\newblock {\em Phys. Rev. E}, 87(3):033302, 2013.

\bibitem{MultiscaleIntegrators}
S.~Delong, Y.~Sun, B.~E. Griffith, E.~Vanden-Eijnden, and A.~Donev.
\newblock {Multiscale temporal integrators for fluctuating hydrodynamics}.
\newblock {\em Phys. Rev. E}, 90:063312, 2014.
\newblock Software available at
  \url{https://github.com/stochasticHydroTools/MixingIBAMR}.

\bibitem{LLNS_S_k}
A.~Donev, E.~Vanden-Eijnden, A.~L. Garcia, and J.~B. Bell.
\newblock {On the Accuracy of Explicit Finite-Volume Schemes for Fluctuating
  Hydrodynamics}.
\newblock {\em Communications in Applied Mathematics and Computational
  Science}, 5(2):149--197, 2010.

\end{thebibliography}

\end{document}